\documentclass[twocolumn,times]{aastex6}

\usepackage{graphicx}
\usepackage{amsmath,amssymb,amstext}
\usepackage{booktabs}

\newcommand{\neii}{\,\hbox{[\ion{Ne}{2}]}}
\newcommand{\neiii}{\,\hbox{[\ion{Ne}{3}]}}
\newcommand{\nev}{\,\hbox{[\ion{Ne}{5}]}}

\newcommand{\siv}{\,\hbox{[\ion{S}{4}]}}
\newcommand{\siii}{\,\hbox{[\ion{S}{3}]}}
\newcommand{\oiv}{\,\hbox{[\ion{O}{4}]}}
\newcommand{\oi}{\,\hbox{[\ion{O}{1}]}}
\newcommand{\oii}{\,\hbox{[\ion{O}{2}]}}
\newcommand{\oiii}{\,\hbox{[\ion{O}{3}]}}
\newcommand{\niii}{\,\hbox{[\ion{N}{3}]}}
\newcommand{\nii}{\,\hbox{[\ion{N}{2}]}}
\newcommand{\cii}{\,\hbox{[\ion{C}{2}]}}
\newcommand{\ci}{\,\hbox{[\ion{C}{1}]}}

\newcounter{magicrownumbers}
\newcommand\rownumber{\stepcounter{magicrownumbers}\arabic{magicrownumbers}}
\shorttitle{Far-infrared line spectra of Seyfert galaxies}
\shortauthors{Fern\'andez-Ontiveros et al.}

\begin{document}

\DeclareGraphicsExtensions{.pdf,.gif,.jpg}

\title{Far-infrared line spectra of active galaxies from the \textit{Herschel}/PACS Spectrometer: the complete database}



\author{Juan Antonio Fern\'andez-Ontiveros\altaffilmark{1} and Luigi Spinoglio}
\affil{Istituto di Astrofisica e Planetologia Spaziali, INAF, Via Fosso del Cavaliere 100, I--00133 Roma, Italy}

\author{Miguel Pereira-Santaella}
\affil{Centro de Astrobiolog\'ia (CSIC/INTA), Ctra. de Torrej\'on a Ajalvir, km 4, E--28850 Torrej\'on de Ardoz, Madrid, Spain}

\author{Matthew A. Malkan}
\affil{Astronomy Division, University of California, Los Angeles, CA 90095-1547, USA}

\author{Paola Andreani}
\affil{European Southern Observatory, Karl-Schwarzschild-Stra{\ss}e 2, D--85748, Garching, Germany}

\and

\author{Kalliopi M. Dasyra\altaffilmark{2}}
\affil{Department of Astrophysics, Astronomy \& Mechanics, Faculty of Physics, University of Athens, Panepistimiopolis Zografos 15784, Greece}

\altaffiltext{1}{j.a.fernandez.ontiveros@iaps.inaf.it / j.a.fernandez.ontiveros@gmail.com}
\altaffiltext{2}{Observatoire de Paris, LERMA (CNRS:UMR8112), 61 Av. de l'Observatoire, F--75014, Paris, France}

\clearpage

\begin{abstract}
We present a coherent database of spectroscopic observations of far-IR fine-structure lines from the \textit{Herschel}/PACS archive for a sample of 170 local AGN, plus a comparison sample of 20 starburst galaxies and 43 dwarf galaxies. Published \textit{Spitzer}/IRS and \textit{Herschel}/SPIRE line fluxes are included to extend our database to the full $10$--$600\, \rm{\micron}$ spectral range. The observations are compared to a set of \textsc{Cloudy} photoionisation models to estimate the above physical quantities through different diagnostic diagrams.
We confirm the presence of a stratification of gas density in the emission regions of the galaxies, which increases with the ionisation potential of the emission lines. The new \oiv$_{25.9\, \rm{\micron}}$/\oiii$_{88\, \rm{\micron}}$ {\it vs} \neiii$_{15.6\, \rm{\micron}}$/\neii$_{12.8\, \rm{\micron}}$ diagram is proposed as the best diagnostic to separate: \textit{i)} AGN activity from any kind of star formation; and \textit{ii)} low-metallicity dwarf galaxies from starburst galaxies. Current stellar atmosphere models fail to reproduce the observed \oiv$_{25.9\, \rm{\micron}}$/\oiii$_{88\, \rm{\micron}}$ ratios, which are much higher when compared to the predicted values. Finally, the (\neiii$_{15.6\, \rm{\micron}}$+\neii$_{12.8\, \rm{\micron}}$)/(\siv$_{10.5\, \rm{\micron}}$+\siii$_{18.7\, \rm{\micron}}$) ratio is proposed as a promising metallicity tracer to be used in obscured objects, where optical lines fail to accurately measure the metallicity. The diagnostic power of mid- to far-infrared spectroscopy shown here for local galaxies will be of crucial importance to study galaxy evolution during the dust-obscured phase at the peak of the star formation and black-hole accretion activity ($1 < z < 4$). This study will be addressed by future deep spectroscopic surveys with present and forthcoming facilities such as \textit{JWST}, ALMA, and \textit{SPICA}.
\end{abstract}


\section{Introduction}\label{intro}

Rest-frame mid- to far-infrared (IR) spectroscopy is a unique tool to study dust-enshrouded galaxies and, in particular, to disentangle the emission originated by star-forming activity from that originated by accretion onto supermassive black holes in the nuclei of active galaxies \citep[e.g.][]{sm92}. Optical/UV lines provide access only to relatively unobscured gas, hampering our ability to investigate the physics in obscured regions. The extinction is however negligible in the mid- to far-IR range, which contains several lines that provide information of the physical conditions even for heavily obscured gas (see Tables\,\ref{tbl_lines} and \ref{tbl_guide}). Therefore, mid- to far-IR lines are the key to investigate not only ``star formation'' in galaxies, but in general all the processes in the majority of galaxies that occur in a dust-embedded phase, and thus are hidden to optical studies. In the dark side of star formation and active galactic nuclei (AGN) activity we find critical aspects of these phenomena: e.g. deeply buried active nuclei, the onset of the star formation and AGN feeding and feedback through gas inflow and outflow. These processes have a major impact on galaxy evolution but cannot be addressed from an optical/UV perspective.

The potential of mid- to far-IR spectroscopy was already proven by the \textit{Infrared Space Observatory} \citep[\textit{ISO};][]{kes96}, using the Short Wavelength Spectrometer \citep[SWS;][]{deg96} and the Long Wavelength Spectrometer \citep[LWS;][]{cle96}. As mid- and far-IR fine-structure emission lines are sensitive to the physical conditions of the interstellar medium (ISM), these transitions can be used to investigate its different phases \citep[e.g.][]{stu00,neg01,spi05} and characterise the primary spectrum of the ionising radiation \citep[e.g.][]{ale99}. \textit{ISO} enabled the development of the first IR diagnostics to separate the AGN, the ISM, and the stellar contributions \citep[e.g.][]{stu02,bra08}, and shed light on the nature of ultra-luminous IR galaxies \citep[ULIRGs;][]{gen98}.

The full exploitation of the mid-IR window was possible thanks to the InfraRed Spectrograph \citep[IRS;][]{hou04} on-board the \textit{Spitzer Space Telescope} \citep{wer04}. The study of large samples of galaxies including ULIRGs, Quasars, Seyfert galaxies, radio, and starburst galaxies \citep{arm07,tom10,vei09,bra06,bs09} showed, e.g., the ability of the \nev$_{14.3, 24.3\, \rm{\micron}}$ and the \oiv$_{25.9\, \rm{\micron}}$ lines, associated to high-density and high-excitation photoionised gas ($n_{\rm e} \gtrsim 10^3\, \rm{cm^{-3}}$, I.P.\,$\gtrsim 54\, \rm{eV}$), to investigate: \textit{i)} the inner part of the narrow-line region (NLR) and identify optically-hidden AGN \citep{arm07}; \textit{ii)} the AGN contribution in ULIRGs \citep{vei09}; \textit{iii)} the AGN nature of radio-galaxies \citep{haa05,lei09}; \textit{iv)} the power of the extinction-free density tracers in the $10^2$--$10^4\, \rm{cm^{-3}}$ range based on the \siii$_{18.7, 33.5\, \rm{\micron}}$ and the \nev$_{14.3, 24.3\, \rm{\micron}}$ lines \citep{tom08,tom10}; \textit{v)} star formation tracers based, e.g., on \neii$_{12.8\, \rm{\micron}}$ and \neiii$_{15.6\, \rm{\micron}}$ \citep{ho07}, or polycyclic aromatic hydrocarbon (PAH) features \citep{odo09,sar09}; \textit{vi)} the warm molecular gas traced by H$_2$ lines \citep[e.g.][]{tom08,bau10}; \textit{vii)} the extreme ultraviolet (EUV) spectra of AGN revealed by high-ionisation fine-structure IR lines \citep{mel11}.

In the far-IR, the \textit{Herschel Space Observatory} \citep{pil10} provided a large gain in spectroscopic sensitivity when compared to \textit{ISO}, allowing to access this spectral range for a larger number of galaxies in the nearby Universe. The Photoconductor Array Camera and Spectrometer \citep[PACS;][]{pog10} and the Spectral and Photometric Imaging REceiver Fourier-transform spectrometer \citep[SPIRE;][]{gri10,nay10} sampled the $50$--$210\, \rm{\micron}$ and $200$--$670\, \rm{\micron}$ ranges, respectively. This spectral region covers the main cooling lines of photo-dissociation regions (PDR), \oi$_{63, 145\, \rm{\micron}}$ and \cii$_{158\, \rm{\micron}}$, which provide information of the cold neutral gas \citep{th85}, as well as the high-\textit{J} CO transitions \citep{kam14,kam15}, allowing the investigation of the excitation of molecular gas; the fine-structure lines of \oi$_{63\, \rm{\micron}}$ and \nii$_{122\, \rm{\micron}}$ as tracers of the IR luminosity and star formation rate in ULIRGs \citep{far13}; the origin of \cii$_{158\, \rm{\micron}}$ emission, mainly associated to the PDR \citep[][hereafter S15]{s15}; X-ray dissociation regions \citep[XDR;][]{spi12b}.


The aim of the present study is to extend for the first time, with a statistical approach, the spectroscopic work done by \textit{Spitzer} in the mid-IR to the longer wavelengths using \textit{Herschel}. Taking advantage of the \textit{Herschel} scientific archive, we extended the previous study presented in S15, limited to a sample of 26 Seyfert galaxies, to a total sample of 170 active galaxies, and a comparison sample of 20 starburst galaxies and 43 dwarf galaxies --\,the latter taken from the Dwarf Galaxy Survey \citep{mad13}. Of particular interest is the development of the diagnostics that will be exploited by future IR observatories, e.g. the \textit{James Webb Space Telescope} \citep[\textit{JWST};][]{gar06} and the \textit{SPace Infrared telescope for Cosmology and Astrophysics} \citep[\textit{SPICA};][]{swi09} for galaxies up to the peak of the star formation rate density (SFRD) at $1 < z < 4$, but also the Atacama Large Millimeter/submillimeter Array (ALMA) for far-IR rest-frame observations of galaxies at high-redshift ($z \gtrsim 3$). The combination of mid- and far-IR lines expands the possible diagnostics to, e.g., the \oiv$_{25.9\, \rm{\micron}}$/\oiii$_{88\, \rm{\micron}}$ ratio (hereafter \oiv$_{25.9}$/\oiii$_{88}$) that has been proposed as a powerful diagnostic to discriminate AGN from star formation activity (S15). The larger statistics in the present work allow us to perform a more robust analysis of the diagnostics already tested in S15, including new line ratios sensitive to metallicity and ionisation \citep{nag11,nag12}. We include in this work the results on the ``compact sample'' of 43 dwarf galaxies presented by \citet{cor15} to investigate how strong star formation activity in low-metallicity environments can be well separated from either AGN and ``normal'' starburst galaxies through mid- and far-IR line ratios. A set of \textsc{Cloudy} photoionisation models were developed using AGN and starburst galaxies as ionisation sources in order to interpret the behaviour of the observed line ratios and their dependence on density, ionisation parameter, and metallicity.


The text is organised as follows. The selection of the sample of galaxies is explained in Section\,\ref{sample}, Section\,\ref{obs} presents the data reduction and describes the catalogs taken from the literature and included in our study, the \textsc{Cloudy} photoionisation models are described in Section\,\ref{models}, the results of this work are detailed in Section\,\ref{results}, and the main conclusions are summarised in Section\,\ref{summ}. Due to the large dataset and associated tables needed to show the results of this study, the majority of the latter appear in their complete form only in the online version of this journal.


\section{Sample of galaxies}\label{sample}

\subsection{AGN sample}
We have assembled the largest sample of active galaxies with far-IR spectroscopy by {\it Herschel}/PACS that was possible from the \textit{Herschel} scientific archive (Table\,\ref{tbl_sample}), in spite of the fact that the {\it Herschel} mission did not observe, during its 3.5 years operational life any systematic and statistically complete sample of active galaxies.

To do so, we have selected all objects in the \citet{ver10} active galaxies catalog that are classified either as QSO or Seyfert galaxies that have been detected with a $S/N > 3$ in at least one of the {\it Herschel}/PACS spectral lines in Table\,\ref{tbl_lines}. The far-IR spectra of these objects were completed with published fine-structure mid-IR emission lines from \textit{Spitzer}/IRS detected with $S/N > 3$ (see Section\,\ref{obs_irs}), mostly in the high-resolution mode (hereafter HR) and in a few cases in low-resolution mode (hereafter LR). These galaxies belong to the following original catalogues: the third and fourth Cambridge catalogues of radio sources \citep[3C and 4C, respectively;][]{edg59,pil65}, the European Southern Observatory ESO/Uppsala survey of the Southern Hemisphere \citep{hol74}, the Fairall catalogue of galaxies \citep{far77}, the New General Catalogue and the Index Catalogues \citep[NGC and IC, respectively;][]{dre88,dre95}, the Infrared Astronomical Satellite (IRAS) point source and faint source catalogues \citep{hel88,mos90}, the Markarian catalogue \citep{mar89}, the Morphological Catalog of Galaxies \citep[MCG;][]{vor62}, the Parkes Radio Sources Catalogue \citep[PKS;][]{wri90}, the Palomar-Green Catalog of Ultraviolet-Excess Stellar Objects \citep[PG;][]{gre09}, the Uppsala General Catalogue of Galaxies \citep[UGC;][]{nil73}, the catalogue of southern peculiar galaxies and associations \citep[AM;][]{am87}, the atlas of peculiar galaxies \citep[Arp;][]{arp66}, and the Zwicky catalog of galaxies \citep{zwi96}.

This AGN sample includes a total of 170 galaxies: 54 Seyfert~1 galaxies (S1), 26 Seyfert~1 galaxies with hidden broad-line regions (S1h; broad lines detected in their IR and/or polarised light spectra), 57 Seyfert~2 galaxies (S2), and 33 LINERs. The redshift distribution of the AGN sample, shown in Fig\,\ref{fig:z_lum}a, covers the range $-0.0008 < z < 1.8264$, although most of the galaxies are located in the Local Universe with a median redshift of $0.0243$ and a median absolute deviation of $0.027$. The far-IR luminosity distribution, based on the flux of the continuum emission adjacent to the \cii$_{158\, \rm{\micron}}$ line, is also shown in Fig.\,\ref{fig:z_lum}b, and has a median value and a median absolute deviation of $\log(L_{\rm FIR} / \rm{erg\,s^{-1}}) = 40.49 \pm 0.61$, covering the $35.54 < log(L_{\rm FIR}/\rm{erg\,s^{-1}}) < 42.52$ range. The metallicities, compiled from the literature (see Table\,\ref{tbl_sample}), span the $0.24 < Z < 2.95\, \rm{Z_\odot}$ range, with a median value and a median absolute deviation of $0.91 \pm 0.21\, \rm{Z_\odot}$.

Apart from the sample of 170 AGN, there are 8 AGN from the \citet{ver10} catalog that were observed by PACS for which no fine-structure emission lines were detected at $S/N > 3$. Consequently, these 8 targets are not included in our main AGN sample, still we include for completeness upper limits for the PACS line fluxes and equivalent widths, PACS continuum measurements, and \textit{Spitzer} line fluxes in Tables\,\ref{tbl_pacs_flux}, \ref{tbl_pacs_width}, \ref{tbl_pacs_cont}, \ref{tbl_irs_flux}.


\subsection{Starburst sample}
As a comparison sample, we compiled a sample of 20 starburst galaxies and reduced the PACS spectroscopic observations from the \textit{Herschel} archive. From the starburst galaxy sample of \citet{bs09}, we have selected the objects classified as ``pure starburst galaxies'', discarding those objects with an AGN contribution. From the sample of \citet{ga09} we have explicitly included only those objects optically classified as ``\textsc{H\,ii}'' region galaxies --\,by their position in the BPT diagram \citep{bpt81}\,-- and without an AGN contribution in the mid-IR range, as provided by these authors. For the mid-IR spectroscopy we used the results from \textit{Spitzer}/IRS high-resolution spectra published in \citet{bs09} and \citet{ga09}. The final sample of 20 starburst galaxies covers a redshift range of $0.0001 < z < 0.0159$ with a median value of $0.0022$ and a median absolute deviation $0.0006$ (Fig.\,\ref{fig:z_lum}a). The far-IR luminosity distribution is based on the flux of the continuum emission adjacent to the \cii$_{158\, \rm{\micron}}$ line and covers the $38.90 < log(L_{\rm FIR}/\rm{erg\,s^{-1}}) < 41.15$ range, with a median value and a median absolute deviation of $\log(L_{\rm FIR} / \rm{erg\,s^{-1}}) = 39.90 \pm 0.40$ (Fig.\,\ref{fig:z_lum}b). The starburst sample span a metallicity range of $0.46 < Z < 2.75\, \rm{Z_\odot}$, with a median value and a median absolute deviation of $1.10 \pm 0.29\, \rm{Z_\odot}$.(see Table\,\ref{tbl_sample}).

\subsection{Dwarf galaxy sample}
\begin{figure*}
  \includegraphics[width=0.50\textwidth]{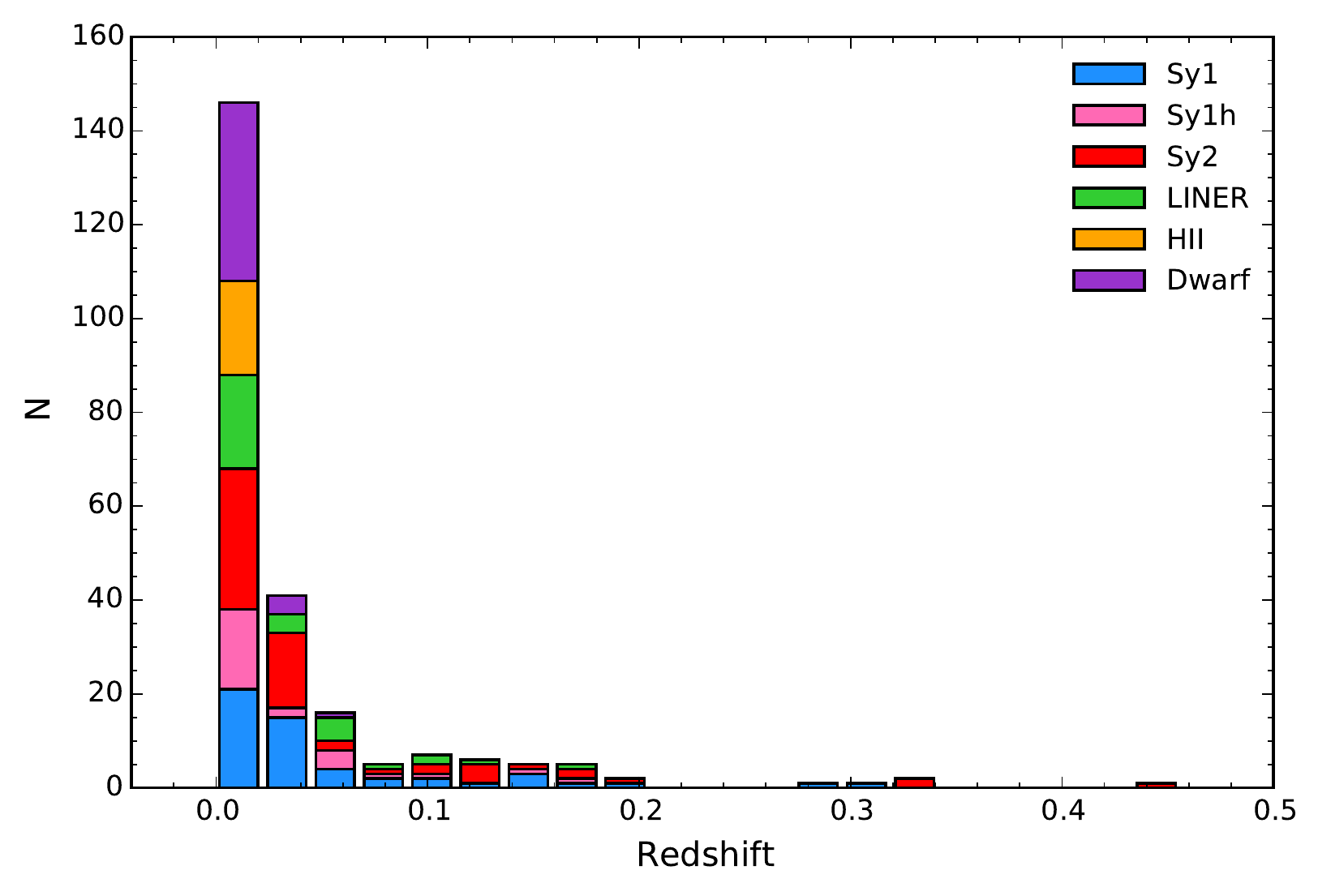}
  \includegraphics[width=0.50\textwidth]{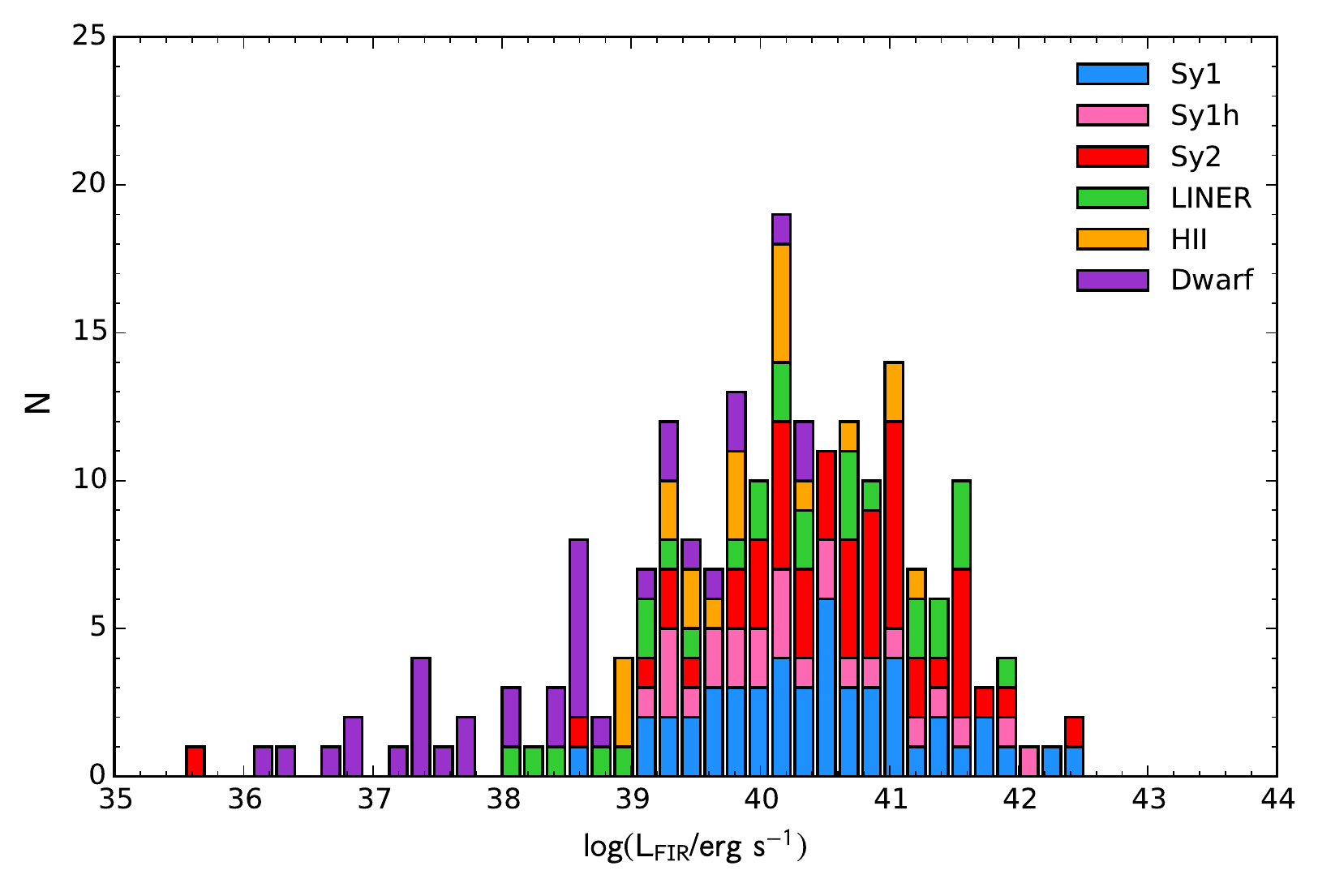}
  \caption{Redshift and luminosity distributions for the AGN sample. The colours correspond to the different spectral types: Seyfert 1 (blue), Seyfert 1 with hidden broad line regions (pink), Seyfert 2 (red), LINER (green), starburst galaxies (yellow), and dwarf galaxies (purple). 
{\bf Left (a):} Redshift distribution. Five objects with $z > 0.5$ are out of the redshift range shown in the plot. 
{\bf Right (b):} Far-IR luminosity ($L_{\rm FIR}$) distribution, based on the continuum flux adjacent to the \cii$_{158\, \rm{\micron}}$ line, observed for most of the galaxies in the AGN and starburst samples, and \textit{Herschel}/PACS photometry at $160\, \rm{\micron}$ for the dwarf galaxy sample from \citet{rem15}.\label{fig:z_lum}}
\end{figure*}

We also compared our samples of AGN and starburst galaxies to the mid- and far-IR fine-structure line dataset for the ``Dwarf Galaxy Survey'' (DGS), presented in \citet{cor15}. The DGS includes blue compact dwarf galaxies with metallicities ranging from $1/50\, \rm{Z_\odot}$ to $1/3\, \rm{Z_\odot}$, selected from the Hamburg/SAO Survey and the First and Second Byurakan Surveys \citep{izo91,ugr03,mad13}. Here we include the 43 dwarf galaxies from the DGS classified as compact \citep{cor15}, i.e. excluding nearby galaxies (NGC\,4214 is the closest one at $D = 2.93\, \rm{Mpc}$). Dwarf galaxies probe a more extreme star formation environment at low metallicities when compared to the starburst galaxies. Their low metallicity might mimic the conditions of chemically unevolved galaxies at high redshift, and thus it is important to include these objects in our diagnostics in order to have a wider view of the star formation process. Far-IR continuum fluxes for the dwarf galaxies are taken from \citet{rem15}. The sample of dwarf galaxies covers a redshift range of $-0.0003 < z < 0.0454$, with a median value of $0.0048$ and a median absolute deviation of $0.0038$ (Fig.\,\ref{fig:z_lum}a). The far-IR luminosity distribution is based on the \textit{Herschel}/PACS photometry at $160\, \rm{\micron}$, published in \citet{rem15}, and covers the $36.13 < \log(L_{\rm FIR}/\rm{erg\,s^{-1}}) < 40.38$ range, with a median value and a median absolute deviation of $\log(L_{\rm FIR} / \rm{erg\,s^{-1}}) = 38.49 \pm 0.91$ (Fig.\,\ref{fig:z_lum}b). The dwarf galaxies included in this study have stellar masses in the $3 \times 10^6$--$3.4 \times 10^{10}\, \rm{M_\odot}$ range, thus overlapping with the masses of nearby starburst galaxies \citep{mad13,rem15}.

\subsection{\textit{ISO}/LWS data}
For comparison, 81 galaxies with far-IR line fluxes observed with \textit{ISO}/LWS, taken from \citet{bra08}, were considered only for those active galaxies present in the original selection \citep{ver10} and for local starburst galaxies, but were not observed by \textit{Herschel}/PACS. The mid-IR line fluxes were compiled by us from published \textit{Spitzer}/IRS spectra. These objects are shown only for comparison and are not included into our main sample, due to the larger aperture of \textit{ISO} compared with that of PACS, therefore the galaxies observed by \textit{ISO}/LWS are represented by open symbols in the figures shown in this work. In Appendix\,\ref{app} we investigate the possible effect of the large \textit{ISO} apertures on the fluxes of the far-IR fine-structure lines. Fig.\,\ref{fig:pacs_iso} shows the comparison of \textit{ISO}/LWS {\it vs} \textit{Herschel}/PACS \cii$_{158\, \rm{\micron}}$ line fluxes for the subsample 45 AGN and starburst galaxies that were observed by both facilities. PACS fluxes are slightly higher with a median value and a median absolute deviation of $F^{\rm LWS}_{\rm [CII]158}/F^{\rm PACS}_{\rm [CII]158} = 0.95 \pm 0.37$, thus the inclusion of \textit{ISO} data in our diagrams is reasonable since significant contamination from extended emission is not expected for our sample of galaxies.

\section{Observations}\label{obs}
The combination of \textit{Herschel}/PACS and \textit{Spitzer}/IRS allows us to cover the fine-structure emission lines from the mid- to the far-IR ($10$--$200\, \rm{\micron}$ in the rest-frame) for all the galaxies in the sample. This database was completed with the \textit{Herschel}/SPIRE published values of the \nii$_{205\, \rm{\micron}}$, and \ci$_{371, 609\, \rm{\micron}}$ line fluxes (mainly from \citealt{kam15}). 
In this Section we describe the reduction of the \textit{Herschel}/PACS observations and the characteristics of the different datasets collected from the literature.

\subsection{\textit{Herschel}/PACS Spectroscopy}\label{obs_pacs}
The far-IR spectra of the galaxies in the AGN and starburst samples were acquired with \textit{Herschel}/PACS, which includes an integral field unit spectrograph observing in the $\sim 50$--$200\, \rm{\micron}$ range, with $5 \times 5$ squared spaxel elements covering a field of view (FOV) of about $47'' \times 47''$ and a spectral resolving power in the range $R = 1000$--$4000$ ($\Delta v = 75$--$300\, \rm{km\,s^{-1}}$), depending on wavelength. The observations were retrieved from the \textit{Herschel Science Archive}. The data were reduced using our own pipeline based on the standard recipes included in the \textsc{hipe}\footnote{\textsc{hipe} is a joint development by the \textit{Herschel} Science Ground Segment Consortium, consisting of ESA, the NASA \textit{Herschel} Science Center, and the HIFI, PACS and SPIRE consortia.} v13.0.0 environment. The data reduction procedure includes outlier flagging, spectral flat-fielding, re-grid of the wavelength sampling, and flux calibration. Most of the targets were acquired using the chop-node mode and reduced with the background normalisation method: the off-source spectra are used to correct the spectral response of the on-source data and to perform the background subtraction. This is the standard procedure in the \textsc{hipe} pipeline for PACS since v13.0.0. Observations acquired with the unchopped mode include a dedicated off-source pointing, which was used to perform the background subtraction.

For each target, a final standard rebinned data cube was produced for the available fine-structure lines listed in Table\,\ref{tbl_lines}. The \cii$_{158\, \rm{\micron}}$ transition shows the best coverage with 159 AGN and 17 starburst galaxies observed and detected, while the \oiii$_{52\, \rm{\micron}}$ transition was detected only for 19 AGN and one of the starburst galaxies. The final spectra are extracted from the innermost $3 \times 3$ spaxels ($28\farcs2 \times 28\farcs2$), to avoid possible flux losses due to the known pointing jitter and offsets. In all cases, a point-source loss correction, implemented within \textsc{hipe}, is applied to the extracted spectra. For a total of 26 sources (all in the AGN sample), one or more lines were not detected in the $3 \times 3$ spectrum, but only in the central spaxel. This is typically the case of faint sources in which the spaxels surrounding the central one are dominated by noise, and thus only the central spaxel has a substantial contribution to the line emission. For these targets, the final spectra and derived fluxes in Table\,\ref{tbl_pacs_flux} are extracted only from the central spaxel (and indicated by a ``c'' table in the column ``Notes'').
A complete flux extraction has been performed for all galaxies using three different methods: \textit{a)} the central spaxel only; \textit{b)} the central $3 \times 3$ array; and \textit{c)} the whole detector area of $5 \times 5$ spaxels.The complete database is reported in Table\,\ref{tbl_pacs_flux_c35}. We note that the $5 \times 5$ fluxes are not corrected for point-source losses (this is not provided by \textsc{hipe} v13.0.0), which are expected to be of the order of $\sim 10\%$ for a point-like source according to the PACS manual.

The analysis was performed using our own routines developed in \textit{Python}\footnote{\url{https://www.python.org}}, and routines from the Astropy\footnote{\url{http://www.astropy.org}} package \citep{rob13}. In order to increase the S/N of the extracted spectra, we used a Wiener filter to remove those spurious spectral features characterised by a width smaller than the instrumental spectral resolution and a peak below $3 \times$\,\textsc{rms}. The latter are produced during the acquisition and/or reduction process and must be removed for a realistic estimate of the flux errors. Line fluxes measured before and after the filtering are in agreement, within the estimated uncertainty, while the S/N ratios improve by a factor of $\sim 3$ on average.

Following the approach in S15, the line fluxes are obtained by numerical integration across the line profiles (blue-shaded area in Figs\,\ref{fig_specIC4329A_c},\ref{fig_specIC4329A_3x3},\ref{fig_specIC4329A_5x5}). The continuum level is estimated from a 1D polynomial fit performed for the spectral channels located in the blue and the red wings adjacent to the line (black-solid line). A single Gaussian was fitted to the profile of each line in order to obtain the central wavelength, standard deviation, and Gaussian flux (red-solid line). The original unfiltered spectrum and its associated uncertainty are shown in grey in Figs\,\ref{fig_specIC4329A_c},\ref{fig_specIC4329A_3x3},\ref{fig_specIC4329A_5x5}. Additionally, $5 \time 5$ spectral maps for each of the lines covered are also provided in the style of Fig.\,\ref{fig_mapIC4329A}, including the Wiener filtered spectrum for the line (in blue) and the adjacent continuum (in black), and the unfiltered spectrum with its associated uncertainty (in grey). In order to facilitate the visualisation and comparison of different line profiles and intensities in these figures, the spectra are shown after continuum subtraction (continuum flux density is quoted in the lower part of the figure as $\sum F_C / \Delta \lambda$) and normalised by a certain factor (quoted in the upper left part of the figure).

In the electronic version of this journal we provide, for each of the far-IR fine-structure lines observed by PACS for the 170 AGN and 20 starburst galaxies in the sample: the line fluxes (Table\,\ref{tbl_pacs_flux}), the equivalent width values (Table\,\ref{tbl_pacs_width}), the continuum level (Table\,\ref{tbl_pacs_cont}), the standard deviation of the best-fit Gaussian profile (Table\,\ref{tbl_pacs_sigma}), and the corresponding figures for the central spaxel (Fig.\,Set\,\ref{fig_specIC4329A_c}), $3 \times 3$ array (Fig.\,Set\,\ref{fig_specIC4329A_3x3}), $5 \times 5$ array extracted spectra (Fig.\,Set\,\ref{fig_specIC4329A_5x5}) and the $5 \times 5$ array line maps (Fig.\,Set\,\ref{fig_mapIC4329A}). The \textit{Herschel}/PACS and \textit{Spitzer}/IRS line fluxes for the sample of dwarf galaxies are taken from \citet{cor15}, and are not reported in this publication. Upper limits are also provided for 8 AGN that were not detected by PACS.

\subsection{\textit{Spitzer}/IRS Spectroscopy}\label{obs_irs}
Table\,\ref{tbl_irs_flux}, in the electronic version of this journal, collects published mid-IR ($10$--$35\, \rm{\micron}$) fine-structure line fluxes measured with \textit{Spitzer}/IRS for our samples of AGN and starburst galaxies. HR observations are favoured, completed with LR spectra in those cases without HR observations (14 galaxies have both IRS-LR spectra and \textit{Herschel}/PACS). These values were complemented with unpublished IRS observations from the \textit{Spitzer} archive and reduced by us for 20 galaxies, following the same procedure as in \citet{p-s10} and \citet{p-s10b}. \textit{Spitzer}/IRS HR line fluxes for the samples of starburst and dwarf galaxies were taken from \citet{bs09} and \citet{ga09}, and \citet{cor15}, respectively.

\subsection{\textit{Herschel}/SPIRE spectroscopy}\label{obs_spire}
In Table\,\ref{tbl_n2c1} we compiled the published \textit{Herschel}/SPIRE fluxes of the \nii$_{205\, \rm{\micron}}$, \ci$_{371\, \rm{\micron}}$ and \ci$_{609\, \rm{\micron}}$ lines, for the samples of AGN, starburst, and dwarf galaxies. These values are mainly provided by \citet{kam15} and complemented with the data published in \citet{isr15}, \citet{p-s13}, \citet{p-s14}, and \citet{kam14}.

\section{Models}\label{models}

The photoionisation code \textsc{Cloudy}\footnote{Available at \url{http://www.nublado.org}} has been used in this study to model the physical conditions of the gas exciting the IR fine-structure lines. Calculations have been performed with its version C13.03, last described by \citet{fer13}, using the py\textsc{Cloudy} library \citep{mor13}. To reproduce the emission of the NLR associated to the nuclei of active galaxies, a grid of constant density models with a plane-parallel geometry and sampling hydrogen density ($n_{\rm H}$) in the $\log (n_{\rm H} / \rm{cm^{-3}}) = 1$ to $6$ range, has been built using AGN spectra as primary radiation sources. The AGN ionising continuum corresponds to a power law extending from the optical to the X-rays with a slope of $\alpha = -1.4$ ($S_\nu \propto \nu^\alpha$), and ionisation parameters ($U$\footnote{Dimensionless ionisation parameter as defined in \citet{fer13}.}) with values of $\log U = -2.0, -2.5, -3.0, -3.5$. For these models we choose a maximum column density of $N_{\rm H} = 10^{23}\, \rm{cm^{-2}}$ as stopping criterion of the spatial integration, representative of the column density found in NLR clouds \citep{moo94}. LINER galaxies typically show a steeper optical-to-UV continuum when compared with brighter AGN (\citealt{ho99,hal96}; see also \citealt{fo12,rmas12}). Following this, a similar grid of models covering the same density and ionisation parameter values but with a steeper ionising continuum ($\alpha = -3.5$), has been computed in order to reproduce the line ratios expected for a weaker ionising spectra, i.e. for a vanishing accretion disc \citep{ho08}. We note however that the excitation mechanism in LINERs is still a matter of debate \citep[][and references therein]{ho08} and different mechanisms might contribute \citep[e.g.][]{dop15}.

The star formation ionising spectrum was simulated using the \textsc{Starburst99} code \citep{lei99} for two cases: \textit{i)} a young burst of star formation with an age of $1\, \rm{Myr}$ and a metallicity of $Z = 0.004$ ($1/5\, \rm{Z_\odot}$), in order to produce a hard UV ionising spectrum in a low metallicity environment; and \textit{ii)} a continuous burst of star formation with an age of $20\, \rm{Myr}$ and solar metallicity (following our previous simulations in S15). In the first case we aim to reproduce the conditions in violent and short lived star formation events, like those in dwarf galaxies, thus we refer to these models as ``dwarf galaxy models'' hereafter. The second case aims to reproduce the more quiescent and continuous star formation in the disk of galaxies, characterised by a softer ionising continuum, accordingly these will be called ``starburst models'' hereafter. The metallicity of the stellar population determines the strength of the ionising continuum, thus we used sub-solar metallicity models for dwarf galaxies\footnote{The majority of dwarf galaxies in the sample show metallicities in the $0.001 \lesssim Z \lesssim 0.008$ range \citep{mad13}.}, and solar metallicities for starburst models. The ionising continuum is also dependent on the age of the stellar population for instant starburst models, but it is fairly independent of the age for continuum starburst models after the first few Myr \citep{lei99}. Thus the main results of this work will not change if a $\gtrsim 5\, \rm{Myr}$ continuum starburst model is considered as reference for the starburst galaxies.

We use models with plane-parallel geometry, constant pressure, initial densities in the $\log(n_{\rm H} / \rm{cm^{-3}}) = 1$ to $6$ range, and ionisation parameters in the $\log U = -2.0$ to $-4.5$ range. In both cases we assumed two intervals for the Kroupa initial mass function (IMF; with exponents $1.3$, $2.3$ and mass boundaries of $0.1$, $0.5$, and $100\, \rm{M_\odot}$), the 1994 Geneva tracks with standard mass-loss rates, and the Pauldrach/Hillier atmospheres, which take into account the effects of non-LTE and radiation driven winds. For the study of the dependence of line ratios with metallicity (Section\,\ref{metal}), the previous models were extended to the following metallicity values: $Z = 0.001$, $0.004$, $0.008$, $0.02$, and $0.04$. These values are based on the metallicities available for the Geneva stellar tracks with standard mass-loss rates, used in the \textsc{Starburst99} models to generate the starburst ionising spectra.

In dwarf galaxy and starburst models we chose a different stopping criteria based on the temperature, instead of the column density. The spatial integration was stopped at the radius where two different temperatures were reached, producing two kinds of models in each case: \textit{i)} models stopped at $T = 1000\, \rm{K}$ that only include the contribution of photoionised gas to the emission lines; and \textit{ii)} models extended to $T = 50\, \rm{K}$ that take into account the contribution from the PDR. In particular, the comparison of $1000\, \rm{K}$ and $50\, \rm{K}$ models allows us to identify which line ratios are affected by the contribution of PDR emission and, e.g., to investigate the origin of \cii$_{158\, \rm{\micron}}$, which is produced by both neutral gas and low-density, low-excitation ionised gas (see also S15). In both cases we used models with constant pressure, with the initial densities mentioned earlier, since they accommodate better the changes in the structure of the cloud when the calculation is extended deeper into the cold neutral gas, down to $50\, \rm{K}$, thus the predicted line ratios for the PDR are more reliable.

For AGN models we did not choose a temperature-based stopping criterion to differentiate the XDR contribution due to: \textit{i)} the simulations do not converge to temperatures lower than $\gtrsim 100\, \rm{K}$, due to the harder spectra of AGN and the X-ray contribution heating the neutral gas; \textit{ii)} further discussion in Sections\,\ref{dens_ion_temp}, \ref{dens_strat} show that the XDR contribution to neutral gas tracers, e.g. \ci$_{609/371}$ and \cii$_{158}$/\oi$_{63}$, is negligible, while the values found for these line ratios are in agreement with PDR emission.

All the models presented here to reproduce AGN (Seyfert and LINERs), dwarf galaxies (with and without PDR), and starburst galaxies (with and without PDR), aim to predict ideal cases for these ionising sources. Most of the objects in our sample are expected to show mixed contributions since our large aperture spectra include emission from the host galaxy and therefore star formation processes contribute to the emission of most lines. Thus, the observed line ratios are expected to lie among these models for most of the diagnostic diagrams.

\section{Results}\label{results}

The final database includes a variety of lines sensitive to different sources of excitation and physical conditions, from the high-ionisation \nev$_{14.3, 24.3\, \rm{\micron}}$ lines --\,indicative of AGN activity, to star formation/\textsc{H\,ii} regions tracers (e.g. \neii$_{12.8\, \rm{\micron}}$, \neiii$_{15.6\, \rm{\micron}}$, \oiii$_{88\, \rm{\micron}}$), and PDR sensitive lines (e.g. \oi$_{63, 145\, \rm{\micron}}$, \ci$_{371, 609\, \rm{\micron}}$, \cii$_{158\, \rm{\micron}}$, the latter also produced in \textsc{H\,ii} regions). In the following sections we will explore the dependence of line ratios on various physical quantities.

\subsection{Density, Ionisation, and Temperature}\label{dens_ion_temp}

\begin{figure*}
  \includegraphics[width=0.505\textwidth]{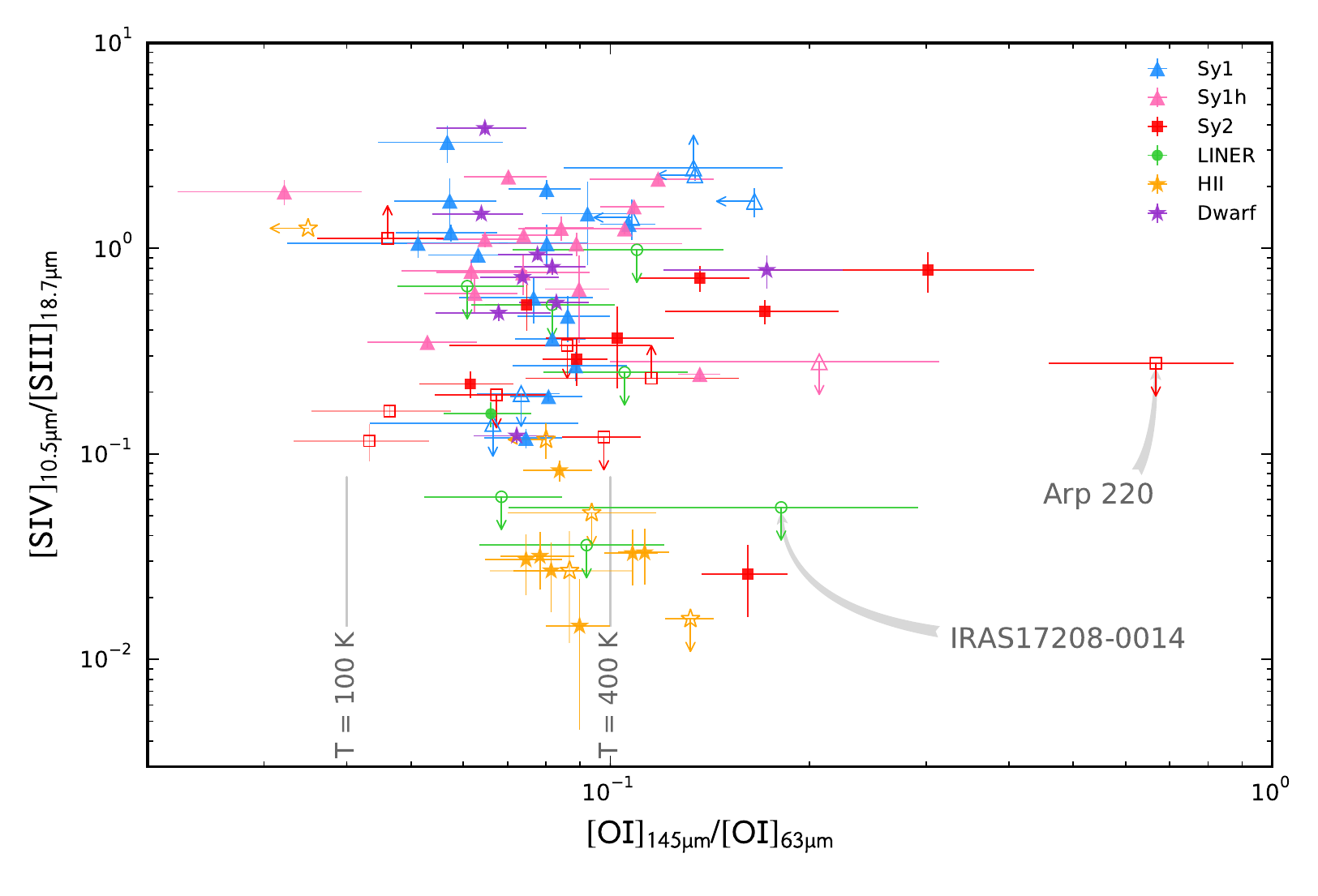}
  \includegraphics[width=0.495\textwidth]{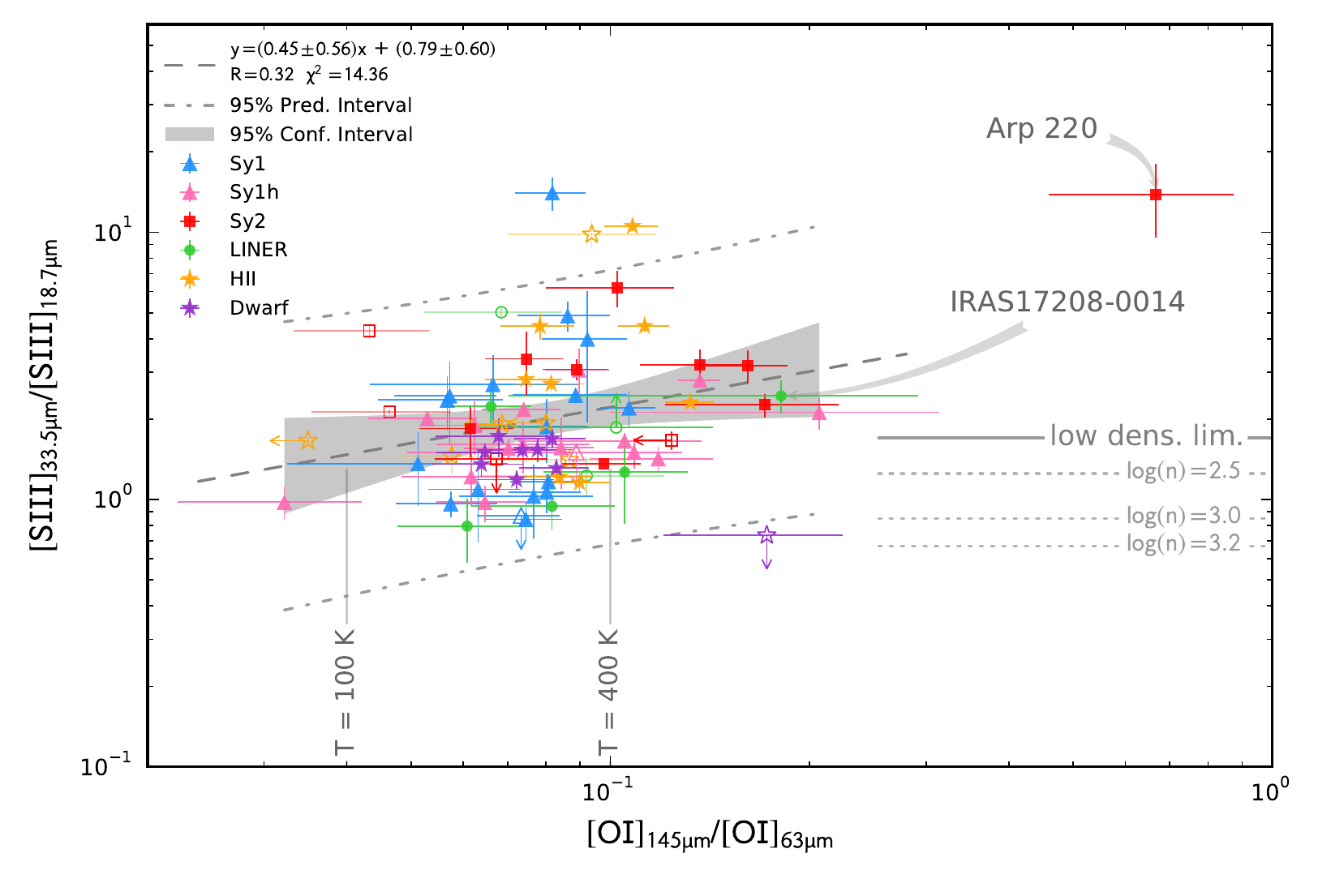}
  \caption{Ionisation and density \textit{vs} temperature. The symbols correspond to the different spectral types: Seyfert 1 (blue triangles), Seyfert 1 with hidden broad line regions (pink triangles), Seyfert 2 (red squares), LINER (green circles), starburst galaxies (yellow stars), dwarf galaxies (purple stars). Open symbols are used for upper and lower limits derived from PACS data, and for \textit{ISO}/LWS data. 
{\bf Left (a):} the [\textsc{O\,i}]$_{145/63}$ ratio {\it vs} the [\textsc{S\,iv}]$_{10.5}$/[\textsc{S\,iii}]$_{18.7}$ ratio. The [\textsc{S\,iv}]$_{10.5}$/[\textsc{S\,iii}]$_{18.7}$ is sensitive to the ionisation, while the [\textsc{O\,i}]$_{145/63}$ line ratio is sensitive to the temperature in the $100$ to $400\, \rm{K}$ range. The line ratios corresponding to these temperature values, assuming a density of $n_{\rm H_2} \approx 10^4\, \rm{cm^{-3}}$ in \citet{lis06}, are marked in the bottom part of the diagram. 
{\bf Right (b):} the [\textsc{O\,i}]$_{145/63}$ ratio {\it vs} the [\textsc{S\,iii}]$_{33.5/18.7}$ line ratio. The latter measures the electron density, whose values --\,assuming a gas in pure collisional regime at $T = 10^4\, \rm{K}$ as in S15\,-- are marked on the right part of the diagram. Arp\,220 shows a strong absorption profile in [\textsc{O\,i}]$_{63\, \rm{\micron}}$ and thus has been excluded from the correlation analysis.}
\label{fig:T_OI}
\end{figure*}

\begin{figure*}
  \centering
  \includegraphics[width=0.505\textwidth]{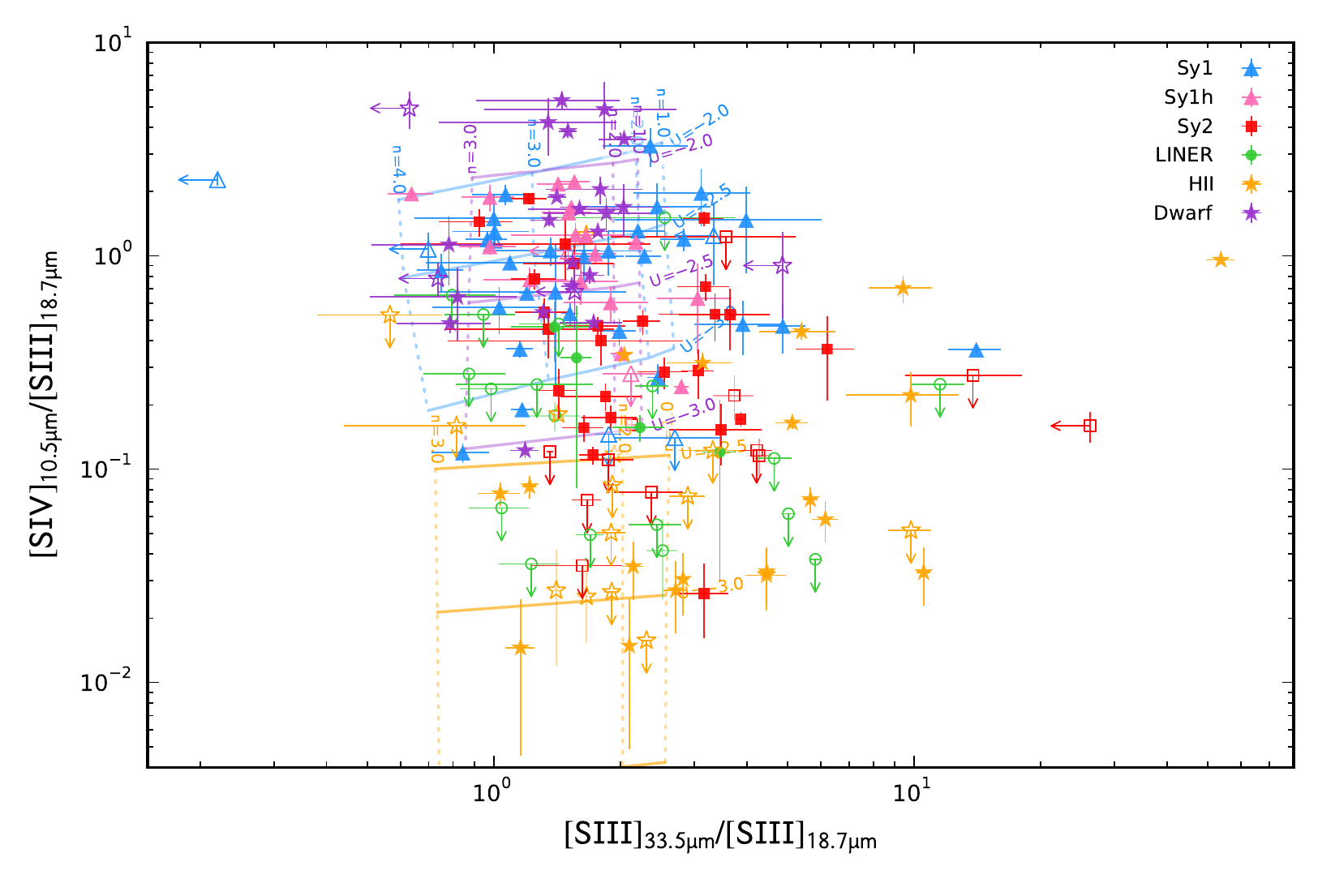}~
  \includegraphics[width=0.495\textwidth]{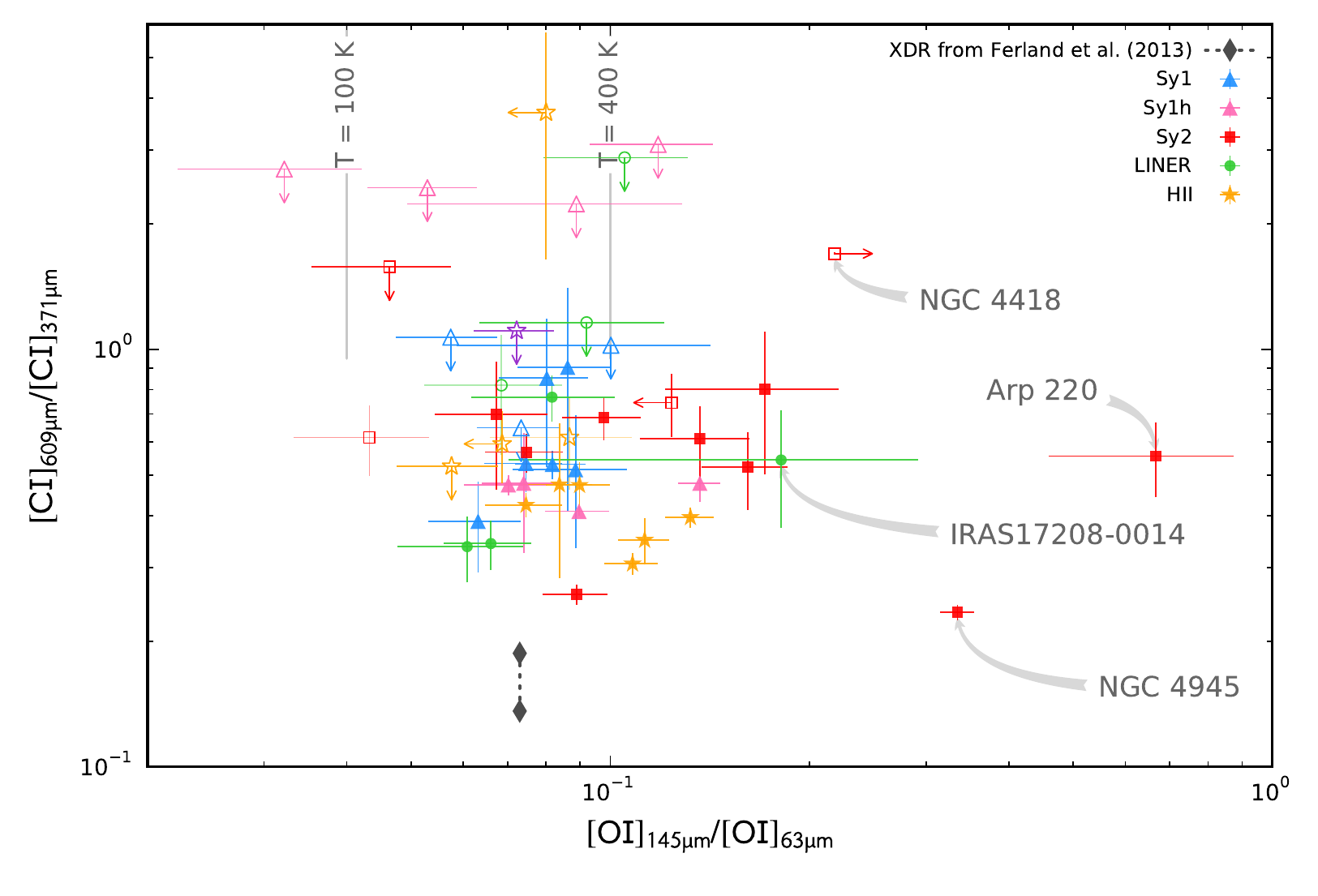}
  \caption{Ionisation, density, and PDR temperature-sensitive line ratios (same notations as in Fig.\,\ref{fig:T_OI}). Photoionisation models of AGN, starburst galaxies, and dwarf galaxies are shown as blue, yellow, and purple grids, respectively. The logarithmic values of the density ($n_{\rm H}$) and ionisation potential ($U$) of the photoionisation models are indicated in the figures. 
  \mbox{\bf Left (a):} The [\textsc{S\,iv}]$_{10.5}$/[\textsc{S\,iii}]$_{18.7}$ is sensitive to the ionisation, while the [\textsc{S\,iii}]$_{33.5/18.7}$ line ratio is sensitive to the electron density. 
  \mbox{\bf Right (b):} The [\textsc{C\,i}]$_{609/371}$ line ratio {\it vs} the [\textsc{O\,i}]$_{145/63}$ ratio, which is sensitive to the temperature in the $100$ to $400\, \rm{K}$ range. Higher [\textsc{C\,i}]$_{609/371}$ ratios correspond to lower temperatures within the $20$--$100\, \rm{K}$ range, according to the models in \citet{mei07}. Black diamonds indicate the predicted line ratios for an XDR, extracted from the table\,2 in \citet{fer13}.}
\label{fig:o1_o1vsc1_c1}
\end{figure*}

As shown in our previous work (S15), emission-line ratios obtained from pairs of mid- and far-IR lines from the same ionic species, e.g. \nii$_{205}$/\nii$_{122}$ (hereafter \nii$_{205/122}$), \siii$_{33.5/18.7}$, \oiii$_{88/52}$, and \nev$_{24.3/14.3}$, have the same ionisation potential but different critical densities (see Table\,\ref{tbl_lines}), and thus can be used to trace the densities of the ionised gas in the $n_{\rm H} \approx 10\, \rm{cm^{-3}}$ to $10^5\, \rm{cm^{-3}}$ range \citep[e.g.][]{rub94}. On the other hand, the cooling lines of the neutral gas, \oi$_{63, 145\, \rm{\micron}}$, \cii$_{158\, \rm{\micron}}$, and \ci$_{371, 609\, \rm{\micron}}$, provide density and temperature diagnostics when they are compared to PDR and XDR models \citep[e.g.][]{th85,lis06,mei07}. In particular, theoretical estimates based on the statistical equilibrium equations predict that the \oi$_{145/63}$ line ratio is, in the optically-thin limit, a good temperature tracer for the neutral gas in the $T \sim 100$--$400\, \rm{K}$ range at $n_{\rm H} \lesssim 10^4\, \rm{cm^{-3}}$ \citep{lis06}. The \ci$_{371, 609\, \rm{\micron}}$ lines, observed by \textit{Herschel}/SPIRE, are formed in the transition layer between ionised carbon and molecular CO gas (C$^+$/C/CO), typically above their critical densities ($n_{\rm H} \gtrsim 10^4\, \rm{cm^{-3}}$). The  line ratio is sensitive to the XDR contribution, since X-rays are able to penetrate deep into the cloud and warm all the neutral carbon, thus lowering the \ci$_{609/371}$ ratio as the temperature increases \citep{mei07,fer13}. In this Section we use our sample to review the possible correlation between density, ionisation, and neutral gas temperature already presented in S15. Thus, Figs\,\ref{fig:T_OI}a,b and \ref{fig:o1_o1vsc1_c1}a are extensions of the work initiated in S15 for a smaller sample of galaxies.


In Fig.\,\ref{fig:T_OI}a, the \siv$_{10.5}$/\siii$_{18.7}$ ratio --\,sensitive to the hardness of the ionising radiation\,-- is compared to the neutral gas temperature sensitive \oi$_{145/63}$ ratio. On the right panel, the \siii$_{33.5/18.7}$ ratio --\,sensitive to the density in the $10^2$--$10^4\, \rm{cm^{-3}}$ range\,-- is compared to the same temperature tracer, \oi$_{145/63}$. As mentioned earlier, open symbols correspond to \textit{ISO}/LWS observations \citep{bra08}, complemented with \textit{Spitzer}/IRS measurements in the literature, for those objects without \textit{Herschel}/PACS spectroscopy. As expected, AGN and starburst are separated by the \siv$_{10.5}$/\siii$_{18.7}$ ratio, sensitive to the ionisation parameter, but no correlation is found between both ratios, i.e. harder radiation fields are not associated with a warmer neutral ISM. On the other hand, density and temperature were tentatively correlated in S15. The extended sample does not show a significant correlation between temperature and density traced by \siii$_{33.5/18.7}$, with $\log \siii_{33.5/18.7} = (0.45 \pm 0.56) \log \oi_{145/63} + (0.79 \pm 0.60)$ and a low Spearman's rank correlation coefficient of $R = 0.32$.

\begin{figure*}
  \figurenum{1}
  \includegraphics[width=\textwidth]{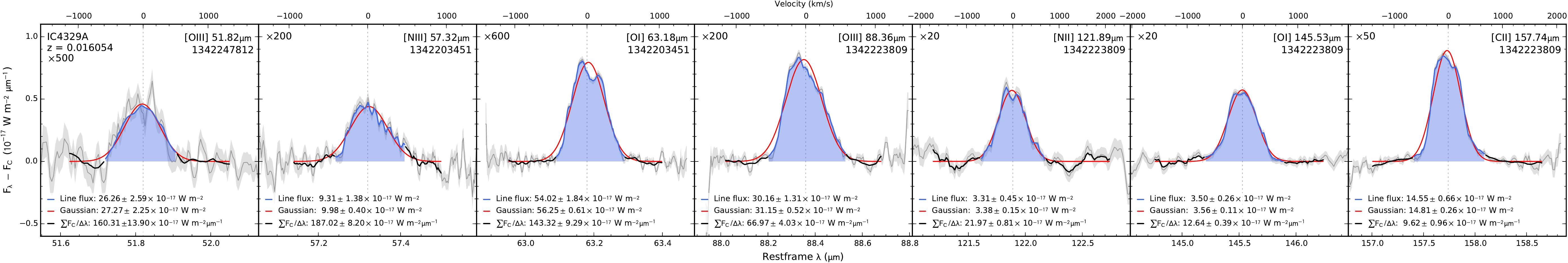}
  \caption{Spectra for the seven fine-structure emission lines covered by \textit{Herschel}/PACS, extracted from the central spaxel, e.g. for IC\,4329A. The continuum-subtracted, unfiltered spectra and its associated uncertainty are show in grey. The Wiener-filtered spectra (filter width: $2 \times$\,instrumental spectral \textsc{fwhm}; height: $3 \times$\textsc{rms}) are shown in blue (line spectral window) and black (continuum spectral window). The best Gaussian fit to the line plus continuum ranges are shown in red. In all cases, the continuum emission has been subtracted by fitting a 1-degree polynomial to the continuum windows, and the spectra are normalised by the factor indicated in the upper-left corner in each frame. Fluxes in the legend correspond to the numerical integration for the line (blue area), the best-fit Gaussian, and the average continuum flux density. The dotted-grey lines mark the theoretical rest-frame wavelengths for the seven transitions. The top-horizontal axis shows the velocity with respect to the theoretical wavelength. Figures for all the galaxies in the AGN and starburst samples are available in the Fig.\,Set\,\ref{fig_specIC4329A_c}.\label{fig_specIC4329A_c}}
\end{figure*}

\begin{figure*}
  \figurenum{2}
  \includegraphics[width=\textwidth]{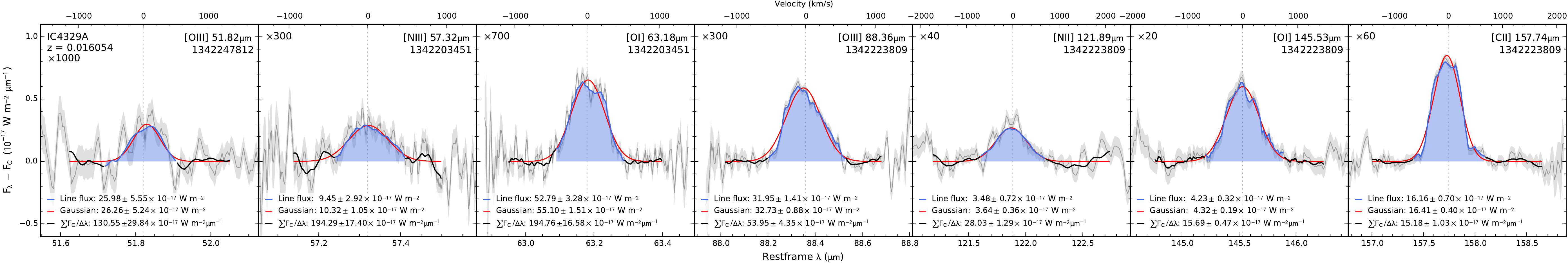}
  \caption{Spectra for the seven fine-structure emission lines covered by \textit{Herschel}/PACS, extracted from the inner $3 \times 3$ spaxel array, e.g. for IC\,4329A (same notations as in Fig.\,\ref{fig_specIC4329A_c}). Figures for all the galaxies in the AGN and starburst samples are available in the Fig.\,Set\,\ref{fig_specIC4329A_3x3}.\label{fig_specIC4329A_3x3}}
\end{figure*}

\begin{figure*}
  \figurenum{3}
  \includegraphics[width=\textwidth]{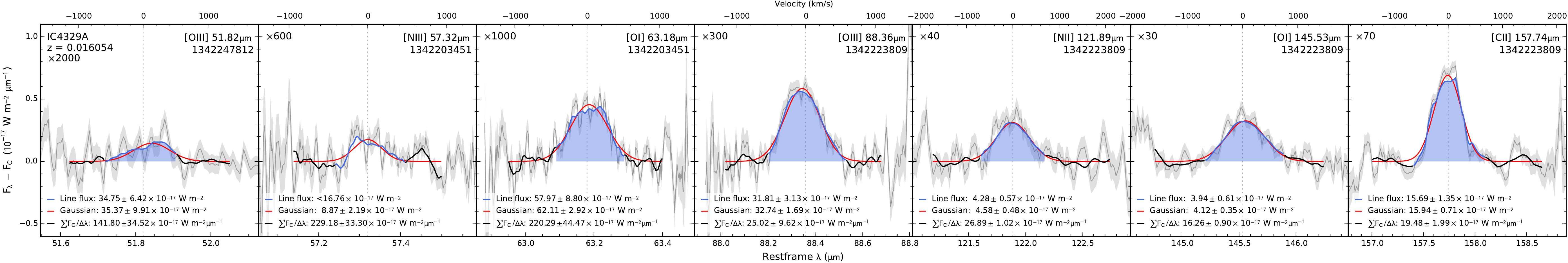}
  \caption{Spectra for the seven fine-structure emission lines covered by \textit{Herschel}/PACS, extracted from the full $5 \times 5$ spaxel array, e.g. for IC\,4329A (same notations as in Fig.\,\ref{fig_specIC4329A_c}). Figures for all the galaxies in the AGN and starburst samples are available in the Fig.\,Set\,\ref{fig_specIC4329A_5x5}.\label{fig_specIC4329A_5x5}}
\end{figure*}

A likely explanation for the lack of correlation between these line ratios is the different ISM phases traced, i.e. ionised gas in \siv$_{10.5}$/\siii$_{18.7}$ and \siii$_{33.5/18.7}$ \textit{vs} neutral gas in \oi$_{145/63}$. Figs\,\ref{fig:T_OI}a,b show that a harder ionising spectrum or a denser ionised gas does not implies a hotter temperature for the associated neutral gas, possibly suggesting a more complex ISM structure. Furthermore, we note that \oi$_{145/63}$ values above $\gtrsim 0.1$ would imply optically-thick emission \citep{th85}, but can be reconciled with optically-thin emission if the effect of heavy element opacity is considered \citep{rub83,rub85}. Specifically, the \oi$_{63\, \rm{\micron}}$ line can be affected by self-absorption even for small amounts of cold foreground gas \citep[$N_{\rm H} \sim 2 \times 10^{20}\, \rm{cm^{-2}}$;][]{lis06}. This effect is particularly dramatic in the cases of Arp\,220 (Figs\,\ref{fig:T_OI}a,b and Fig.\,\ref{fig:o1_o1vsc1_c1}b), NGC\,4945 and NGC\,4418 (Fig.\,\ref{fig:o1_o1vsc1_c1}b), and IRAS17208--0014 (Figs\,\ref{fig:T_OI}b). Their spectra for the central spaxel array --\,available as online material in Fig.\,Set\,\ref{fig_specIC4329A_c}.77 (NGC\,4418), \ref{fig_specIC4329A_c}.95 (NGC\,4945), \ref{fig_specIC4329A_c}.135 (IRAS17208--0014), \ref{fig_specIC4329A_c}.127 (Arp\,220)\,--  show a prominent absorption component in the \oi$_{63\, \rm{\micron}}$ line \citep[see also fig.\,11 in][for the cases of Arp\,220 and NGC\,4418]{gon12}, or even in the \oiii$_{88\, \rm{\micron}}$ line (IRAS17208--0014). A self-absorption contribution to the \oi$_{63\, \rm{\micron}}$ line would also play against the correlations explored in this Section. Furthermore, shocks might also play a role in the \oi$_{63\, \rm{\micron}}$ line emission \citep[see][]{hol89,lut03}.

Fig.\,\ref{fig:o1_o1vsc1_c1}a shows the  \siv$_{10.5}$/\siii$_{18.7}$ {\it vs} \siii$_{33.5/18.7}$ line ratios, i.e. ionisation {\it vs} density of the ionised gas. S1 galaxies, S1h galaxies, and dwarf galaxies, where the ionisation is stronger, show tentatively lower \siii$_{33.5/18.7}$ ratios with median values and median absolute deviations of $1.76 \pm 1.02$, $1.59 \pm 0.50$, $1.53 \pm 0.33$, respectively, when compared to starburst galaxies with $2.86 \pm 2.40$. However the difference is not statistically significant due to the high dispersion in the \siii$_{33.5/18.7}$ ratio, thus no correlation is found between ionisation and density. This lack of correlation can be explained by the combination of different ionisation parameter values and densities in the \textsc{Cloudy} photoionisation models shown in Fig.\,\ref{fig:o1_o1vsc1_c1}a. Most of the S1, S1h, and dwarf galaxies in our sample are in agreement with photoionisation models in the $-3.0 < \log U < -2.0$ and $1.0 < \log(n/\rm{cm^{-3}}) < 4.0$ ranges, starburst galaxies span the $-3.5 < \log U < -2.5$ and $1.0 < \log(n/\rm{cm^{-3}}) < 3.0$ ranges, while S2 galaxies and LINERs are found in both the AGN and the starburst domains. Note that galaxies with \siii$_{33.5/18.7}$ ratios $\gtrsim 2$ are above the low-density limit, thus \siii$_{33.5/18.7}$ is no longer sensitive to density for those cases.

Fig.\,\ref{fig:o1_o1vsc1_c1}b shows the \ci$_{609/371}$ {\it vs} the \oi$_{145/63}$ line ratios, both sensitive to the neutral gas temperature in the PDR, where \ci$_{609/371}$ decreases and \oi$_{145/63}$ increases with increasing temperature. Again, our sample does not show any correlation between these two ratios, probably due to the different density ranges probed by the \oi$_{145/63}$ ($n_{\rm H} \lesssim 10^4\, \rm{cm^{-3}}$) and the \ci$_{609/371}$ ($n_{\rm H} \gtrsim 10^4\, \rm{cm^{-3}}$) ratios, and the above mentioned self-absorption in the \oi$_{63\, \rm{\micron}}$ line. The \ci$_{371, 609\, \rm{\micron}}$ lines probe a denser and colder gas inside the cloud when compared to the \oi$_{63, 145\, \rm{\micron}}$ lines. Overall, most of the galaxies show the typical values expected for PDR emission, including AGN galaxies which show similar ratios when compared to starburst galaxies. A median \ci$_{609/371}$ ratio of $0.45$ with a median absolute deviation of $0.11$ are measured for the eight starburst galaxies detected, comparable to the median $0.53 \pm 0.21$ obtained for Seyfert and LINER galaxies. This implies that XDR emission, whose ratio is expected to be around \ci$_{609/371} \lesssim 0.19$ (see table\,2 in \citealt{fer13}; also simulations by \citealt{mei07}), does not have an important contribution --\,within the \textit{Herschel}/SPIRE aperture of $\sim 35''$\,-- to the neutral carbon emission for the AGN galaxies in our sample, which show \ci$_{609/371}$ ratios typical for PDR and similar to starburst galaxies.


\subsection{Density stratification}\label{dens_strat}

The observed properties of the emission lines are closely linked to the physical conditions of the region where they are originated. Lines with a high ionisation potential such as \nev$_{14.3, 24.3\, \rm{\micron}}$ ($97.1\, \rm{eV}$) can only be produced in the innermost region of the NLR, as they are powered by the strong ionising spectrum of the AGN. Thus they are expected to trace a denser gas when compared with \oiii$_{52, 88\, \rm{\micron}}$ ($35.1\, \rm{eV}$) and \siii$_{18.7, 33.5\, \rm{\micron}}$ ($23.3\, \rm{eV}$), produced in the outer NLR and \textsc{H\,ii} regions, and \nii$_{122, 205\, \rm{\micron}}$ ($14.5\, \rm{eV}$), produced also in low-density \textsc{H\,ii} regions. This density stratification was shown by S15 and is reviewed here using our larger sample.

\begin{figure*}
\centering
\includegraphics[width=8cm]{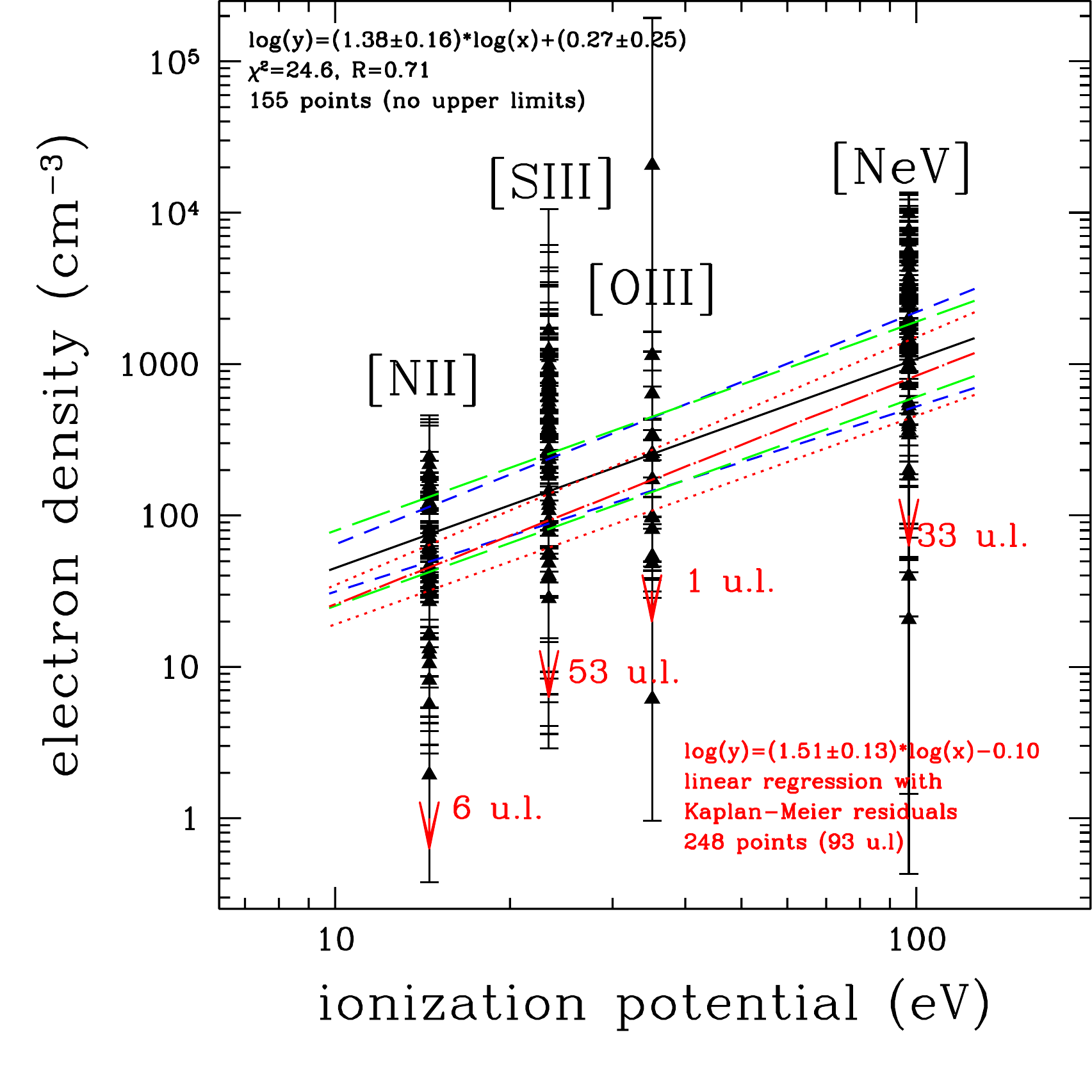}
\caption{Density stratification. Derived values for the electron density {\it vs} the ionisation potential for the galaxies in the sample with two detected lines of the same species. A weighted least-square fit (black-solid line) shows a correlation between density and ionisation. Blue- and green-dashed lines show the uncertainty associated with the slope and the intercept of the fit, respectively. The red dot-dashed line is the result of the linear regression using the Kaplan-Meier residuals, and the red dotted lines correspond to its associated uncertainty.}
\label{fig:ion_dens}
\end{figure*}

Table\,\ref{tbl_denstrat} shows, for each object in the sample, the \nev$_{24.3/14.3}$, \oiii$_{88/52}$, \siii$_{33.5/18.7}$, and \nii$_{205/122}$ line ratios, where available, and their associated uncertainties, the average electron densities derived for each ratio --\,assuming purely collisionally excited gas at a temperature of $10^4\, \rm{K}$, as in S15\,-- and the density error due to the line ratio uncertainty. Fig.\,\ref{fig:ion_dens} shows the electron density {\it vs} the ionisation potential for all the galaxies in the sample with at least a pair of lines detected of the same species. Our extended sample confirms the correlation found by S15: lines ionised by a harder radiation field originate in gas with higher densities ($\sim 10^3$--$10^4\, \rm{cm^{-3}}$), while softer radiation is associated with lower densities ($10$--$10^2\, \rm{cm^{-3}}$). This correlation is quantified by: \textit{a)} a weighted least-square fit of the form $\log y = (1.38 \pm 0.16) \log x + (0.27 \pm 0.25)$, $\chi^2 = 24.6$, correlation coefficient $R = 0.71$ (black-solid line; blue-dashed lines indicate the uncertainty associated with the fit); \textit{b)} a linear regression using the Kaplan-Meier residuals\footnote{The linear regression was performed using the \textsc{buckleyjames} routine, available in the \textsc{stsdas} data analysis package: \\ \mbox{\url{http://stsdas.stsci.edu/cgi-bin/gethelp.cgi?buckleyjames.hlp}}} of the form $\log y = (1.51 \pm 0.13) \log x - 0.10$ (red dot-dashed line; red-dotted lines indicate the associated uncertainty). The Kaplan-Meier residuals method, included in the \textsc{asurv} package\footnote{Astronomy Survival Analysis Package, described in: \\ \mbox{\url{http://stsdas.stsci.edu/cgi-bin/gethelp.cgi?survival.hlp}}}, is further described in \citet{iso86}. A total of $155$ pairs of lines were used for the weighted least-square fit, while $248$ pairs of lines --\,$93$ of them as upper limits\,-- were included in the Kaplan-Meier residuals fit.
We note that the upper limits on the density are not caused by a detection limit in the line observations, but rather due to the limited range where the line ratio is sensitive to density (see Table\,\ref{tbl_denstrat}, also fig.\,2 in S15). The upper limits reported in Table\,\ref{tbl_denstrat} and Fig.\,\ref{fig:ion_dens} derived from the \nii$_{205/122}$, \siii$_{33.5/18.7}$, and \nev$_{24.3/14.3}$ line ratios denote density values below the corresponding low-density limits of each of these three line ratios: $1\, \rm{cm^{-3}}$,  $10\, \rm{cm^{-3}}$, and $100\, \rm{cm^{-3}}$ for \nii$_{205/122}$, \siii$_{33.5/18.7}$, and \nev$_{24.3/14.3}$, respectively. Only in the case of NGC\,7172, for which the \nii$_{122\, \rm{\micron}}$ was extracted from the PACS central spaxel, the upper limit on the density based on the \nii$_{205/122}$ line ratio seems to be caused by the different apertures between PACS ($9.4''$) and SPIRE ($\sim 17''$).

The slope of the linear regression obtained using these two different methods is consistent within the uncertainties, and also with the results in S15, being the error here a factor of $\sim 2$ smaller due to the improved statistics.

An additional line ratio, \neiii$_{36.0/15.6}$, was considered for this analysis. This ratio traces gas at very high densities in the $3 \lesssim \log(n_{\rm e}/\rm{cm^{-3}}) \lesssim 6$ range, even higher than those probed by \nev$_{24.3/14.3}$, with a lower excitation ($\sim 41\, \rm{eV}$). Unfortunately, the \neiii$_{36.0\, \rm{\micron}}$ is out of the \textit{Spitzer}/IRS-SH spectral coverage for most of the galaxies in our sample. We compiled a few \textit{Spitzer} measurements given by \citet{bs09} and \citet{ram13}, plus \textit{ISO}/SWS and LWS measurements\footnote{As discussed in Appendix\,\ref{app}, aperture effect on \textit{ISO}/LWS line fluxes are not expected to be important in our sample of galaxies.} from \citet{stu02}, \citet{ver03}, and \citet{sat04}. From a total of 29 galaxies (12 of them are upper limits in \neiii$_{36.0\, \rm{\micron}}$), 28 have \neiii$_{36.0/15.6}$ ratios (16 galaxies) or upper limits (12 galaxies) below the low-density limit, i.e. $\lesssim 10^3\, \rm{cm^{-3}}$. This is consistent with the scenario proposed in S15 and confirmed here, i.e. the gas with the highest densities is traced by the lines with the highest ionisation potential. The measured \neiii$_{36.0/15.6}$ ratios are associated to a lower density gas when compared to the \nev$_{24.3/14.3}$ ratio.

\subsection{Diffuse low-excitation photoionised gas and dense neutral gas}\label{neut}

The UV radiation, emitted by the \textsc{H\,ii} regions in galactic star forming regions or by the accretion disk in AGN, can heat the nearby gas clouds via photoelectric emission by dust grains \citep[e.g.][]{th85}. This originates the PDR, a transition region in the cloud between ionised and molecular gas ($N_{\rm H} \lesssim 10^{22}\, \rm{cm^{-2}}$), in which \cii$_{158\, \rm{\micron}}$ and \oi$_{63\, \rm{\micron}}$ are the most important cooling lines, mainly excited by collisions with H$_2$ molecules and \textsc{H\,i} \citep[$n_{\rm H} \gtrsim 10^3$--$10^4\, \rm{cm^{-3}}$;][]{th85,lis06}. Although most of the \cii$_{158\, \rm{\micron}}$ emission comes from neutral gas, it might also have a non-negligible contribution ($\sim 15\%$) from low-excitation ionised gas \citep[e.g. S15,][]{cor15}.

\begin{figure*}
  \includegraphics[width=0.5\textwidth]{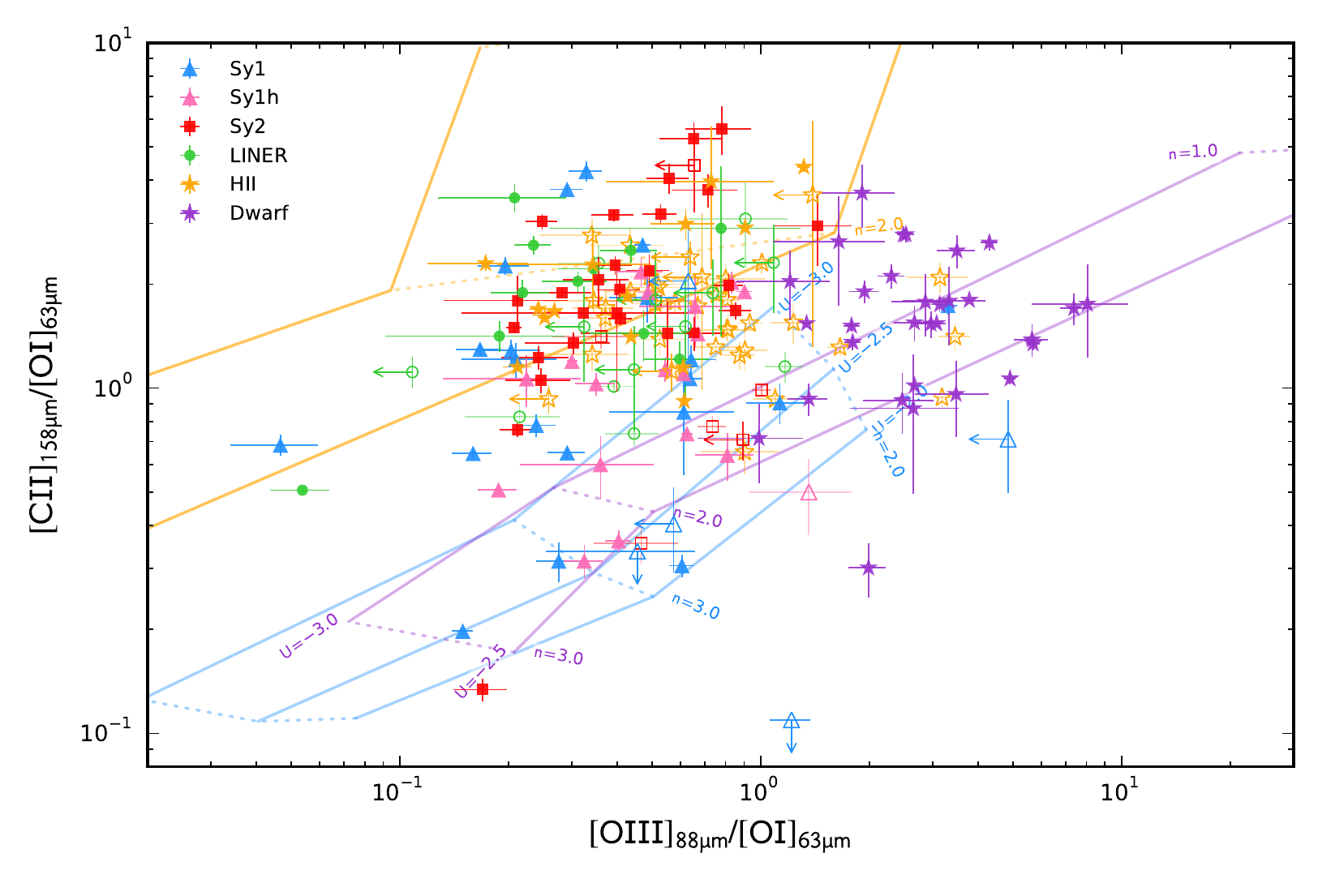} 
  \includegraphics[width=0.5\textwidth]{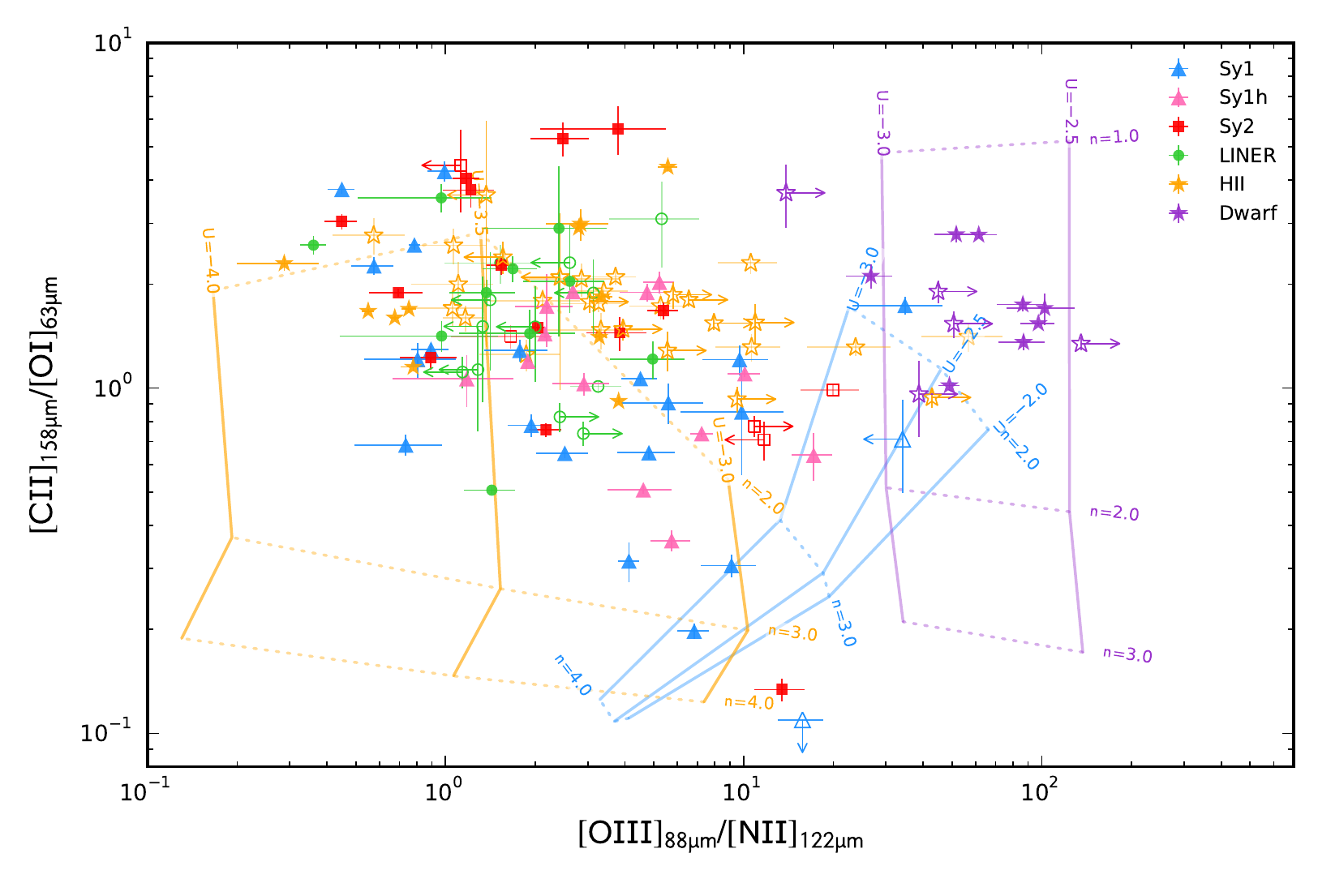}
  \caption{Line ratio diagrams (same notations as in Fig.\,\ref{fig:T_OI}). 
{\bf Left (a):} the [\textsc{C\,ii}]$_{158}$/[\textsc{O\,i}]$_{63}$ line ratio {\it vs} the [\textsc{O\,iii}]$_{88}$/[\textsc{O\,i}]$_{63}$ ratio. 
{\bf Right (b):} the [\textsc{C\,ii}]$_{158}$/[\textsc{O\,i}]$_{63}$ line ratio {\it vs} the ionisation-sensitive [\textsc{O\,iii}]$_{88}$/[\textsc{N\,ii}]$_{122}$ ratio.}
\label{fig:c2_o1a}
\end{figure*}

\begin{figure*}
  \includegraphics[width=0.5\textwidth]{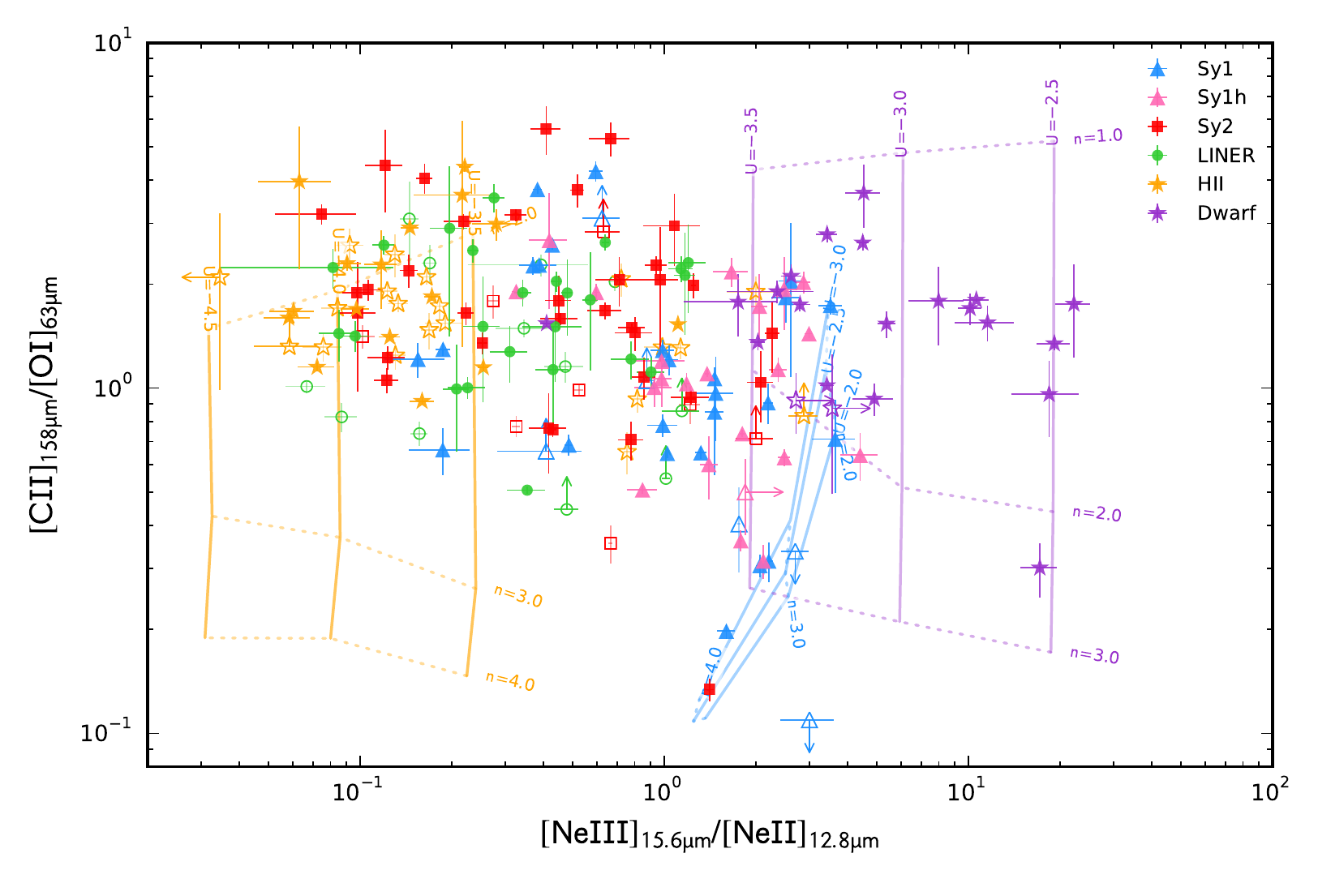} 
  \includegraphics[width=0.5\textwidth]{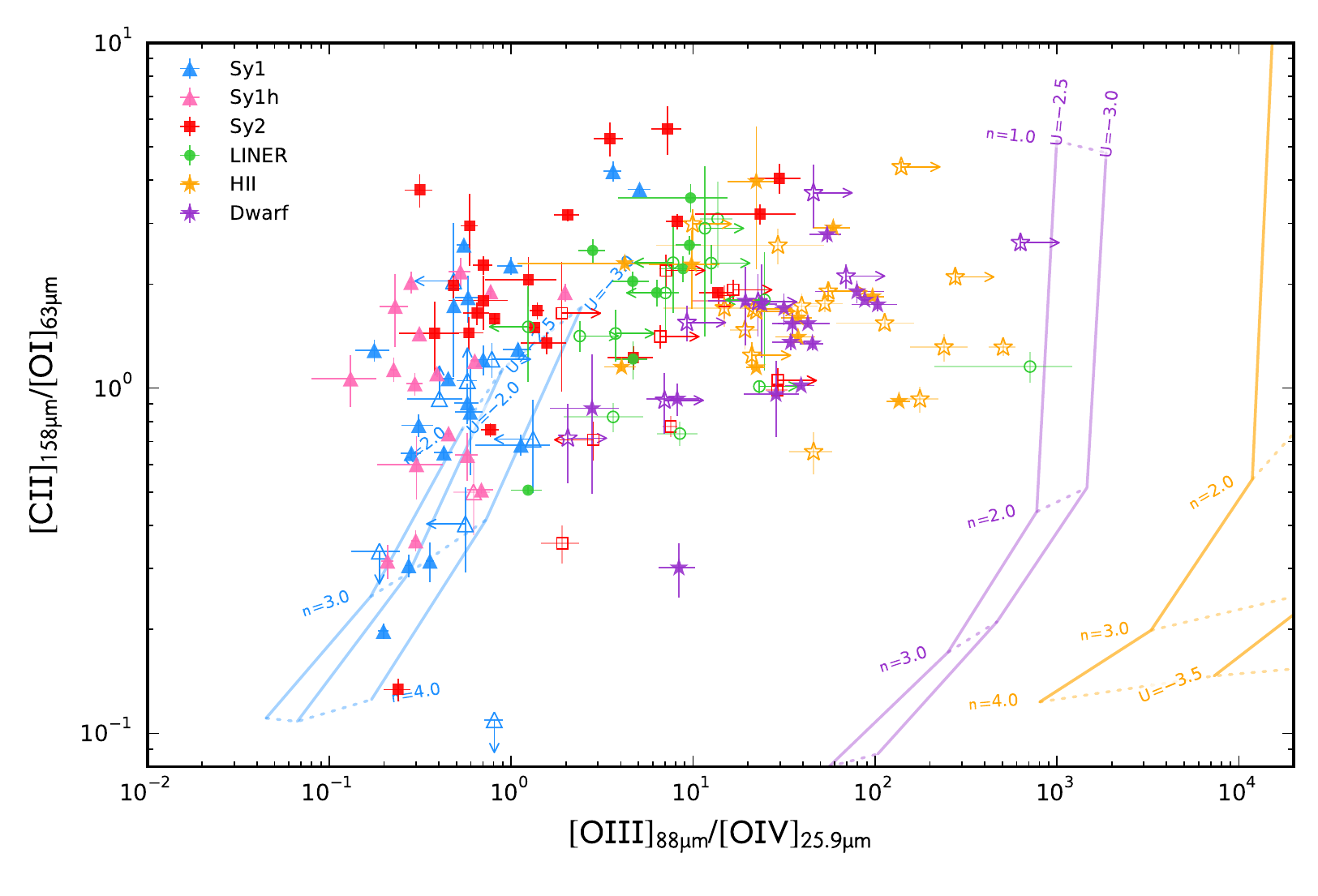} 
  \caption{Line ratio diagrams (same notations as in Fig.\,\ref{fig:T_OI}). 
{\bf Left (a):} the [\textsc{C\,ii}]$_{158}$/[\textsc{O\,i}]$_{63}$ line ratio {\it vs} the ionisation-sensitive [Ne\,\textsc{iii}]$_{15.5}$/[Ne\,\textsc{ii}]$_{12.8}$ ratio. 
{\bf Right (b):} the [\textsc{C\,ii}]$_{158}$/[\textsc{O\,i}]$_{63}$ line ratio {\it vs} the ionisation-sensitive [\textsc{O\,iii}]$_{88}$/[\textsc{O\,iv}]$_{25.9}$ ratio.}
\label{fig:c2_o1b}
\end{figure*}

The Figs\,\ref{fig:c2_o1a}a,b and \ref{fig:c2_o1b}a,b show the \cii$_{158}$/\oi$_{63}$ ratio {\it vs} the \oiii$_{88}$/\oi$_{63}$, the \oiii$_{88}$/\nii$_{122}$, the \neiii$_{15.6}$/\neii$_{12.8}$, and \oiii$_{88}$/\oiv$_{25.9}$ line ratios, respectively. The \cii$_{158}$/\oi$_{63}$ ratio in PDR depends mostly on density, although optical depth effects in \oi$_{63\, \rm{\micron}}$ and \textsc{H\,ii} region contribution to the \cii$_{158\, \rm{\micron}}$ line might limit the ability of this ratio as density tracer \citep{abe07}. No statistically significant difference has been found for the \cii$_{158}$/\oi$_{63}$ ratio between the different sub-classes. As suggested by the \ci$_{609/371}$ line ratio in Section\,\ref{dens_ion_temp}, the XDR does not seem to contribute significantly, since the \cii$_{158}$/\oi$_{63}$ line ratio is similar for both AGN and starburst galaxies.

The \oiii$_{88}$/\oi$_{63}$ ratio in Fig.\,\ref{fig:c2_o1a}a is almost insensitive to ionisation, being dwarf galaxies the only class that is clearly distinguished with values of $\gtrsim 1$. Our models confirm this behaviour, with a few S1 and S1h located in the AGN grid, while the rest of Seyfert galaxies, LINERs, and starburst galaxies are found within the grid of starburst models. The \oiii$_{88}$/\nii$_{122}$ ratio in Fig.\,\ref{fig:c2_o1a}b is mildly sensitive to ionisation, it is not able to separate Seyfert's from starburst galaxies, while dwarf galaxies are clearly distinguished due to their strong \oiii$_{88\, \rm{\micron}}$ emission. \textsc{Cloudy} models in Fig.\,\ref{fig:c2_o1a}b also show this dependency on ionisation, with a superposition for AGN and dwarf galaxy models at low densities ($\lesssim 100\, \rm{cm^{-3}}$) and high ionisation parameter values ($\log U \gtrsim -3$). A high-excitation photoionised gas tracer, such as the \oiii$_{88}$/\oiv$_{25.9}$ line ratio in Fig.\,\ref{fig:c2_o1b}b, is able to separate starburst galaxies, Seyfert nuclei, and dwarf galaxies. \textsc{Cloudy} models in Fig.\,\ref{fig:c2_o1b}a are in agreement with the observed \cii$_{158}$/\oi$_{63}$ and \neiii$_{15.6}$/\neii$_{12.8}$ line ratios, and explain the higher \neiii$_{15.6}$/\neii$_{12.8}$ ratios shown by dwarf galaxies. We note in Fig\,\ref{fig:c2_o1b}b that neither starburst models (in yellow; most of the grid falls outside the figure due to the high \oiii$_{88}$/\oiv$_{25.9}$ predicted ratios) nor dwarf galaxy models (in purple) are able to reproduce the observed \oiii$_{88}$/\oiv$_{25.9}$ line ratios, which are overestimated by more than an order of magnitude. The models predictions for oxygen and neon high-ionisation ratios will be further discussed in Section\,\ref{tracers}.

\begin{figure*}
  \includegraphics[width=0.5\textwidth]{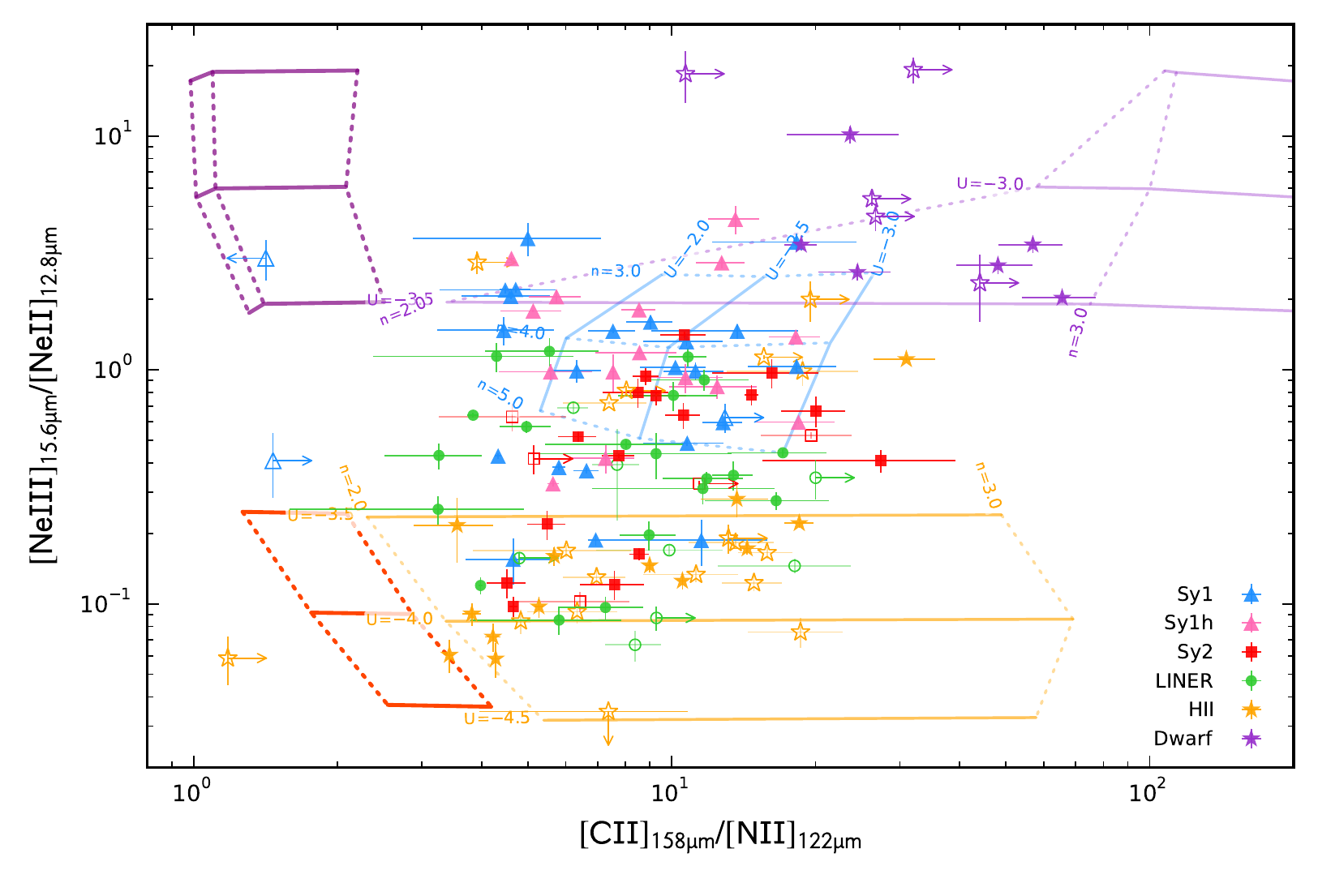}
  \includegraphics[width=0.5\textwidth]{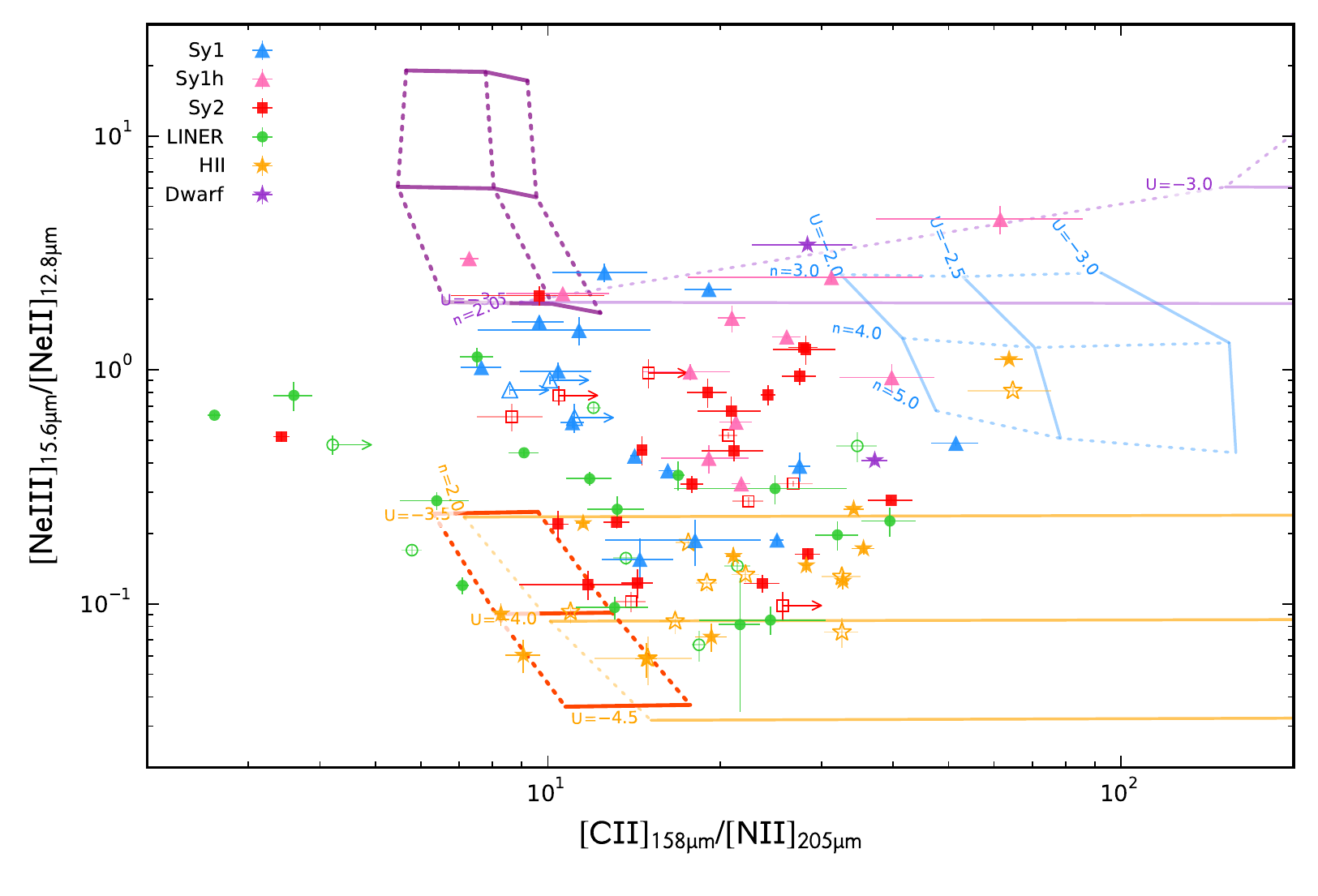}
  \caption{Line ratio diagrams (same notations as in Fig.\,\ref{fig:T_OI}). The grid of models in yellow colour shows the usual starburst+PDR models, with the integration going down to the temperature of $50\, \rm{K}$, while the grid in orange colour shows the pure photoionised models ($T_{\rm stop} = 1000\, \rm{K}$). Similarly, the dark purple grid correspond to dwarf galaxy models stopped at $T_{\rm stop} = 1000\, \rm{K}$, while the light purple grid show the dwarf+PDR models, stopped at $50\, \rm{K}$. 
{\bf Left (a):} the [Ne\,\textsc{iii}]$_{15.5}$/[Ne\,\textsc{ii}]$_{12.8}$ line ratio {\it vs} the [\textsc{C\,ii}]$_{158}$/[\textsc{N\,ii}]$_{122}$ ratio. 
{\bf Right (b):} the [Ne\,\textsc{iii}]$_{15.5}$/[Ne\,\textsc{ii}]$_{12.8}$ line ratio {\it vs} the [\textsc{C\,ii}]$_{158}$/[\textsc{N\,ii}]$_{205}$ ratio.}
\label{fig:c2_n2vsne3_ne2}
\end{figure*}

Figs\,\ref{fig:c2_n2vsne3_ne2}a,b show the \neiii$_{15.6}$/\neii$_{12.8}$ ratio {\it vs} the \cii$_{158}$/\nii$_{122}$ and \cii$_{158}$/\nii$_{205}$ ratios, respectively. The latter ratios are sensitive to the fraction of PDR to low-excitation photoionised gas emission, since \cii$_{158\, \rm{\micron}}$ is mostly originated in PDR while \nii$_{122, 205\, \rm{\micron}}$ are emitted by ionised gas. The PDR emission is inherently included in the \textsc{Cloudy} models whose radial integration have been extended to the low temperature of $T = 50\, \rm{K}$ (yellow grid in Fig.\,\ref{fig:c2_n2vsne3_ne2}), while models stopped at $T = 1000\, \rm{K}$ denote the ratios predicted for pure ionised gas (dark orange grid). Similarly, we compare dwarf galaxy models stopped at $T = 1000\, \rm{K}$ (dark violet grid) and those extended to $T = 50\, \rm{K}$, thus including the PDR emission (light violet grid). The difference in the \cii$_{158}$/\nii$_{122, 205}$ ratios between pure photoionised models and those including the PDR is even more extreme in the case of dwarf galaxies. When compared with the models, the observed ratios suggest that most of the \cii$_{158\, \rm{\micron}}$ emission is originated in the PDR, in agreement with the results of \citet{cor15} for the same sample of dwarf galaxies included here. These authors estimate that only $\approx 15\%$ of \cii$_{158\, \rm{\micron}}$ is originated is \textsc{H\,ii} regions. The observed line ratios in Fig.\,\ref{fig:c2_n2vsne3_ne2}a also reveal a possible dependence, for starburst galaxies, of the \cii$_{158}$/\nii$_{122}$ ratio on the ionisation, traced by the \neiii$_{15.6}$/\neii$_{12.8}$ ratio. A harder ionising spectrum might produce a larger amount of N$^{2+}$ in the difuse ionised gas, thus decreasing the relative contribution of the \nii$_{122\, \rm{\micron}}$ emission. Consequently, the high \cii$_{158}$/\nii$_{122}$ ratios found in dwarf galaxies are explained by the high excitation in these objects \citep{cor15}. This is also the case of NGC\,1222, a near-solar metallicity starburst that possibly experienced a merger with a low-metallicity companion \citep{bec07}. In this regard, the \cii$_{158}$/(\nii$_{122}$+\niii$_{57}$) ratio might be more appropriate to evaluate the relative PDR to ionised gas contribution. Unfortunately, the \niii$_{57\, \rm{\micron}}$ line has been observed only for six starburst galaxies (four detections), thus we do not have enough statistics to test the proposed PDR-to-ionised gas ratio.

\begin{figure*}
  \includegraphics[width=0.5\textwidth]{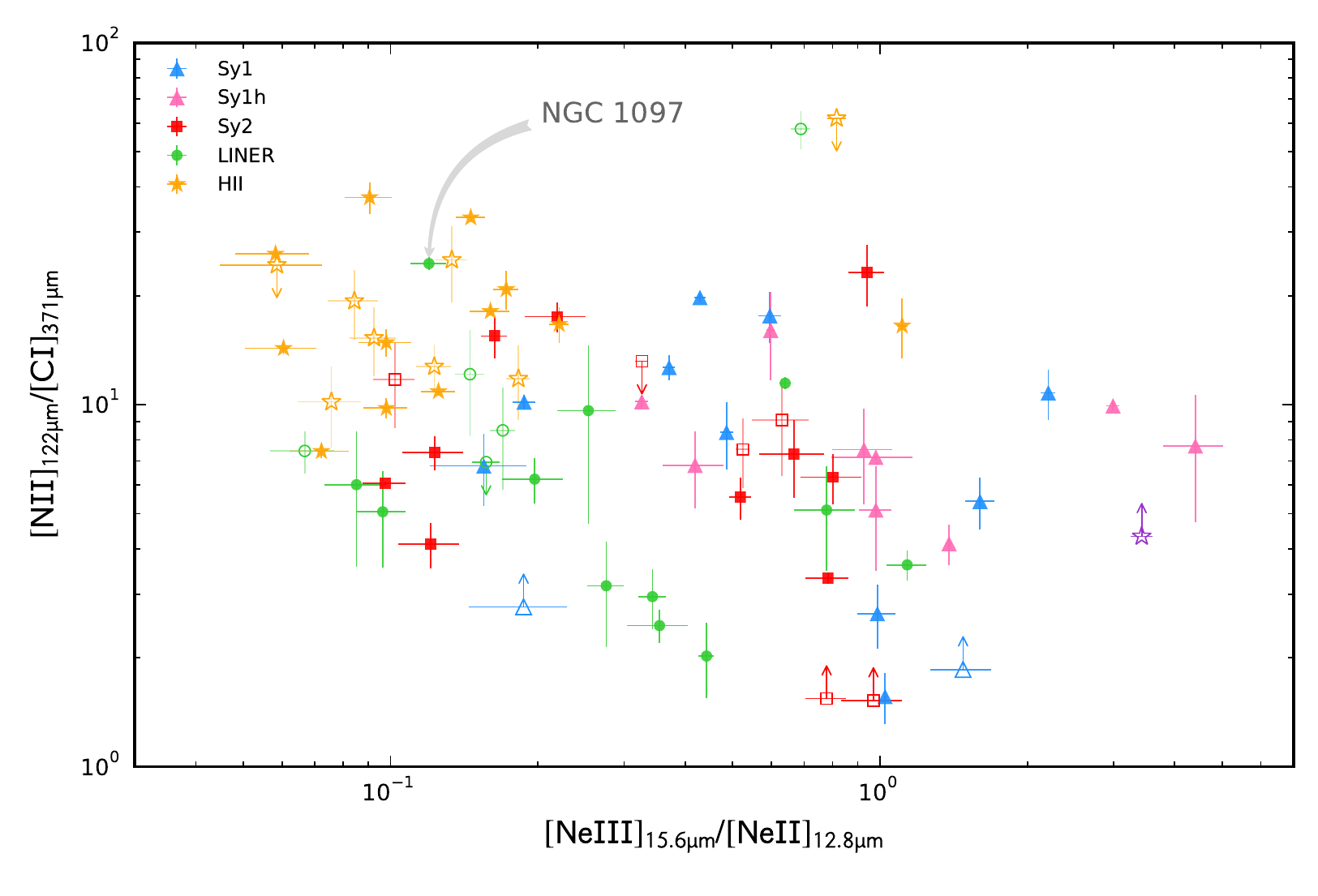}
  \includegraphics[width=0.5\textwidth]{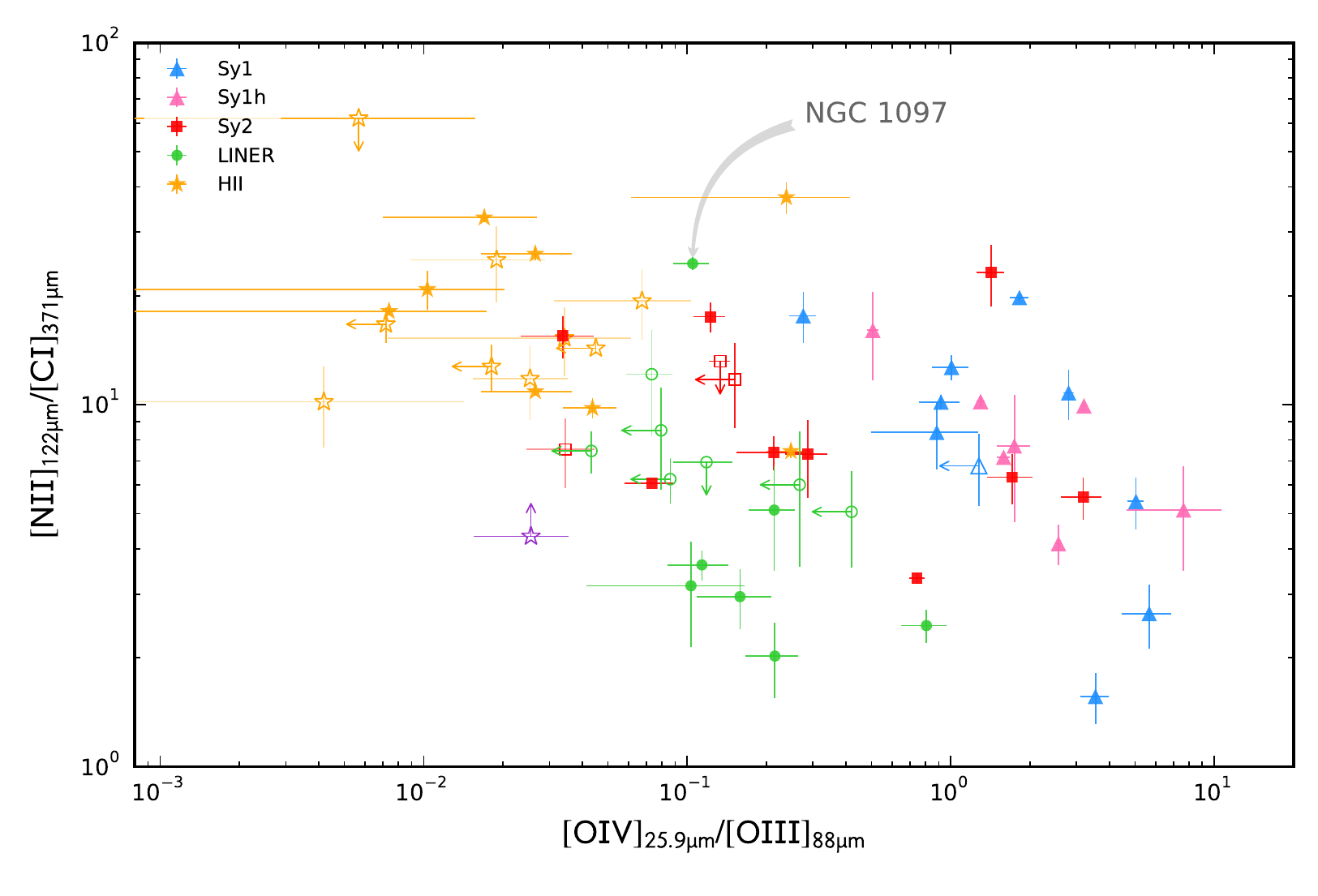}
  \caption{Line ratio diagrams (same notations as in Fig.\,\ref{fig:T_OI}). 
{\bf Left (a):} the [N\,\textsc{ii}]$_{122}$/[Ne\,\textsc{ii}]$_{12.8}$ line ratio {\it vs} the [\textsc{N\,ii}]$_{122}$/[\textsc{C\,i}]$_{371}$ ratio. 
{\bf Right (b):} the [\textsc{O\,iv}]$_{25.9}$/[\textsc{O\,iii}]$_{88}$ line ratio {\it vs} the [\textsc{N\,ii}]$_{122}$/[\textsc{C\,i}]$_{371}$ ratio.}
\label{fig:n2_c1}
\end{figure*}

Fig.\,\ref{fig:n2_c1} shows the \nii$_{122}$/\ci$_{371}$ line ratio {\it vs} the ionisation sensitive ratios \neiii$_{15.6}$/\neii$_{12.8}$ and \oiv$_{25.9}$/\oiii$_{88}$. The \nii$_{122}$/\ci$_{371}$ ratio accounts for the relative contribution of low-excited photoionised gas to neutral gas in the PDR. Overall, starburst galaxies seem to be biased toward higher values of the \nii$_{122}$/\ci$_{371}$ ratio, being always found above $> 7$, with a median value of $16.6$ and a median absolute deviation of $7.4$. LINERs show tentatively lower values with an average of \nii$_{122}$/\ci$_{371}$ $\sim 5.1$ and a deviation of $3.0$. A possible explanation would be the presence of an XDR in LINERs, which would contribute with a brighter \ci$_{371\, \rm{\micron}}$ line emission. However the \ci$_{609}$/\ci$_{371}$ ratio in Fig.\,\ref{fig:o1_o1vsc1_c1}b showed typical PDR values, suggesting that the XDR contribution to the atomic carbon lines is negligible in our sample at the angular resolution of the SPIRE data. A decreasing \nii \ contribution in favour of \niii \ emission at higher excitation would explain the lower \nii$_{122}$/\ci$_{371}$ ratios found in LINERs, if a harder ionising spectrum is assumed for LINERs when compared to starburst galaxies. However, Fig.\,\ref{fig:c2_n2vsne3_ne2} does not show a dependence of the \cii$_{158}$/\nii$_{122}$ ratio with the ionisation parameter for LINERs, as it was the case for starburst galaxies. An alternative explanation for the different \nii$_{122}$/\ci$_{371}$ ratios is that starburst galaxies have a larger filling factor of diffuse low-excitation photoionised gas within the SPIRE aperture when compared to LINER galaxies. In this regard, we note that most of the LINERs in Fig.\,\ref{fig:n2_c1} are found in early-type spirals, while the LINER with the highest ratio, NGC\,1097 ($25 \pm 1$), has a bright circumnuclear starburst ring unresolved within the SPIRE aperture of $\sim 35''$. A larger sample of LINERs and starburst galaxies would be needed in order to prove this behaviour.


\subsection{Discriminating AGN and Starburst}\label{tracers}

\subsubsection{Oxygen and Neon line ratios}\label{o_ne_rat}
High to low excitation IR line ratios such as \nev$_{14.3}$/\neii$_{12.8}$ and \oiv$_{25.9}$/\neii$_{12.8}$ have been commonly used to measure the relative contributions of AGN and star formation in active galaxies \citep{stu02,arm07,mel08,tom08}. Since photoionisation by young stars cannot have a significant contribution to transitions with a high ionisation potential ($\gtrsim 50\, \rm{eV}$), as the AGN does, these line ratios are sensitive to the relative strength of the AGN compared to the starburst. So far, these diagnostics have been tested by comparing samples of active nuclei with starburst galaxies \citep[e.g.][]{tre10}. In this work we compare AGN and starburst galaxies but also low-metallicity dwarf galaxies, and therefore we propose a new diagnostic including the long-wavelength \oiii$_{88\, \rm{\micron}}$ line measured by \textit{Herschel}/PACS.

In S15 we showed that the \oiv$_{25.9}$/\oiii$_{88}$ line ratio is a good tracer of the relative AGN to star formation contribution. 
These oxygen lines have critical densities in the $10^2$--$10^4\, \rm{cm^{-3}}$ range and their ratio is sensitive to the hardness of the ionising radiation in the $54.94$--$35.12\, \rm{eV}$ range ($4$--$2.6\, \rm{Ry}$). Thus, the \oiii$_{88\, \rm{\micron}}$ line can be excited by both young stars and AGN activity, while the \oiv$_{25.9\, \rm{\micron}}$ transition becomes much more intense under the presence of an AGN \citep{stu02,d-s09,spi12a}. The \oiv$_{25.9\, \rm{\micron}}$ line is brighter than the \nev$_{14.3, 24.3\, \rm{\micron}}$ lines, and can still be detected in star formation environments. This makes this line ideal to identify the location of the different populations, i.e. AGN and starburst as well as dwarf galaxies, in the diagnostic diagrams.

\begin{figure*}
  \includegraphics[width=0.5\textwidth]{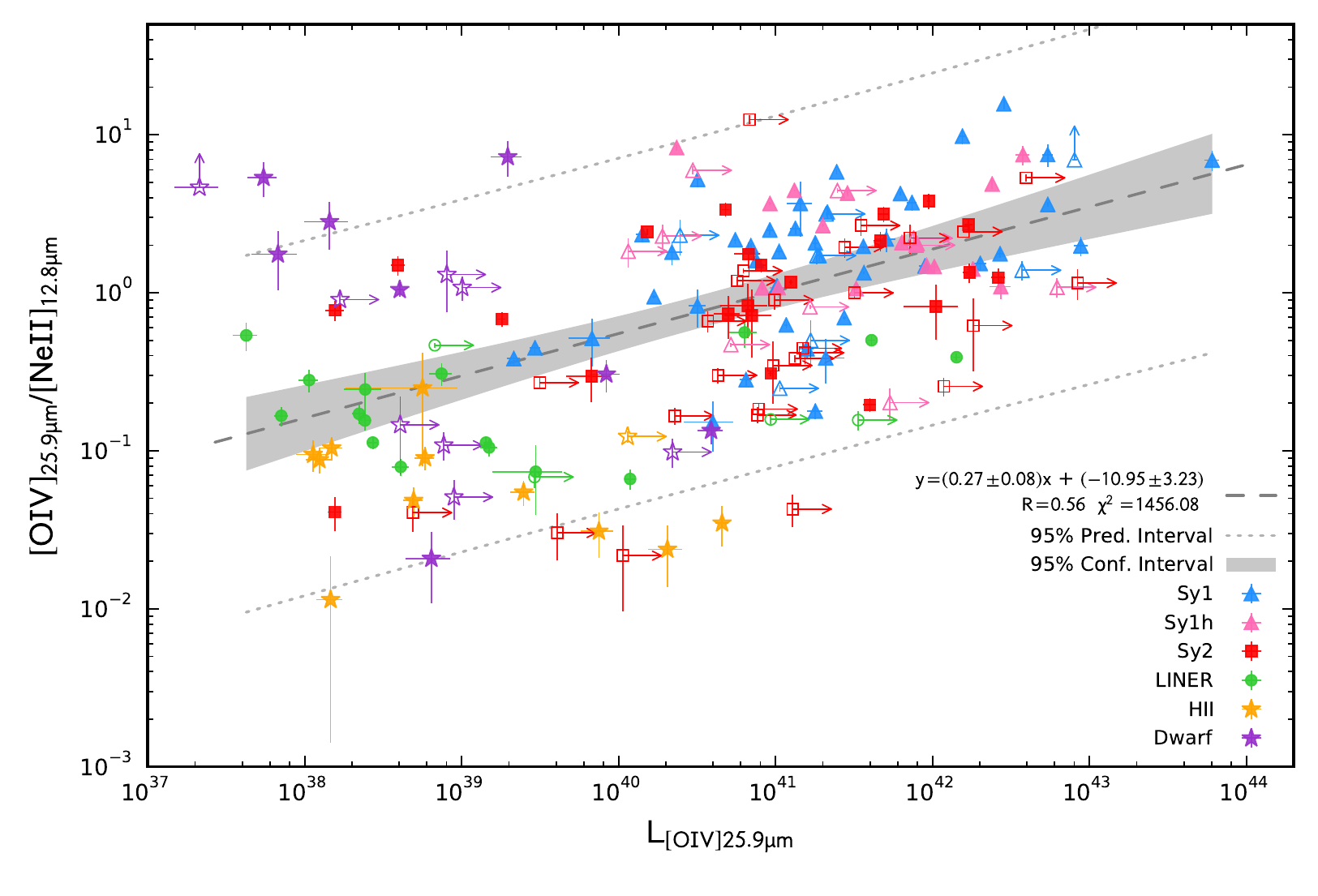}
  \includegraphics[width=0.5\textwidth]{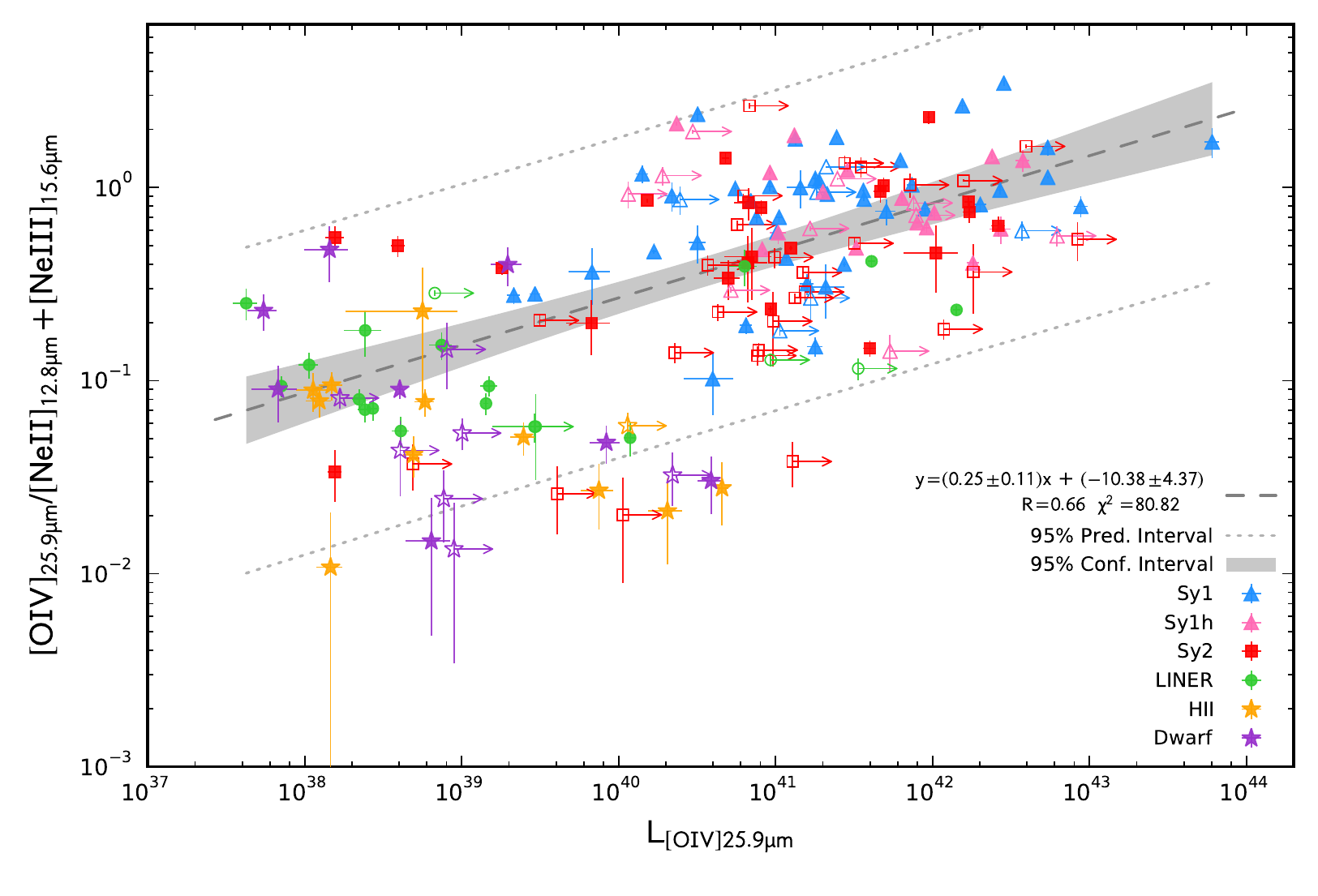}
  \caption{Line ratios sensitive to the relative AGN/starburst contribution (same notations as in Fig.\,\ref{fig:T_OI}). 
{\bf Left (a):} the [\textsc{O\,iv}]$_{25.9}$/[Ne\,\textsc{ii}]$_{12.8}$ line ratio {\it vs} the [\textsc{O\,iv}]$_{25.9\, \rm{\micron}}$ luminosity. 
{\bf Right (b):} the [\textsc{O\,iv}]$_{25.9}$/([Ne\,\textsc{ii}]$_{12.8}$+[Ne\,\textsc{iii}]$_{15.6}$) line ratio {\it vs} the [\textsc{O\,iv}]$_{25.9\, \rm{\micron}}$ luminosity.}
\label{fig:o4_ne23}
\end{figure*}

In Fig.\,\ref{fig:o4_ne23}a,b we show the usual \oiv$_{25.9}$/\neii$_{12.8}$ ratio \citep[e.g.][]{gen98,stu02,mel08} and the \oiv$_{25.9}$/(\neii$_{12.8}$+\neiii$_{15.6}$) ratio, respectively, as a function of the \oiv$_{25.9\, \rm{\micron}}$ luminosity ($L_{\rm [OIV] 25.9}$). If only AGN and starburst galaxies were taken into account in Fig.\,\ref{fig:o4_ne23}a, a gradient would be present from lower values of \oiv$_{25.9}$/\neii$_{12.8}$ $\sim 2 \times 10^{-2}$ and relatively low \oiv$_{25.9\, \rm{\micron}}$ luminosities of $\sim 10^{38}\, \rm{erg\, s^{-1}}$, where star formation dominates the ratio in \textsc{H\,ii} region galaxies, to \oiv$_{25.9}$/\neii$_{12.8}$ $\sim 10$ and high $L_{\rm [OIV] 25.9} \sim 10^{43}\, \rm{erg\, s^{-1}}$, where AGN activity is the dominant source of radiation (blue and pink triangles, and red squares). LINERs (green dots) are found in the intermediate region between star formation and Seyfert's, while five S2 galaxies (red squares) appear to be dominated by star formation. However, this correlation no longer applies if we consider the sample of dwarf galaxies, which are able to reach the AGN-dominated regime in this diagram, showing \oiv$_{25.9}$/\neii$_{12.8}$ line ratios similar to those of Seyfert's, but at low \oiv$_{25.9\, \rm{\micron}}$ luminosities compared to AGN. The least squares regression still shows a trend with higher line ratios found at higher luminosities,
\begin{equation}
\begin{aligned}
  \log \left( \frac{\rm [OIV]_{25.9}}{\rm [NeII]_{12.8}} \right) = & (0.27 \pm 0.08) \log L_{\rm [OIV] 25.9} \\
  & + (-10.95 \pm 3.32),
\end{aligned}
\end{equation}
but the large dispersion introduced by dwarf galaxies, which results in a relatively low Spearman's rank correlation coefficient of $R = 0.56$, limits the validity of the \oiv$_{25.9}$/\neii$_{12.8}$ line ratio as an AGN/starburst tracer.

The lower metallicity in dwarf galaxies allows the presence of a hotter main sequence and thus a harder radiation field, increasing the filling factor of low-density ionised gas \citep{sch92,cor15}. In particular, the ionising spectra of dwarf galaxies is much harder just after the Lyman break, which also increases the \neiii$_{15.6}$/\neii$_{12.8}$ line ratio (and \siv$_{10.5}$/\siii$_{18.7}$, see below) when compared to a solar-metallicity starburst \citep[][see Fig.\,\ref{fig:c2_n2vsne3_ne2}a]{oha06,hao09}. Our Fig.\,\ref{fig:o4_ne23}a demonstrates that the \oiv$_{25.9}$/\neii$_{12.8}$ line ratio does not probe the relative AGN-to-starburst contribution at low metallicities, since a large fraction of the star formation component is probably emitted in the form of \neiii$_{15.6\, \rm{\micron}}$ \citep[see][]{ho07}. This casts doubts on our ability to separate AGN and starburst components, e.g. in composite objects with extreme star formation bursts or mergers involving a chemically unevolved dwarf companion, and prompts us to find a more suitable diagnostic ratio, which also should be valid for these scenarios that are expected to be even more common at higher redshift \citep[e.g.][]{ala14,ate14,ate15}.

Following the previous argument, we show the \oiv$_{25.9}$/(\neii$_{12.8}$+\neiii$_{15.6}$) ratio in Fig\,\ref{fig:o4_ne23}b, since the sum of the two Neon lines would be a more reliable tracer of the star formation contribution also for dwarf galaxies. In this case, both dwarf and star forming galaxies have in average lower \oiv$_{25.9}$/(\neii$_{12.8}$+\neiii$_{15.6}$) ratios of $\sim 4 \times 10^{-2}$, when compared to Seyfert galaxies ($\sim 1$), although some overlap between dwarfs/starburst galaxies and Seyfert galaxies still remains in the range \oiv$_{25.9}$/(\neii$_{12.8}$+\neiii$_{15.6}$) $\sim 0.1$ to $0.5$. The least square regression confirms the trend of increasing rations with increasing $L_{\rm [OIV] 25.9}$,
\begin{equation}
\begin{aligned}
  \log \left( \frac{\rm [OIV]_{25.9}}{\rm [NeII]_{12.8} + [NeIII]_{15.6}} \right) = & (0.25 \pm 0.11) \log L_{\rm [OIV] 25.9} \\
  & + (-10.38 \pm 4.37),
\end{aligned}
\end{equation}  
and the correlation coefficient improves ($R = 0.66$), with regard to the \oiv$_{25.9}$/\neii$_{12.8}$ ratio. 

\begin{figure*}
  \includegraphics[width=0.5\textwidth]{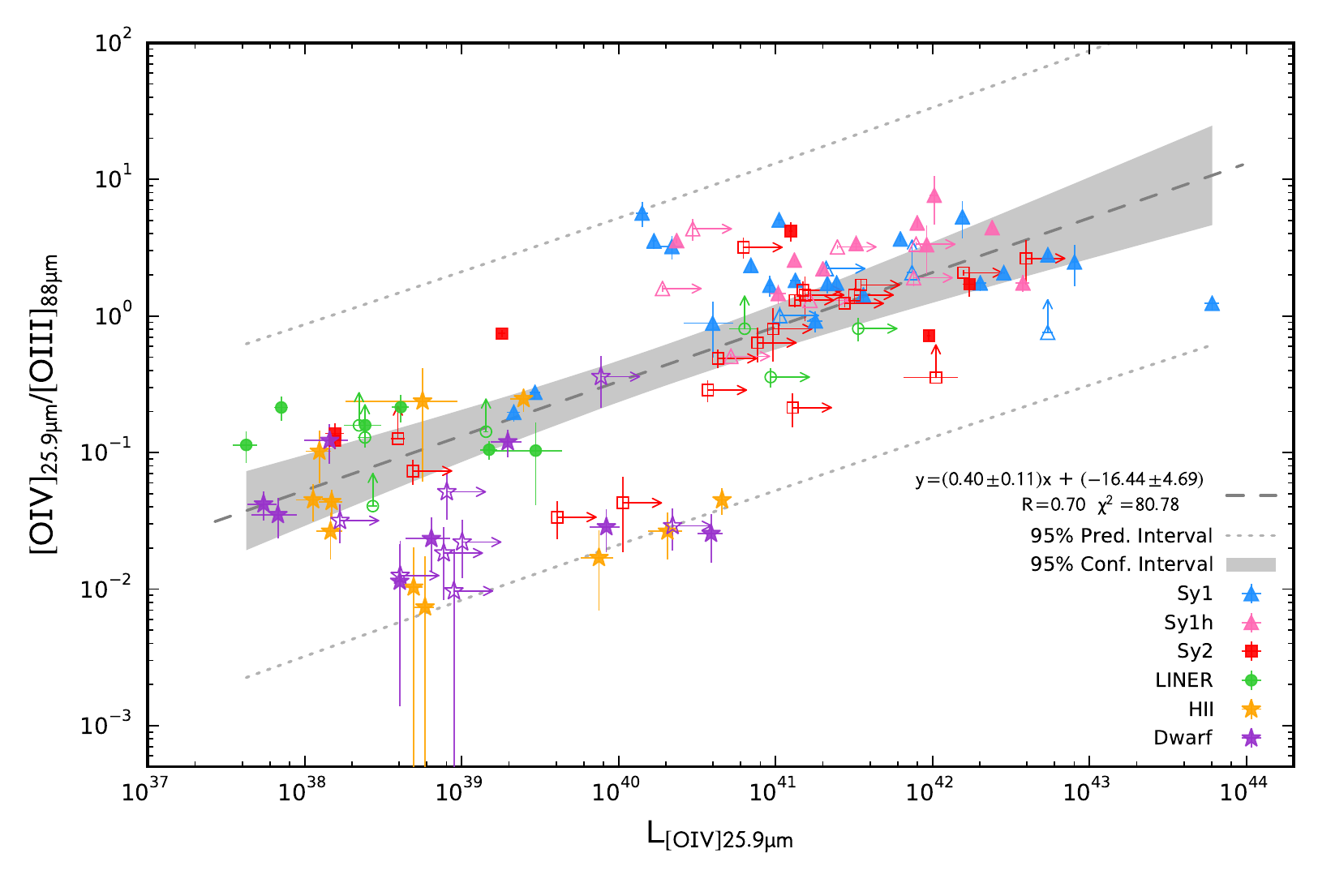}
  \includegraphics[width=0.5\textwidth]{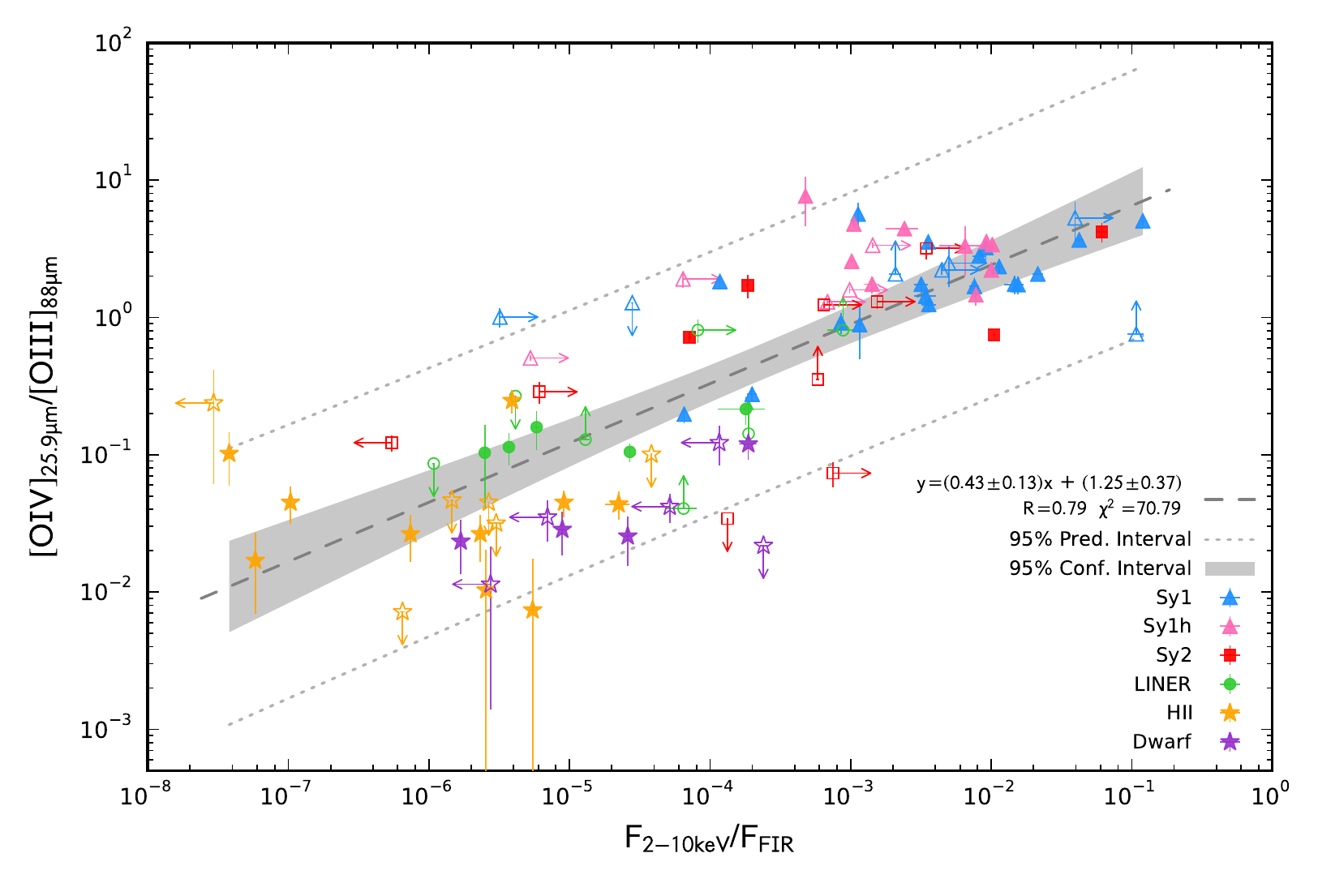}
  \caption{Line ratios sensitive to the relative AGN/starburst contribution (same notations as in Fig.\,\ref{fig:T_OI}). 
{\bf Left (a):} the [\textsc{O\,iv}]$_{25.9}$/[\textsc{O\,iii}]$_{88}$ line ratio {\it vs} the [\textsc{O\,iv}]$_{25.9\, \rm{\micron}}$ luminosity. 
{\bf Right (b):} the [\textsc{O\,iv}]$_{25.9}$/[\textsc{O\,iii}]$_{88}$ line ratio {\it vs} the ratio of absorption-corrected $2$--$10\, \rm{keV}$ X-ray flux to far-IR flux (derived from the continuum adjacent to the [\textsc{C\,ii}]$_{158\, \rm{\micron}}$ line). Compton-thick objects are shown as lower limits in the horizontal axis.}
\label{fig:o4_o3}
\end{figure*}

In Fig.\,\ref{fig:o4_o3}a we show the ability of the \oiv$_{25.9}$/\oiii$_{88}$ line ratio to separate the contribution of AGN from starburst and dwarf galaxies in our sample, as a function of the \oiv$_{25.9\, \rm{\micron}}$ luminosity. The least squares regression is steeper when compared to the previous line ratios,
\begin{equation}
\begin{aligned}
  \log \left( \frac{\rm [OIV]_{25.9}}{\rm [OIII]_{88}} \right) = & (0.40 \pm 0.11) \log L_{\rm [OIV] 25.9} \\
  & + (-16.44 \pm 4.69),
\end{aligned}
\end{equation}
with a correlation coefficient of $R = 0.7$. This is the best diagnostic ratio to distinguish Seyfert activity from star formation, with a minimum overlap between AGN dominated ratios (\oiv$_{25.9}$/\oiii$_{88}$ $\gtrsim 0.3$) and those dominated by starburst or dwarf galaxies. We note that dwarf galaxies are indistinguishable from starburst galaxies in this diagram, since their ionising spectra are not hard enough in the $54$--$35\, \rm{eV}$ range to produce a higher fraction of \oiv$_{25.9}$/\oiii$_{88}$. A few AGN are dominated by the starburst component, thus their \oiv$_{25.9}$/\oiii$_{88}$ ratios are in the same range as starburst galaxies. This will be discussed further in Section\,\ref{new_bpt}. LINERs appear to show a dual behaviour, with the three brightest sources in the Seyfert domain and the rest of them located in the starburst domain, with a line ratio of $5 \times 10^{-3} \lesssim$ \oiv$_{25.9}$/\oiii$_{88}$ $\lesssim 0.3$. This suggests that the \oiii$_{88\, \rm{\micron}}$ line for the fainter LINERs in our sample might include an important contribution from their host galaxies.

Finally, Fig.\,\ref{fig:o4_o3}b shows the \oiv$_{25.9}$/\oiii$_{88}$ ratio {\it vs} the absorption-corrected X-ray $2$--$10\, \rm{keV}$ flux\footnote{Collected from values published in the literature, see Table\,\ref{tbl_sample}. Compton thick objects are labelled with ``Y'' in the CT column.} to far-IR flux ratio, the latter measured as the continuum adjacent to the \cii$_{158\, \rm{\micron}}$ line (Table\,\ref{tbl_pacs_cont}). X-ray to far-IR flux ratios for Compton thick objects appear as lower limits. This diagram confirms the behaviour observed in Fig.\,\ref{fig:o4_o3}a by using an independent estimate of the accretion to the total energy output. S1 and S1h galaxies are dominated by the AGN, and thus show the highest values for both the X-ray to far-IR ratio and the \oiv$_{25.9}$/\oiii$_{88}$ line ratios. Starburst galaxies are very faint in the $2$--$10\, \rm{keV}$ range but bright far-IR emitters, and so they show low \oiv$_{25.9}$/\oiii$_{88}$ ratios. The least squares regression between both ratios is
\begin{equation}
\begin{aligned}
  \log \left( \frac{\rm [OIV]_{25.9}}{\rm [OIII]_{88}} \right) = & (0.43 \pm 0.13) \log \left( \frac{\rm F_{2-10 keV}}{\rm F_{FIR}} \right) \\
  & + (1.25 \pm 0.37),
\end{aligned}
\end{equation}
with a correlation coefficient of $R = 0.79$.

We note that most of the \oiii$_{88\, \rm{\micron}}$ fluxes are extracted from the $3 \times 3$ pixel array ($28\farcs2 \times 28\farcs2$), an area which is a factor of $2.5$ larger than the \textit{Spitzer}/IRS-LH slit area ($11\farcs1 \times 22\farcs3$). A possible contamination by star formation in the host galaxy would decrease the measured ratio for the central AGN. To minimise this effect we have built the same diagrams using the fluxes extracted from PACS central pixel ($9\farcs4 \times 9\farcs4$), i.e. using an aperture $\sim 3$ smaller than the \textit{Spitzer}/IRS slit for the \oiv$_{25.9\, \rm{\micron}}$ line. The variations in the line ratios are within the errors for most of the objects (see Fig.\,\ref{fig:pacs_iso} and discussion in Appendix\,\ref{app}), thus we discard a strong contamination by star formation in the \oiii$_{88\, \rm{\micron}}$ line outside of the innermost $\sim 10\arcsec$ for the galaxies in our sample.

\subsubsection{A new AGN/star formation diagnostic}\label{new_bpt}
The global view of the AGN-starburst diagnostics that has been discussed in the previous Section is fully represented in the diagnostic diagrams of Figs\,\ref{fig:o4_o3vsne3_ne2}a,b, which show the \neiii$_{15.6}$/\neii$_{12.8}$ ($41$--$22\, \rm{eV}$ or $3.0$--$1.6\, \rm{Ry}$) and the \siv$_{10.5}$/\siii$_{18.7}$ ($35$--$23\, \rm{eV}$ or $2.6$--$1.7\, \rm{Ry}$) ratios, respectively, {\it vs} the \oiv$_{25.9}$/\oiii$_{88}$ ratio ($35$--$55\, \rm{eV}$ or $2.6$--$4.0\, \rm{Ry}$). The filled symbols have been coloured based on their relative \nev$_{14.3\, \rm{\micron}}$ line flux to far-IR flux ratio, as a proxy of the relative AGN-to-far-IR luminosity. In the vertical axis, the Neon and Sulphur ratios probe the steepness of the ionising continuum between $\sim 21\, \rm{eV}$, just after the Lymann break ($13.6\, \rm{eV}$), and $41\, \rm{eV}$, before the He\,\textsc{ii} ionisation edge ($54\, \rm{eV}$). The stellar spectra are very sensitive to the metallicity in this range \citep{sch92,mok04}, thus dwarf galaxies exhibit large \neiii$_{15.6}$/\neii$_{12.8}$ and \siv$_{10.5}$/\siii$_{18.7}$ ratios, even larger than Seyfert galaxies. Therefore, the vertical axis in Figs\,\ref{fig:o4_o3vsne3_ne2}a,b is especially sensitive to the excitation originated by thermal processes. A large dropout in the spectra of dwarf galaxies occurs shortward of the He\,\textsc{ii} ionisation edge, thus their \oiv$_{25.9}$/\oiii$_{88}$ ratios are far from those exhibited by Seyfert galaxies, and very similar to the values found in starburst galaxies. LINERs are able to produce both higher \oiv$_{25.9}$/\oiii$_{88}$ and \neiii$_{15.6}$/\neii$_{12.8}$ (\siv$_{10.5}$/\siii$_{18.7}$) line ratios with regard to starburst galaxies. The colour scale confirms that AGN with starburst-like \oiv$_{25.9}$/\oiii$_{88}$ line ratios are indeed galaxies in which the starburst contribution dominates both the far-IR and \oiii$_{88\, \rm{\micron}}$ emission. The \oiv$_{25.9}$/\oiii$_{88}$ can be considered as a proxy for the relative contribution of the non-thermal photoionisation.


In Figs\,\ref{fig:o4_o3vsne3_ne2}a,b, we compare the observed line ratios with our \textsc{Cloudy} photoionisation models for AGN (blue grid), LINERs (green grid), starburst galaxies (yellow grid), and dwarf galaxies (purple grid). From this comparison we conclude that:

\begin{enumerate}
  \item Observed Seyfert galaxies with the highest \oiv$_{25.9}$/\oiii$_{88}$ and \neiii$_{15.6}$/\neii$_{12.8}$ ratios (or \siv$_{10.5}$/\siii$_{18.7}$) are consistent with AGN models with densities in the $\log (n_{\rm H}/\rm{cm^3}) = 2.0$ to $4.0$ range and ionisation parameters in the $\log U = -2.0$ to $-3.0$ range.
  
  \item The transition from AGN dominated galaxies to star-forming galaxies defines a tail in the two diagrams that can be explained by: \textit{i)} decreasing ionisation ratios (Fig.\,\ref{fig:o4_o3vsne3_ne2}b); \textit{ii)} decreasing power-law index, as shown by LINER models; or \textit{iii)} increasing contribution to the \oiii$_{88\, \rm{\micron}}$ line by star formation in the host galaxy, as show by the \nev$_{14.3\, \rm{\micron}}$ to far-IR flux ratio.
    
  \item Extreme dwarf galaxy models with very high and thus unlikely densities ($\gtrsim 10^4\, \rm{cm^{-3}}$) are needed in order to reproduce the \oiv$_{25.9}$/\oiii$_{88}$ ratios measured in dwarf galaxies. These same ratios cannot be reproduced by the softer ionising spectra of starburst models, which underestimate the \oiv$_{25.9}$/\oiii$_{88}$ ratio by at least an order of magnitude even at the highest values of density and ionisation parameter. A solar metallicity starburst model with $1\, \rm{Myr}$ would increase the \neiii$_{15.6}$/\neii$_{12.8}$ line ratio by $\sim 40\%$ relative to the $20\, \rm{Myr}$ starburst model, but would not increase significantly the \oiv$_{25.9}$/\oiii$_{88}$ line ratio, since the contribution of the younger population does not increase considerably above $\gtrsim 54\, \rm{eV}$.

  \item LINER ratios Figs\,\ref{fig:o4_o3vsne3_ne2}a,b can be explained by a much steeper power law when compared to Seyfert nuclei with $\alpha \gtrsim -3.5$ and $-2.5 \lesssim \log U \lesssim -3.5$, in agreement with the absence of the big blue bump in their spectra \citep{ho96,ho08} and the predicted recession of the innermost part of the accretion disk in radiatively inefficient AGN \citep{nar95}. Additionally, shocks might play an important role in the excitation of IR lines in LINERs, as shown by \citet{stu06}.
\end{enumerate}
    
To reproduce the observed line ratios in dwarf galaxies for a gas with $\log (n_{\rm H}/\rm{cm^3}) = 2.0$--$3.0$ would require a ionising spectrum in the $\sim 55$--$35\, \rm{eV}$ range harder than that provided by our \textsc{Starburst99} model ($1\, \rm{Myr}$ and $Z = 0.004$), which rely on current stellar population synthesis models. The lack of ionising photons above $\approx 40\, \rm{eV}$, known as the ``\neiii'' problem, has been discussed in the literature \citep[e.g.][]{sel96,sim08,zas13}, specially at low metallicities where high-excitation lines are particularly bright \citep{sta15}. However, Figs\,\ref{fig:o4_o3vsne3_ne2}a,b prove that this disagreement is even more dramatic for starburst galaxies: the observations show that starburst galaxies are able to produce a much brighter \oiv$_{25.9\, \rm{\micron}}$ emission when compared to the models predictions. This line has an IP slightly above the He\,\textsc{ii} ionisation edge at $54\, \rm{eV}$, which cannot be produced by hot stars in the main sequence. \oiv$_{25.9\, \rm{\micron}}$ was previously detected in starburst galaxies by \citet{lut98}, and its origin has been associated to either X-ray emission produced by shocks in the stellar atmospheres or X-ray binaries \citep{sel96,izo12,sta15}, or to strong winds produced during the Wolf-Rayet (WR) phase of massive stars with $\sim 3$--$5\, \rm{Myr}$ \citep{sch99}. The latter could be the case for dwarf galaxies, which usually show other WR spectral features like He\,\textsc{ii} $4686\, \rm{\text{\AA}}$ emission, but it seems hardly the explanation for low-excitation starburst galaxies.

Starburst models with ages in the $3.5$--$5\, \rm{Myr}$, at the peak of the WR feedback contribution, are able to reproduce the observed \oiv$_{25.9}$/\oiii$_{88}$ ratios, but at the cost of over-predict \neiii$_{15.6}$/\neii$_{12.8}$ ratios by a factor of $\sim 30$, due to their harder ionising spectra. Considering this, and the short duration of the WR phase in a the lifetime of a star forming region ($\lesssim 1.5\, \rm{Myr}$), we consider unlikely the origin of high \oiv$_{25.9}$/\oiii$_{88}$ line ratios as caused by WR stars, specially in our samples of starburst galaxies. The similar \oiv$_{25.9}$/\oiii$_{88}$ ratios found for dwarf and starburst galaxies in our sample suggest that the strength of the ionising spectrum at $\sim 54\, \rm{eV}$ in star-forming regions scales with the \oiii$_{88\,\rm{\micron}}$ emission, as this line ratio does not vary with the metallicity of the stellar population.

High ionisation lines above $54\, \rm{eV}$ are expected to be more sensitive to the contribution of shocks in a starburst galaxy than lower ionisation lines, since the stellar contribution is negligible above that value. Among the low ionisation and neutral lines, shocks can contribute significantly to the \oi$_{63\, \rm{\micron}}$ emission \citep{hol89}, as shown by \citet{lut03} for the case of NGC\,6240, a LIRG merger. Furthermore, the \oi$_{63\, \rm{\micron}}$ line can be affected by self-absorption, as discussed in Section\,\ref{dens_ion_temp}, and depends on the PDR temperature. All these contributions make it difficult to evaluate the possible effect of shocks in our starburst sample, and also limit the utility of this line in the diagnostic diagrams. We expect that the shock contribution to the mid- and far-IR lines in this study, if present, would be significant only for line ratios involving the \oiv$_{25.9\, \rm{\micron}}$, the \nev$_{14.3, 24.3\, \rm{\micron}}$, and/or the \oi$_{63\, \rm{\micron}}$ lines.

\subsection{Metallicity}\label{metal}
\begin{figure*}
  \includegraphics[width=0.5\textwidth]{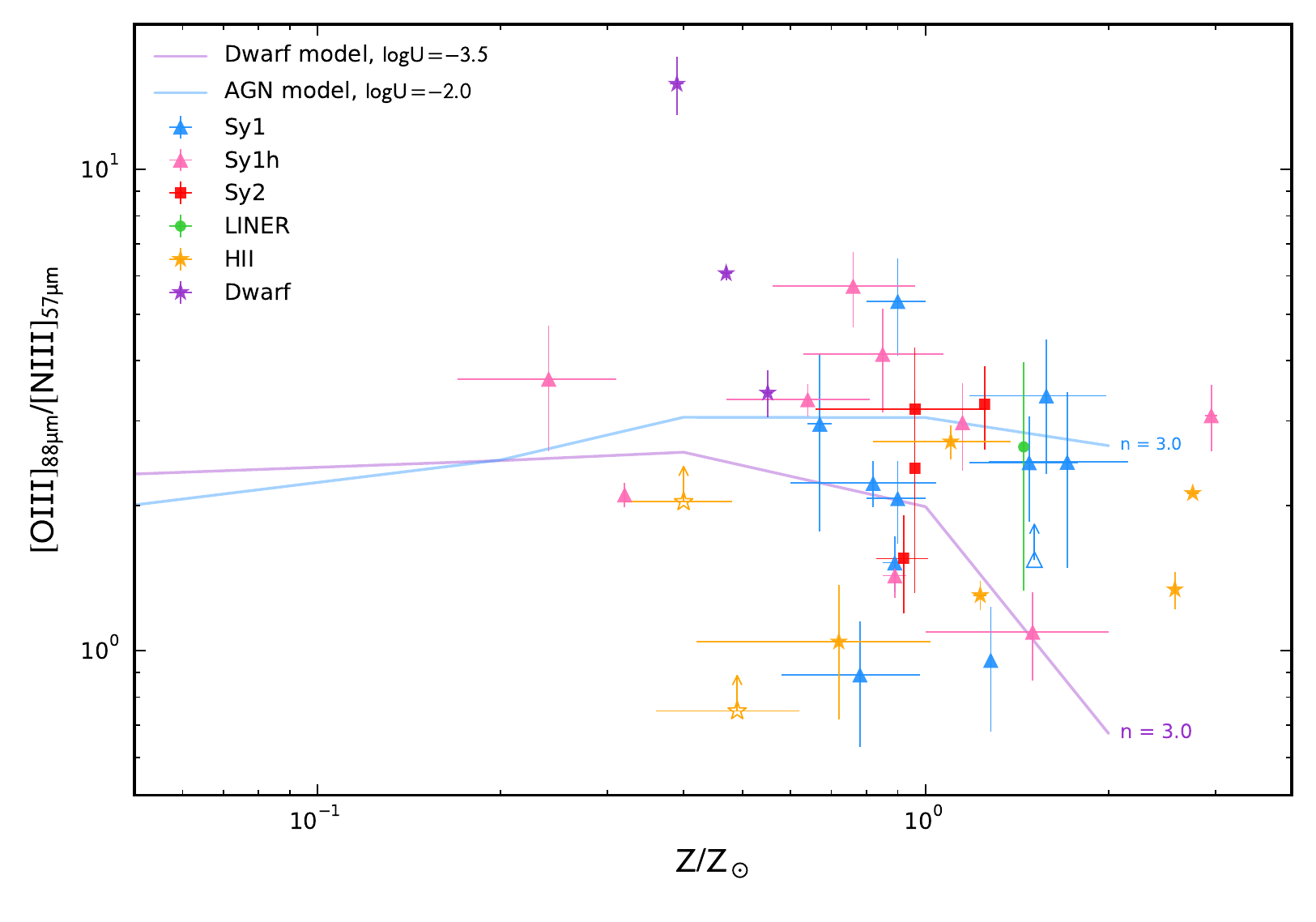}
  \includegraphics[width=0.5\textwidth]{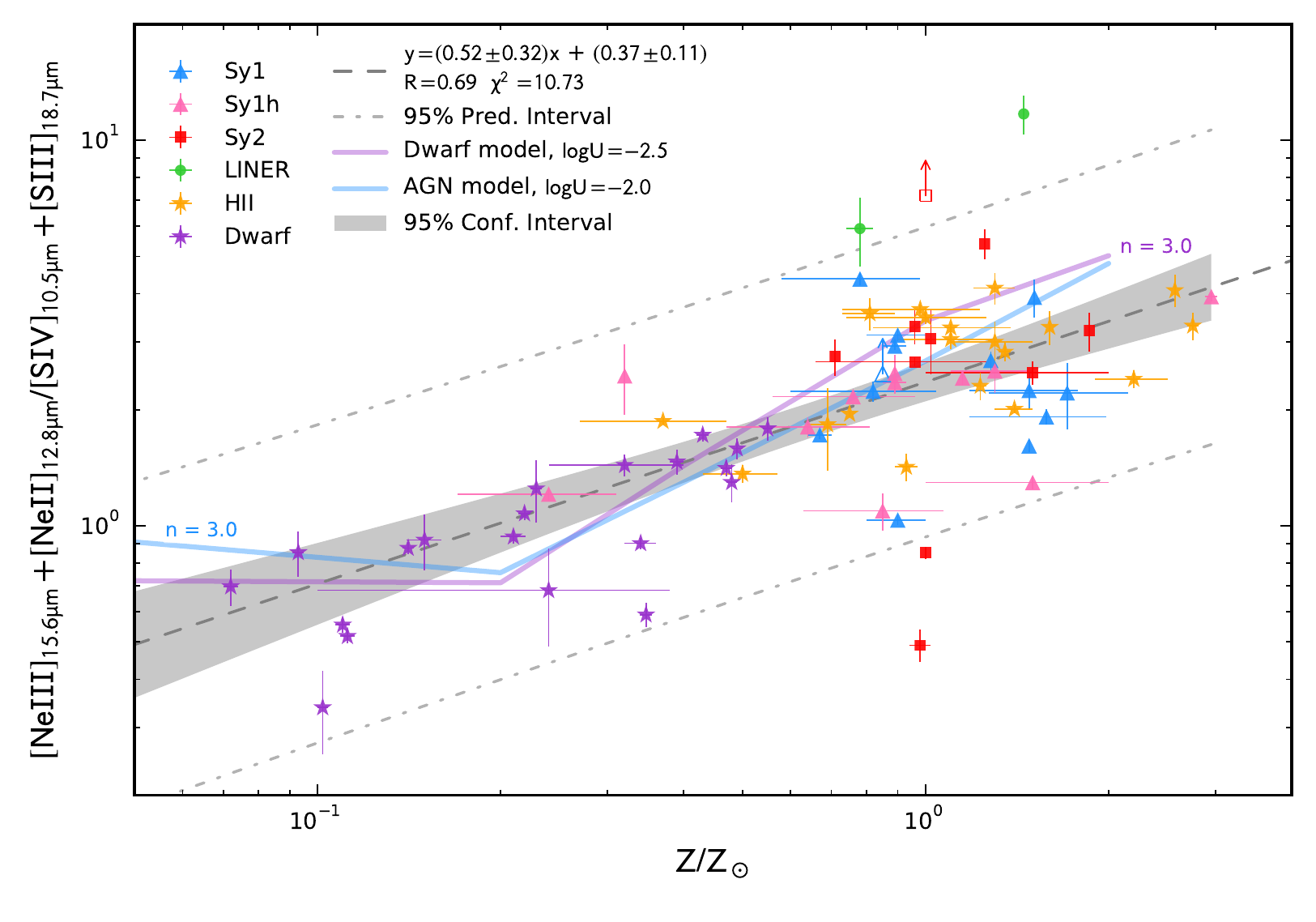} \\
  \caption{Metallicity-sensitive line ratios {\it vs} metallicities derived from optical line ratios (same notations as in Fig.\,\ref{fig:T_OI}). AGN models (in blue; $\log U = -2.0$, $\log (n_{\rm H}/\rm{cm^{-3}}) = 3$) and dwarf galaxy models (in purple; $\log U = -3.5$, $\log (n_{\rm H}/\rm{cm^{-3}}) = 3$) are also shown for different metallicities in the $1/20$--$2\, \rm{Z_\odot}$ range. 
{\bf Left (a):} [\textsc{O\,iii}]$_{88}$/[\textsc{N\,iii}]$_{57}$ line ratio {\it vs} optical metallicities.  
{\bf Right (b):} ([Ne\,\textsc{iii}]$_{15.6}$+[Ne\,\textsc{ii}]$_{12.8}$)/([\textsc{S\,iv}]$_{10.5}$+[\textsc{S\,iii}]$_{18.7}$) line ratio {\it vs} optical metallicities. Note that the AGN and dwarf galaxy models shown here include the effect of Sulphur depletion (see Section\,\ref{metal}).
}
\label{fig:metal}
\end{figure*}

As mentioned in Sect.\,\ref{intro}, to use diagnostics based on mid- and far-IR lines is a very useful tool for the study of galaxies at high-redshift. Of special interest are the metallicity-sensitive line ratios proposed by \citet{nag11,nag12}, which allow metallicity determinations in dust-embedded sources, e.g. ULIRGs (Rigopoulou 2015 priv. comm.), where optical line ratios cannot be measured reliably due to the high extinction. In this Section we test two metallicity-sensitive diagnostics by comparing the IR fine-structure line ratios with metallicity estimates based on optical lines: the \oiii$_{88}$/\niii$_{57}$ line ratio (Fig.\,\ref{fig:metal}a), proposed by \citet{nag11}, and a new diagram based on the (\neiii$_{15.6}$+\neii$_{12.8}$ )/(\siv$_{10.5}$+\siii$_{18.7}$) ratio (Fig.\,\ref{fig:metal}b), presented here for the first time.

Most of the optical metallicities shown in Figs\,\ref{fig:metal}a,b were compiled from the literature and are referenced in Table\,\ref{tbl_sample}. For a few objects without published metallicities but with published optical line fluxes (\nii$_{6548,6584\, \rm{\text{\AA}}}$, \oii$_{3727\, \rm{\text{\AA}}}$, \oiii$_{4959,5007\, \rm{\text{\AA}}}$, H$\alpha$, H$\beta$), we derived the optical metallicities using the \citet{pet04} and \citet{pil05} calibrations. 
The blue lines show the metallicity dependence predicted by our \textsc{Cloudy} AGN models for $\log U = -2.0$ and $\log (n_{\rm H}) = 3$ (blue-solid line). The purple-solid lines correspond to the dwarf galaxy models with $\log U = -3.5$ and $\log (n_{\rm H}) = 3$.

Fig.\,\ref{fig:metal}a shows a large dispersion with the observed \oiii$_{88}$/\niii$_{57}$ ratios distributed around the model predictions. No clear trend is found for the individual populations, although the low number of star-forming galaxies with observations of both lines (three dwarfs plus seven starburst galaxies) limits the assessment of this diagnostic. Our AGN models show a relatively constant ratio, while dwarf galaxy models show a decreasing ratio at super-solar metallicities, in agreement with the models in \citet{nag11}. This is related to the softening of the ionising spectrum at higher metallicities, i.e. the \oiii$_{88}$/\niii$_{57}$ in our models is mainly driven by the metallicity of the ionising stellar population, instead of the metallicity of the ionised gas. Continuous starburst models (not shown) show a behaviour very similar to that of dwarf galaxy models, with a slightly flatter dependency on metallicity. The dispersion of the observed ratios could be explained by different density and ionisation values, which should be determined independently in order to use the \oiii$_{88}$/\niii$_{57}$ ratio as a diagnostic for metallicity.

The (\neiii$_{15.6}$+\neii$_{12.8}$)/(\siv$_{10.5}$+\siii$_{18.7}$) ratio in Fig.\,\ref{fig:metal}b shows a clear correlation ($R = 0.69$) with optical metallicities, of the form
\begin{equation}
\begin{aligned}
  \log \left( \frac{\rm [NeIII]_{15.6} + [NeII]_{12.8}}{\rm [SIV]_{10.5} + [SIII]_{18.7}} \right) = & (0.53 \pm 0.32) \log \left( \tfrac{Z}{\rm Z_\odot} \right) \\
  & + (0.37 \pm 0.11)
\end{aligned}
\end{equation}
If only the dwarf and starburst galaxies are considered, the correlation is maintained with a higher correlation coefficient of $R = 0.89$. As shown in Sect.\,\ref{tracers}, both Neon and Sulphur ratios are very sensitive to the metallicity-dependent excitation of the stellar population ionising the gas. Thus, the (\neiii$_{15.6}$+\neii$_{12.8}$)/(\siv$_{10.5}$+\siii$_{18.7}$) ratio should cancel most of this dependency since \neii$_{12.8\, \rm{\micron}}$ (\siii$_{18.7\, \rm{\micron}}$) will turn into \neiii$_{15.6\, \rm{\micron}}$ (\siv$_{10.5\, \rm{\micron}}$) with increasing ionisation. The dependency of the (\neiii$_{15.6}$+\neii$_{12.8}$)/(\siv$_{10.5}$+\siii$_{18.7}$) ratio could be explained by the depletion of Sulphur on dust grains \citep[e.g.][]{ver03}, which causes an increasing Sulphur deficiency with increasing metallicity. In this scenario, the total (\siv$_{10.5}$+\siii$_{18.7}$) emission does not change significantly, since depletion maintains the Sulphur abundance roughly constant with increasing metallicity \citep[see fig.\,8 in][]{ver03}, while the total (\neiii$_{15.6}$+\neii$_{12.8}$) increases with the Neon abundance. In order to test this effect, we modified our \textsc{Cloudy} models by assuming a constant Sulphur abundance above $Z = 0.004$, while the Neon abundance increases with the metallicity. Instant starburst models (purple-solid line in Fig.\,\ref{fig:metal}), equivalent to dwarf galaxy models at $Z = 0.004$, show a similar trend as that shown by the observations, i.e. the (\neiii$_{15.6}$+\neii$_{12.8}$)/(\siv$_{10.5}$+\siii$_{18.7}$) ratio increases with metallicity due to the Sulphur depletion. Seyfert galaxies show line ratios in agreement with those of starburst galaxies, as well as AGN models (blue-solid line). This suggests that the (\neiii$_{15.6}$+\neii$_{12.8}$)/(\siv$_{10.5}$+\siii$_{18.7}$) ratio is a robust extinction-free metallicity tracer for both AGN and starburst galaxies. This is very important for the study of the dusty obscured galaxy evolution at redshift $1 < z < 4$, at the peak of the star formation and black hole accretion activity, that will be addressed by future space missions, e.g. \textit{SPICA} \citep{swi09}.

\section{Summary and Conclusions}\label{summ}

In this work we have presented the complete database of far-IR fine-structure lines observed by \textit{Herschel}/PACS during its 3.5 years operational life for 170 AGN-classified galaxies in the \citet{ver10} catalog. In order to complete the full mid- to far-IR spectra in the $10$--$600\, \rm{\micron}$ range we collected published fine-structure line fluxes measured with \textit{Spitzer}/IRS and \textit{Herschel}/SPIRE. As a comparison sample, we compiled an equivalent database for starburst galaxies extracted from \citet{bs09} and \citet{ga09}. Additionally, we included 43 dwarf galaxies from the Dwarf Galaxy Survey \citep{mad13,cor15}, in order to probe a more extreme star formation environment at low metallicities.

The photoionisation code \textsc{Cloudy} has been used to reproduce the physical conditions of the gas exciting the mid- to far-IR fine-structure lines. We produced four models using different ionising spectra: a pure AGN model ($\alpha = -1.4$; $S_\nu \propto \nu^{\alpha}$), a LINER model ($\alpha = -3.5$), a starburst galaxy model ($20\, \rm{Myr}$ continuum star formation with $Z = \rm{Z_\odot}$), and a dwarf galaxy model ($1\, \rm{Myr}$ instant starburst with $Z = 1/5\, \rm{Z_\odot}$). Calculations for the starburst and the dwarf galaxy models have been extended down to $T = 50\, \rm{K}$ in order to include the PDR emission region.

The main results of this study are:
\begin{itemize}
  
  \item The \oi$_{145/63}$ line ratio, used as temperature tracer in the $100$--$400\, \rm{K}$ range for the neutral gas in the PDR, does not show a clear correlation with the ionisation nor the density of the ionised gas, traced by the \siv$_{10.5}$/\siii$_{18.7}$ and the \siii$_{33.5/18.7}$ ratios, respectively. High \oi$_{145/63}$ ratios of $\gtrsim 0.1$ are found in objects with self-absorption in the \oi$_{63\, \rm{\micron}}$ line profile.
  
  \item The \ci$_{609/371}$ line ratio, sensitive to the temperature in the PDR ($20$--$100\, \rm{K}$), show similar median values for AGN ($0.53 \pm 0.21$) and starburst galaxies ($0.45 \pm 0.11$). This is much higher than the ratios predicted from XDR simulations ($\sim 0.15$--$0.19$), suggesting that the neutral carbon lines in our AGN sample are likely dominated by relatively low-density PDR emission originated in the ISM of the host galaxies.
    
  \item The density stratification found in S15 is confirmed here with 155 pairs of lines from \nii$_{205/122}$, \siii$_{33.5/18.7}$, \oiii$_{88/52}$, and \nev$_{24.3/14.3}$ line ratios, plus 93 upper limits. Both the least squares fit and the Kaplan-Meier residuals fit are consistent with an increasing gas density --\,measured by the line ratios\,-- with the ionisation potential (IP) of the transition. This suggests that harder radiation fields are traced by the higher density gas found in the innermost part of the NLR.


  \item The line ratios of S1 and S1h are always very similar, suggesting that both AGN types are indistinguishable from an IR perspective, thus S1h correspond to optically obscured S1 \citep[e.g.][]{tom10}.

  \item The \cii$_{158}$/\nii$_{122, 205}$ line ratios, sensitive to the relative contributions of the PDR and the low excitation photoionised gas, suggest a major contribution ($\gtrsim 80\%$) of PDR emission to the \cii$_{158\, \rm{\micron}}$ line. The observed ratios \cii$_{158}$/\nii$_{122, 205}$ $\gtrsim 3$ are in agreement with the simulations stopped at $T_{\rm stop} = 50\, \rm{K}$, which include PDR emission. Values lower than those observed are predicted for pure photoionised gas ($T_{\rm stop} = 1000\, \rm{K}$). An additional correlation of \cii$_{158}$/\nii$_{122, 205}$ with ionisation potential (traced by \neiii$_{15.6}$/\neii$_{12.8}$) might also be present, as \nii \ emission decreases in favour of \niii.


  \item The inclusion in the diagnostic diagrams of dwarf galaxies demonstrates that classical line ratios such as \oiv$_{25.9}$/\neii$_{12.8}$ fail as AGN/starburst tracers at low metallicities. Instead, the \oiv$_{25.9}$/\oiii$_{88}$ ratio is an excellent tracer to discriminate between non-thermal excitation in AGN and thermal ionisation produced by any kind of star formation activity. Since it probes a relatively hard range of the spectrum ($54.94$--$35.12\, \rm{eV}$, $4$--$2.6\, \rm{Ry}$) this line ratio is ideal for composite objects with extreme star formation bursts and mergers involving a chemically unevolved companion, a scenario that seems to be more common with increasing redshift. Alternatively, the \oiv$_{25.9}$/(\neiii$_{15.6}$+\neii$_{12.8}$) ratio can also be used as a proxy for the relative AGN to starburst contribution.

  \item A new AGN/star formation diagnostic diagram is proposed, based on the \oiv$_{25.9}$/\oiii$_{88}$ and \neiii$_{15.6}$/\neii$_{12.8}$ (or alternatively \siv$_{10.5}$/\siii$_{18.7}$). The \oiv$_{25.9}$/\oiii$_{88}$ ratio is sensitive to the relative contributions from AGN and star formation, while the \neiii$_{15.6}$/\neii$_{12.8}$ ratio (\siv$_{10.5}$/\siii$_{18.7}$) is sensitive to the ionisation, mostly driven by the metallicity in the stellar population. The combination of these line ratios allows to clearly separate the different populations in the diagram, i.e. AGN, dwarf galaxies, and starburst galaxies, and provides an extinction-free diagnostic ideal for future mid- and far-IR spectroscopic surveys.

  \item Photoionisation by current stellar atmosphere models cannot reproduce the observed \oiv$_{25.9}$/\oiii$_{88}$ line ratios due to the lack of predicted photons above the He\,\textsc{ii} ionisation edge ($\gtrsim 54\, \rm{eV}$, $4\, \rm{Ry}$). Furthermore, both dwarf and starburst galaxies show similar \oiv$_{25.9}$/\oiii$_{88}$ line ratios in the $5 \times 10^{-3}$--$3 \times 10^{-1}$ range, with no apparent dependency on the metallicity of the stellar population. An improved treatment of stellar atmosphere models above $54\, \rm{eV}$ will be necessary in order to explain the observed line ratios, e.g. by including the contribution of shocks and non-thermal emission in the EUV.

  \item LINERs show intermediate line ratios of \oiv$_{25.9}$/\oiii$_{88}$, \neiii$_{15.6}$/\neii$_{12.8}$, and \siv$_{10.5}$/ \siii$_{18.7}$, between Seyfert and starburst galaxies. These line ratios can be explained by assuming a softer UV spectrum with a slope of $\alpha \approx -3.5$, in line with the receding accretion disc scenario expected at low-luminosities and/or low-accretion efficiencies.

  \item The (\neiii$_{15.6}$+\neii$_{12.8}$)/(\siv$_{10.5}$+\siii$_{18.7}$) line ratio is sensitive to the gas metallicity for AGN, starburst and dwarf galaxies, this is confirmed by the models including the effect of Sulphur depletion. Thus, this ratio is proposed as a powerful extinction-free metallicity tracer, ideal for dust regions and obscured objects where optical determinations are not possible. Future facilities, such as \textit{JWST} for the Local Universe and \textit{SPICA} along Cosmic history, at the peak of star formation and black hole accretion activity ($1 < z < 4$), will be able to trace the build up of heavy elements during galaxy evolution.

\end{itemize}


\acknowledgments
The authors would like to thank the refere for his/her valuable comments which helped to improve the manuscript. The authors would also like to thank M. Melendez for his suggestions on \textsc{Cloudy} models. PACS has been developed by a consortium of institutes led by MPE (Germany) and including UVIE (Austria); KU Leuven, CSL, IMEC (Belgium); CEA, LAM (France); MPIA (Germany); INAF-IFSI/OAA/OAP/OAT, LENS, SISSA (Italy); IAC (Spain). This development has been supported by the funding agencies BMVIT (Austria), ESA-PRODEX (Belgium), CEA/CNES (France), DLR (Germany), ASI/INAF (Italy), and CICYT/MCYT (Spain). This research made use of Astropy, a community-developed core Python package for Astronomy (Astropy Collaboration, 2013).



\vspace{5mm}
\facilities{Herschel(PACS and SPIRE),Spitzer(IRS)}
\software{python,cloudy,pycloudy,astropy}

\clearpage



\figsetstart
\figsetnum{1}
\figsettitle{Extracted sprectra from the central spaxel for AGN and starburst galaxies.}

\figsetgrpstart
\figsetgrpnum{1.110}
\figsetgrptitle{IC\,4329A}
\figsetplot{./figset_spec/IC4329A_c.pdf}
\figsetgrpnote{Extracted spectrum from the central spaxel for IC\,4329A.}
\figsetgrpend

\figsetgrpstart
\figsetgrpnum{1.173}
\figsetgrptitle{NGC\,891}
\figsetplot{./figset_spec/NGC\,891_c.pdf}
\figsetgrpnote{Extracted spectrum from the central spaxel for NGC\,891.}
\figsetgrpend

\figsetend

\figsetstart
\figsetnum{2}
\figsettitle{Extracted sprectra from the $3 \times 3$ spaxel array for AGN and starburst galaxies.}

\figsetgrpstart
\figsetgrpnum{2.110}
\figsetgrptitle{IC\,4329A}
\figsetplot{./figset_spec/IC4329A_3x3.pdf}
\figsetgrpnote{Extracted spectrum from the $3 \times 3$ spaxel array for IC\,4329A.}
\figsetgrpend

\figsetgrpstart
\figsetgrpnum{2.173}
\figsetgrptitle{NGC\,891}
\figsetplot{./figset_spec/NGC\,891_3x3.pdf}
\figsetgrpnote{Extracted spectrum from the $3 \times 3$ spaxel array for NGC\,891.}
\figsetgrpend

\figsetend

\figsetstart
\figsetnum{3}
\figsettitle{Extracted sprectra from the $5 \times 5$ spaxel array for AGN and starburst galaxies.}

\figsetgrpstart
\figsetgrpnum{3.110}
\figsetgrptitle{IC\,4329A}
\figsetplot{./figset_spec/IC4329A_u5x5.pdf}
\figsetgrpnote{Extracted spectrum from the $5 \times 5$ spaxel array for IC\,4329A.}
\figsetgrpend

\figsetgrpstart
\figsetgrpnum{3.173}
\figsetgrptitle{NGC\,891}
\figsetplot{./figset_spec/NGC\,891_u5x5.pdf}
\figsetgrpnote{Extracted spectrum from the $5 \times 5$ spaxel array for NGC\,891.}
\figsetgrpend

\figsetend

\figsetstart
\figsetnum{4}
\figsettitle{Spectral maps obtained from the $5 \times 5$ spaxel array for the far-IR fine structure lines observed in the samples of AGN and starburst galaxies.}

\figsetgrpstart
\figsetgrpnum{4.627}
\figsetgrptitle{NGC\,891 [\textsc{O\,i}]$_{63\, \rm{\micron}}$}
\figsetplot{./figset_map/NGC\,891_[OI]63_map.pdf}
\figsetgrpnote{Spectral map of the $5 \times 5$ spaxel array for the [\textsc{O\,i}]$_{63\, \rm{\micron}}$ line in NGC\,891.}
\figsetgrpend

\figsetgrpstart
\figsetgrpnum{4.628}
\figsetgrptitle{NGC\,891 [\textsc{N\,ii}]$_{122\, \rm{\micron}}$}
\figsetplot{./figset_map/NGC\,891_[NII]122_map.pdf}
\figsetgrpnote{Spectral map of the $5 \times 5$ spaxel array for the [\textsc{N\,ii}]$_{122\, \rm{\micron}}$ line in NGC\,891.}
\figsetgrpend

\figsetend

\begin{figure*}
  \figurenum{4}
  \includegraphics[width=\textwidth]{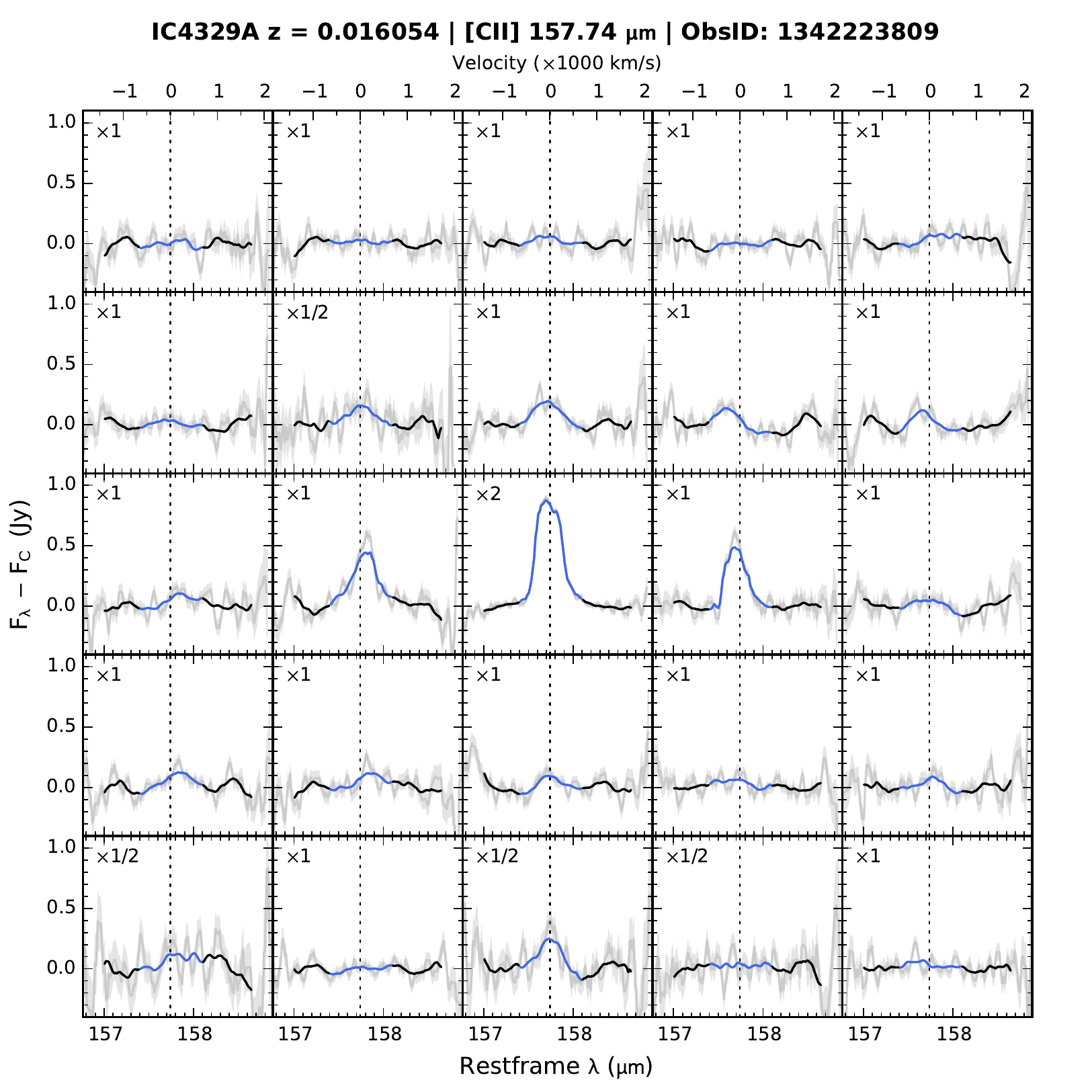}
  \caption{Spectral map of the full array of $5 \times 5$ spaxels, e.g. for the [\textsc{C\,ii}] transition in IC\,4329A. The continuum-subtracted, unfiltered spectra and its associated uncertainty are show in grey. The Wiener-filtered spectra (filter width: $2 \times$\,instrumental spectral \textsc{fwhm}; height: $3 \times$\textsc{rms}) are shown in blue (line spectral window) and black (continuum spectral window). In all cases, the continuum emission has been subtracted by fitting a 1-degree polynomial to the continuum windows, and the spectra are normalised by the factor indicated in the upper-left corner in each frame. Maps obtained for the far-IR fine structure lines observed in the samples of AGN and starburst galaxies are available in the Fig.\,Set\,\ref{fig_mapIC4329A}.\label{fig_mapIC4329A}}
\end{figure*}


\begin{figure*}
  \centering
  \includegraphics[width=0.8\textwidth]{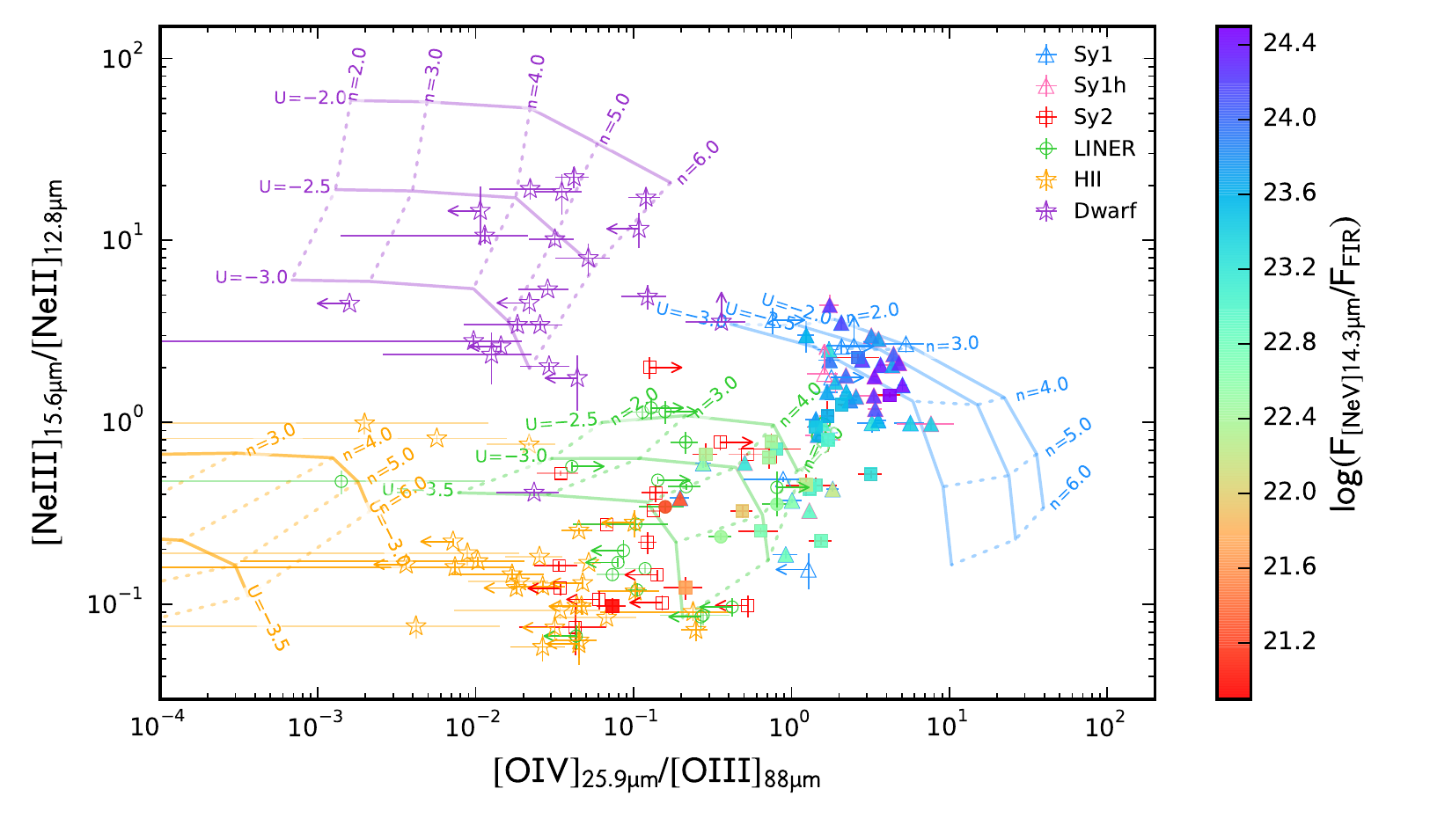}\\
  \includegraphics[width=0.8\textwidth]{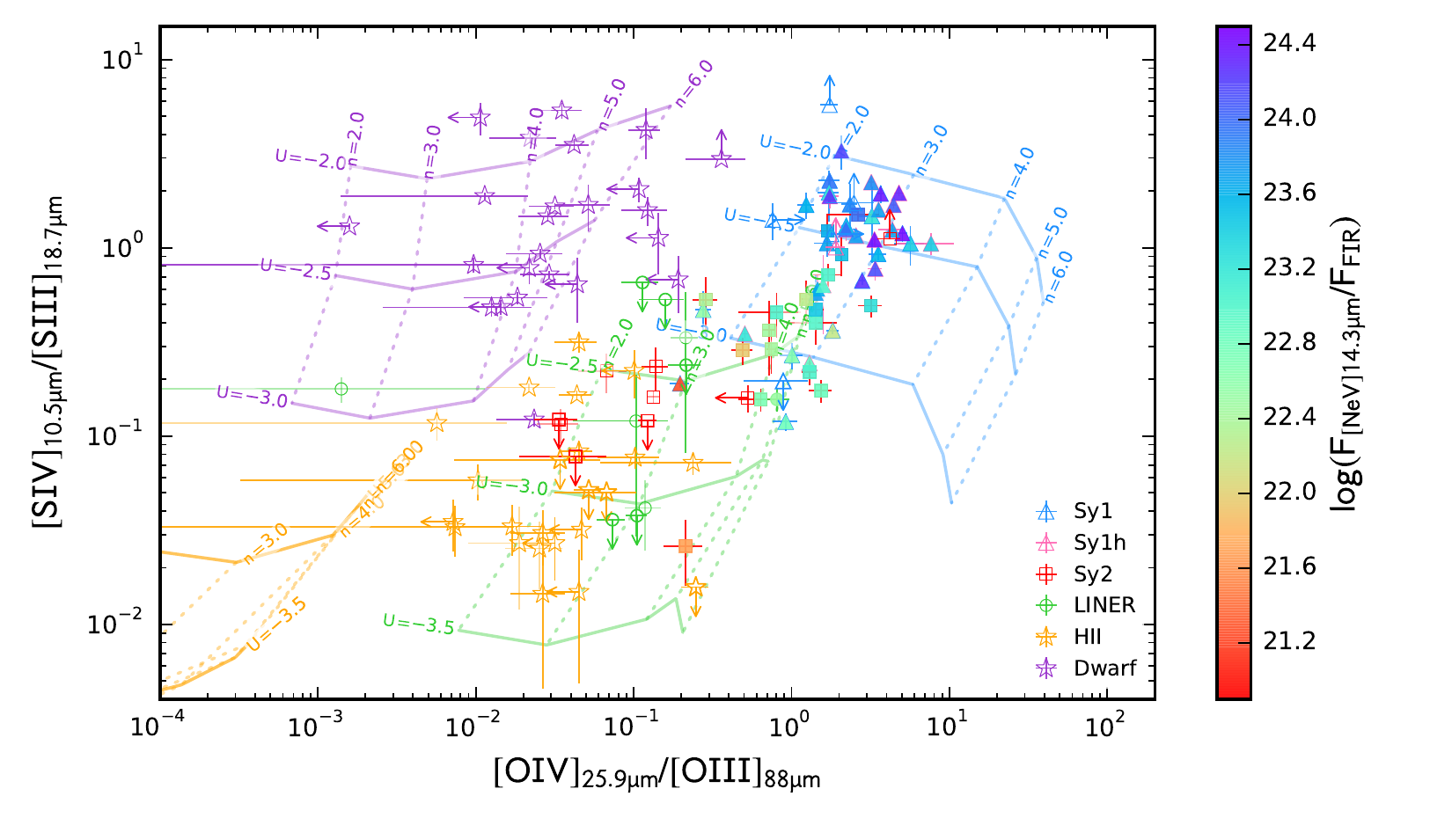}\\
\caption{Ionisation-sensitive line ratios (same notations as in Fig.\,\ref{fig:T_OI}). Photoionisation models of AGN, LINER, starburst galaxies, and dwarf galaxies are shown as blue, green, yellow, and purple grids, respectively. The logarithmic values of the density ($n_{\rm H}$) and ionisation potential ($U$) of the photoionisation models are indicated in the figures. Symbols are colour-coded according to their $\rm F_{\rm [Ne\,V]14.3}/F_{\rm FIR}$ flux ratio, when available (see colour bar). 
{\bf Top (a):} the [Ne\,\textsc{iii}]$_{15.6}$/[Ne\,\textsc{ii}]$_{12.8}$ line ratio {\it vs} the [\textsc{O\,iv}]$_{25.9}$/[\textsc{O\,iii}]$_{88}$ ratio. 
{\bf Bottom (b):} the [\textsc{S\,iv}]$_{10.5}$/[\textsc{S\,iii}]$_{18.7}$ line ratio {\it vs} the [\textsc{O\,iv}]$_{25.9}$/[\textsc{O\,iii}]$_{88}$ ratio.}
\label{fig:o4_o3vsne3_ne2}
\end{figure*}

\clearpage

\begin{deluxetable}{lcccccccccc}
\tabletypesize{\footnotesize}
\setlength{\tabcolsep}{0.3em}
\tablecolumns{11} 
\tablecaption{Fine-structure lines in the mid- to far-IR range. \label{tbl_lines}}
\tablehead{\colhead{Line} & \colhead{$\lambda$} & \colhead{$\nu$} &  \colhead{I.P.} &  \colhead{$E$} &  \colhead{$n_{\rm cr}$} & \colhead{Spec. Res.} & \colhead{Ang. Res.} & \colhead{\# of AGN} & \colhead{\# of SB} &  \colhead{\# of DW}  \\
  & ($\mu$m)                   & (GHz)                    & (eV) & (K) & ($\rm{cm^{-3}}$) &(km~s$^{-1}$)                        & (arcsec)                                  &                                  &                    &                       \\}
\startdata
\siv$^2$P$_{3/2}$--$^2$P$_{1/2}$   & \phn10.51 & 28524.50 & 34.79 & 1369 & $5.39 \times 10^4$ & $\sim 500$ & $4.7 \times 11.3$ & 112 & 19 & 30 \\
\neii$^2$P$_{1/2}$--$^2$P$_{3/2}$  & \phn12.81 & 23403.00 & 21.56 & 1123 & $7.00 \times 10^5$ & $\sim 500$ & $4.7 \times 11.3$ & 172 & 20 & 21 \\
\nev$^3$P$_2$--$^3$P$_1$          & \phn14.32 & 20935.23 & 97.12 & 1892 & $3 \times 10^4$ & $\sim 500$ & $4.7 \times 11.3$ & 115 & \phn0 & \phn1 \\
\neiii$^3$P$_1$--$^3$P$_2$        & \phn15.56 & 19266.87 & 40.96 & 925 & $2.68 \times 10^5$ & $\sim 500$ & $4.7 \times 11.3$ & 172 & 19 & 29 \\
\siii$^3$P$_2$--$^3$P$_1$         & \phn18.71 & 16023.11 & 23.34 & 769 & $2.22 \times 10^4$ & $\sim 500$ & $4.7 \times 11.3$ & 129 & 20 & 26 \\
\nev$^3$P$_1$--$^3$P$_0$          & \phn24.32 & 12326.99 & 97.12 & 596 & $5.0 \times 10^5$ & $\sim 500$ & $11.1 \times 22.3$ & \phn90 & \phn0 & \phn0 \\
\oiv$^2$P$_{3/2}$--$^2$P$_{1/2}$   & \phn25.89 & 11579.47 & 54.94 & 555 & $10^4$ & $\sim 500$ & $11.1 \times 22.3$ & 147 & 12 & 17 \\
\siii$^3$P$_1$--$^3$P$_0$          & \phn33.48 & \phn8954.37 & 23.34 & 430 & $7.04 \times 10^3$ & $\sim 500$ & $11.1 \times 22.3$ & 110 & 20 & 21 \\
\,\hbox{[\ion{Si}{2}]}$^2$P$_{3/2}$--$^2$P$_{1/2}$   & \phn34.81 & \phn8612.25 & \phn8.15 & 413 & $3.4 \times 10^5$\tablenotemark{a},$10^3$ & $\sim 500$ & $11.1 \times 22.3$ & 100 & 20 & 20 \\
\oiii$^3$P$_2$--$^3$P$_1$          & \phn51.81 & \phn5787.57 & 35.12 & 441 & $3.6 \times 10^3$ & $\sim 105$ & \phn9.4\tablenotemark{b} & \phn19 & \phn1 & \phn0 \\
\niii$^2$P$_{3/2}$--$^2$P$_{1/2}$   & \phn57.32 & \phn5230.43 & 29.60 & 251 & $3.0 \times 10^3$ & $\sim 105$  & \phn9.4 &  \phn32 & \phn5 & \phn3 \\ 
\oi$^3$P$_2$--$^3$P$_1$            & \phn63.18 & \phn4744.77 & -- & 228 & $4.7 \times 10^5$\tablenotemark{a} & $\sim \phn86$ & \phn9.4 & 115 & 20 & 31 \\
\oiii$^3$P$_1$--$^3$P$_0$          & \phn88.36 & \phn3393.01 & 35.12 & 163 & $510$ & $\sim 124$ & \phn9.4 &  \phn83 & 16 & 38 \\
\nii$^3$P$_2$--$^3$P$_1$           &   121.90  & \phn2459.38 & 14.53 & 118 & $310$ & $\sim 290$  & \phn9.4 &  \phn79 & 17 & \phn8 \\
\oi$^3$P$_1$--$^3$P$_0$            &   145.52  & \phn2060.07 & -- & 98 & $9.5 \times 10^4$\tablenotemark{a} & $\sim 256$ & 10.3 & \phn66 &  \phn8 & 12 \\
\cii$^2$P$_{3/2}$--$^2$P$_{1/2}$    &   157.74  & \phn1900.54 & 11.26 & 91 & $50$,$2.8 \times 10^3$\tablenotemark{a} & $\sim 238$ &   11.2 & 159 & 17 & 41 \\
\nii$^3$P$_1$--$^3$P$_0$           &   205.3  & \phn1460.27 & 14.53 & 70 & $48$ & $\sim 297$ & $\sim16.8$ & \phn60 & 13 & \phn2 \\
\ci$^3$P$_{2}$--$^3$P$_{1}$        &   370.37  & \phn\phn809.44 & -- & 38.9 & $2.8 \times 10^{3}$\tablenotemark{a} & $\sim 536$ & $\sim 35.0$ & \phn59 & 13 & \phn1 \\
\ci$^3$P$_{1}$--$^3$P$_{0}$         &   609.7  & \phn\phn491.70 & -- & 23.6 &$4.7 \times 10^{2}$\tablenotemark{a} & $\sim 882$ & $\sim 35.0$ & \phn32 & 12 & \phn0 \\
\enddata
\tablenotetext{a}{Critical density for collisions with hydrogen atoms.}
\tablenotetext{b}{The beam size for \textit{Herschel}/PACS is dominated by the spaxel size ($9\farcs4$) below $\sim 120\, \rm{\micron}$.}
\tablecomments{The columns correspond to the central wavelength, frequency, ionisation potential, excitation temperature, critical density, spectral and spatial resolution of the data presented in this work, and the number of AGN, starburst, and dwarfs galaxies with line detections above $3 \times$\,\textsc{rms}. Critical densities and excitation temperatures from: \citet{lau77,th85,gre93,stu02,cor12,far13}.}

\end{deluxetable}

\clearpage

\begin{deluxetable}{ccc}
\tabletypesize{\footnotesize}
\tablecaption{Summary of the line ratios discussed in the paper and their associated diagnostic.\label{tbl_guide}}
\tablehead{
\colhead{Line Ratio} & \colhead{Diagnostic} & \colhead{Parameter range} \\
}
\startdata
    \nii$_{205/122}$ & Density of ionised gas & $1$--$10^3\, \rm{cm^{-3}}$ \\
    \oiii$_{88/52}$ & Density of ionised gas & $10$--$10^{4.5}\, \rm{cm^{-3}}$ \\
    \siii$_{33.5/18.7}$ & Density of ionised gas & $10$--$10^5\, \rm{cm^{-3}}$ \\
    \nev$_{24.3/14.3}$ & Density of ionised gas & $100$--$10^6\, \rm{cm^{-3}}$ \\
    \siv$_{10.5}$/\siii$_{18.7}$ & Ionisation parameter & $34.8$--$23.3\, \rm{eV}$ \\
    \oiii$_{88}$/\oi$_{63}$ & Ionised/neutral gas ratio & -- \\
    \oiii$_{88}$/\nii$_{122}$ & Ionisation parameter & $35.1$--$14.5\, \rm{eV}$ \\
    \neiii$_{15.6}$/\neii$_{12.8}$ & Ionisation parameter & $41.0$--$21.6\, \rm{eV}$ \\
    \oiv$_{25.9}$/\oiii$_{88}$ & Ionisation parameter, AGN/Starburst & $54.9$--$35.1\, \rm{eV}$ \\
    \oi$_{145/63}$ & Temperature of neutral gas (PDR) & $100$--$400\, \rm{K}$ \\
    \ci$_{609/371}$ & Temperature of neutral gas (PDR) & $20$--$100\, \rm{K}$ \\
    \cii$_{158}$/\oi$_{63}$ & PDR density & -- \\
    \cii$_{158}$/\nii$_{122}$ & PDR/low-excitation ionised gas contribution & -- \\
    \nii$_{122}$/\ci$_{371}$ & Low-excitation ionised gas/neutral gas contribution & -- \\
    \nii$_{122}$/\ci$_{371}$ & Low-excitation ionised gas/neutral gas contribution & -- \\
    \oiv$_{25.9}$/\neii$_{12.8}$ & AGN/Starburst contribution & -- \\
    \oiv$_{25.9}$/(\neii$_{12.8}$+\neiii$_{15.6}$) & AGN/Starburst contribution & -- \\
    \oiii$_{88}$/\niii$_{57}$ & Ionisation parameter and metallicity & $35.1$--$29.6\, \rm{eV}$ \\
    (\neii$_{12.8}$+\neiii$_{15.6}$)/(\siii$_{18.7}$+\siv$_{10.5}$) & Metallicity & $\sim 0.07$--$2.5\, \rm{Z_\odot}$ \\
\enddata
\end{deluxetable}


\floattable
\begin{deluxetable}{rlccccccccccccccc}
\rotate  
\tabletypesize{\footnotesize}
\setlength{\tabcolsep}{0.3em}
\tablecolumns{17}
\tablewidth{0pt}
\tablecaption{Samples of AGN, starburst, and dwarf galaxies observed with the PACS spectrometer.\label{tbl_sample}}
\tablehead{
\colhead{\#} & \colhead{Name} & \colhead{R.A. (2000)} & \colhead{Dec. (2000)}  & \colhead{$z$} & \colhead{Dist.} & \colhead{RefD\tablenotemark{a}} & \colhead{Type} & \colhead{$L_{\rm{2-10keV}}$} & \colhead{CT} & \colhead{RefX\tablenotemark{b}} & \colhead{$L_{\rm{[NeV]14.3}}$} & \colhead{$L_{\rm{[NeV]24.3}}$} & \colhead{$L_{\rm{[OIV]25.9}}$} & \colhead{$Z$} & \colhead{RefZ\tablenotemark{c}} & \colhead{Alt.} \\[1pt]
 & & (h:m:s) & (d:m:s) &  & (Mpc) & & & ($\rm{erg/s})$ & &   &   ($\rm{erg/s}$) &  ($\rm{erg/s}$) & ($\rm{erg/s}$) & ($\rm{Z_\odot}$) & & }

\startdata
\multicolumn{17}{c}{\bf AGN sample} \\
\cline{1-17}
\setcounter{magicrownumbers}{0}
\rownumber & Mrk334         & 00:03:09.6038 & +21:57:36.8064 & 0.021945 & 96.6 & 29 & S1.8 & \nodata &   &    & 41.16    & 41.09    & 41.22    & & & UGC00006 \\
\rownumber & Mrk938         & 00:11:06.5412 & -12:06:27.6624 & 0.019617 & 81.8 &    & S2   & 42.33   & N & 25 & $<$40.24 & $<$39.47 & $<$39.72 & & & NGC17, NGC34 \\
\rownumber & IRAS00182-7112 & 00:20:34.7210 & -70:55:26.2488 & 0.326999 & 1656.9 &    & S2   & 43.90   & Y & 44 & $<$41.63 & $<$41.82 & $<$42.06 & & & \\

\cline{1-17} \\
\multicolumn{17}{c}{\bf Starburst sample} \\
\cline{1-17}
\setcounter{magicrownumbers}{0}
\rownumber & NGC253 & 00:47:33.0727 & -25:17:18.9960 & 0.000811 & 3.66 & 11 & \textsc{H\,ii} & 39.96 & N & 43 & $<$38.52 & $<$39.07 & 39.39    & 0.72$\pm$0.30 & 13 & Sculptor \\
\rownumber & M74    & 01:36:41.7384 & +15:47:00.8592 & 0.002192 & 9.46 & 23 & \textsc{H\,ii} & 38.12 &   & 26 & $<$37.21 & $<$37.69 & $<$37.14 & 0.46          &  8 & NGC628 \\
\rownumber & NGC891 & 02:22:32.8464 & +42:20:52.6740 & 0.001761 & 10.8 & 31 & \textsc{H\,ii} & 40.46 &   & 67 & $<$36.75 & $<$37.94 & $<$38.16 & 1.6           &  6 & UGC01831 \\

\cline{1-17} \\
\multicolumn{17}{c}{\bf Dwarf sample} \\
\cline{1-17}
\setcounter{magicrownumbers}{0}
\rownumber & HS0017+1055 & 00:20:21.4000 & +11:12:21.0000 &  0.018846 &  78.6 & & Dwarf & \nodata &   &    & \nodata  & \nodata & \nodata  & 0.09$\pm$0.02 & 10 & \\ 
\rownumber & Haro11      & 00:36:52.4544 & -33:33:16.7652 &  0.020598 &  86.0 & & Dwarf & 40.98   & N & 23 & $<$39.67 & \nodata & 40.59    & 0.47$\pm$0.01 & 10 & \\
\rownumber & HS0052+2536 & 00:54:56.3647 & +25:53:08.0052 &  0.045385 & 193.0 & & Dwarf & \nodata &   &    & $<$40.52 & \nodata & $<$40.14 & 0.24$\pm$0.14 & 10 & \\

\cline{1-17} \\
\multicolumn{17}{c}{\bf No lines detected by PACS} \\
\cline{1-17}
\setcounter{magicrownumbers}{0}
\rownumber & 3C33         & 01:08:52.8818 & +13:20:14.1684 & 0.0597   & 256.6 & & S1h  & 43.86 & N & 64 & 41.20 & 41.10 & 41.80 & 0.89 & 3 & 4C+13.07 \\
\rownumber & PG1114+445   & 11:17:06.3950 & +44:13:33.2076 & 0.143862 & 655.5 & & S1.0 & 43.95 & N & 51 & \nodata & $<$41.94 & 42.04 & & & \\
\rownumber & ESO506-G27   & 12:38:54.5820 & -27:18:28.1376 & 0.025024 & 104.8 & & S2   & 43.06 & N & 70 & \nodata & \nodata & 40.70 & & & AM1236-270 \\

\enddata
\tablenotetext{a}{References for redshift-independent distances:
(1)\,\citet{ada12};
(2)\,\citet{cor08};
(3)\,\citet{dk02}.}
\tablenotetext{b}{References for X-ray absorption-corrected luminosities:
(1)\,\citet{awa00};
(2)\,\citet{bal11};
(3)\,\citet{bas99}.}
\tablenotetext{c}{References for optical metallicities:
(1)\,\citet{con12};
(2)\,\citet{dav13};
(3)\,\citet{dor15}.}

\tablecomments{The columns correspond to the galaxy name, coordinates taken from 2MASS Point Source Catalog, redshift, redshift-independent distance, reference for the distance, spectral type, absorption-corrected $2$--$10\, \rm{keV}$ luminosity, Compton thick (Y: yes; N: no), reference for the X-ray luminosity, \nev$_{14.3\, \rm{\micron}}$, \nev$_{24.3\, \rm{\micron}}$, and \oiv$_{25.9\, \rm{\micron}}$ luminosities, metallicity based on optical lines, reference for the metallicity, and alternative name(s). Table\,\ref{tbl_sample} is published in its entirety in the machine readable format. A portion is shown here for guidance regarding its form and content.}
\end{deluxetable}

\clearpage

\begin{deluxetable}{rlcccccccc}
\tabletypesize{\footnotesize}
\tablecolumns{10}
\tablewidth{0pt}
\tablecaption{Far-IR fine structure line fluxes from \textit{Herschel}/PACS.\label{tbl_pacs_flux}}
\tablehead{
 &  &  \multicolumn{7}{c}{Line fluxes ($10^{-17}\, \rm{W\,m^{-2}}$)} &   \\ 
\cline{3-9}
\colhead{\#} & \colhead{Name} & \colhead{\oiii} & \colhead{\niii}  & \colhead{\oi} & \colhead{\oiii} & \colhead{\nii} & \colhead{\oi} & \colhead{\cii} & \colhead{Notes} \\ [1pt]
 & & $51.81\, \rm{\micron}$ & $57.32\, \rm{\micron}$ & $63.18\, \rm{\micron}$ & $88.36\, \rm{\micron}$ & $121.90\, \rm{\micron}$ & $145.52\, \rm{\micron}$ & $157.74\, \rm{\micron}$ & }

\startdata
\multicolumn{10}{c}{\bf AGN} \\
\cline{1-10}
\setcounter{magicrownumbers}{0}
\rownumber & Mrk334 & \nodata & \nodata & \nodata & \nodata & \nodata & \nodata & 76.66$\pm$1.77 & 3$\times$3 \\
\rownumber & Mrk938 & \nodata & \nodata & 78.12$\pm$6.36 & 19.24$\pm$3.58 & \nodata & 5.26$\pm$0.92 & 82.15$\pm$1.51 & 3$\times$3 \\
\rownumber & IRAS00182-7112 & \nodata & \nodata & 7.10$\pm$1.67 & $<$2.04 & \nodata & \nodata & \nodata & c \\

\cline{1-10} \\
\multicolumn{10}{c}{\bf Starburst galaxies}  \\
\cline{1-10}
\setcounter{magicrownumbers}{0}
\rownumber & NGC253 & 389.21$\pm$183.61 & 599.54$\pm$178.33 & 2947.6$\pm$85.7 & 625.21$\pm$53.54 & 802.72$\pm$28.14 & 389.61$\pm$30.99 & 3392.7$\pm$22.9 & 3$\times$3 \\
\rownumber & M74 & \nodata & \nodata & 20.58$\pm$10.84 & $<$14.16 & 5.85$\pm$2.70 & \nodata & 43.12$\pm$1.82 & 3$\times$3 \\
\rownumber & NGC891 & \nodata & \nodata & 85.54$\pm$11.08 & \nodata & 73.28$\pm$2.19 & \nodata & \nodata & 3$\times$3 \\

\cline{1-10} \\
\multicolumn{10}{c}{\bf No lines detected by PACS}  \\
\cline{1-10}
\setcounter{magicrownumbers}{0}
\rownumber & 3C33 & \nodata & \nodata & \nodata & \nodata & \nodata & $<$1.54 & \nodata & 3$\times$3 \\
\rownumber & PG1114+445 & \nodata & \nodata & $<$3.06 & \nodata & \nodata & \nodata & $<$1.3 & 3$\times$3 \\
\rownumber & ESO506-G27 & \nodata & \nodata & \nodata & \nodata & \nodata & $<$0.96 & \nodata & 3$\times$3 \\

\enddata

\tablecomments{Far-IR fine structure line fluxes for the samples of AGN and starburst galaxies, from {\it Herschel} PACS observations. Table\,\ref{tbl_pacs_flux} is published in its entirety in the machine readable format. A portion is shown here for guidance regarding its form and content.}

\end{deluxetable}

\clearpage

\begin{deluxetable}{rlcccccccc}
\tabletypesize{\footnotesize}
\tablecolumns{10} 
\tablewidth{0pt}
\tablecaption{Equivalent width for far-IR lines from \textit{Herschel}/PACS.\label{tbl_pacs_width}}
\tablehead{
 & & \multicolumn{7}{c}{Equivalent Width ($\rm{\micron}$)} & \\
\cline{3-9}  \\
\colhead{\#} & \colhead{Name} & \colhead{\oiii} & \colhead{ \niii}  & \colhead{\oi} & \colhead{\oiii} & \colhead{\nii} & \colhead{\oi} & \colhead{\cii} & \colhead{Notes} \\ [1pt]
 & & $51.81\, \rm{\micron}$ & $57.32\, \rm{\micron}$ & $63.18\, \rm{\micron}$ & $88.36\, \rm{\micron}$ & $121.90\, \rm{\micron}$ & $145.52\, \rm{\micron}$ & $157.74\, \rm{\micron}$ & }

\startdata
\multicolumn{10}{c}{\bf AGN} \\
\cline{1-10}
\setcounter{magicrownumbers}{0}
\rownumber & Mrk334 & \nodata & \nodata & \nodata & \nodata & \nodata & \nodata & 2.252$\pm$0.193 & 3$\times$3 \\
\rownumber & Mrk938 & \nodata & \nodata & 0.057$\pm$0.005 & 0.026$\pm$0.005 & \nodata & 0.032$\pm$0.006 & 0.656$\pm$0.016 & 3$\times$3 \\
\rownumber & IRAS00182-7112 & \nodata & \nodata & 0.064$\pm$0.015 & $<$0.049 & \nodata & \nodata & \nodata & c \\

\cline{1-10} \\
\multicolumn{10}{c}{\bf Starburst galaxies} \\
\cline{1-10}
\setcounter{magicrownumbers}{0}
\rownumber & NGC253 & 0.008$\pm$0.004 & 0.007$\pm$0.002 & 0.039$\pm$0.001 & 0.013$\pm$0.001 & 0.037$\pm$0.001 & 0.031$\pm$0.003 & 0.374$\pm$0.003 & 3$\times$3 \\
\rownumber & M74 & \nodata & \nodata & 0.017$\pm$0.009 & \nodata & 0.014$\pm$0.006 & \nodata & 0.430$\pm$0.026 & 3$\times$3 \\
\rownumber & NGC891 & \nodata & \nodata & 0.035$\pm$0.005 & \nodata & 0.080$\pm$0.002 & \nodata & \nodata & 3$\times$3 \\

\cline{1-10} \\
\multicolumn{10}{c}{\bf No lines detected by PACS} \\
\cline{1-10}
\setcounter{magicrownumbers}{0}
\rownumber & 3C33 & \nodata & \nodata & \nodata & \nodata & \nodata & $<$0.352 & \nodata & 3$\times$3 \\
\rownumber & PG1114+445 & \nodata & \nodata & \nodata & \nodata & \nodata & \nodata & \nodata & 3$\times$3 \\
\rownumber & ESO506-G27 & \nodata & \nodata & \nodata & \nodata & \nodata & $<$0.051 & \nodata & 3$\times$3 \\

\enddata

\tablecomments{Equivalent width values derived for the far-IR fine structure lines observed by \textit{Herschel}/PACS. Table\,\ref{tbl_pacs_width} is published in its entirety in the machine readable format. A portion is shown here for guidance regarding its form and content.}

\end{deluxetable}

\clearpage

\begin{deluxetable}{rlcccccccc}
\tabletypesize{\footnotesize}
\tablecolumns{10} 
\tablewidth{0pt}
\tablecaption{Far-IR continuum fluxes from \textit{Herschel}/PACS spectra.\label{tbl_pacs_cont}}
\tablehead{
 & & \multicolumn{7}{c}{Continuum Flux ($\rm{Jy}$)} & \\
\cline{3-9} \\
\colhead{\#} & \colhead{Name} & \colhead{\oiii} & \colhead{\niii}  & \colhead{\oi} & \colhead{\oiii} & \colhead{\nii} & \colhead{\oi} & \colhead{\cii} & \colhead{Notes} \\ [1pt]
 & & $51.81\, \rm{\micron}$ & $57.32\, \rm{\micron}$ & $63.18\, \rm{\micron}$ & $88.36\, \rm{\micron}$ & $121.90\, \rm{\micron}$ & $145.52\, \rm{\micron}$ & $157.74\, \rm{\micron}$ & }

\startdata
\multicolumn{10}{c}{\bf AGN}  \\
\cline{1-10}
\setcounter{magicrownumbers}{0}
\rownumber & Mrk334 & \nodata & \nodata & \nodata & \nodata & \nodata & \nodata & 2.83$\pm$0.23 & 3$\times$3 \\
\rownumber & Mrk938 & \nodata & \nodata & 18.11$\pm$0.51 & 19.23$\pm$0.31 & \nodata & 11.54$\pm$0.16 & 10.39$\pm$0.17 & 3$\times$3 \\
\rownumber & IRAS00182-7112 & \nodata & \nodata & 1.49$\pm$0.04 & 1.09$\pm$0.03 & \nodata & \nodata & \nodata & c \\

\cline{1-10} \\
\multicolumn{10}{c}{\bf Starburst galaxies}  \\
\cline{1-10}
\setcounter{magicrownumbers}{0}
\rownumber & NGC253 & 450.23$\pm$14.16 & 936.35$\pm$8.24 & 1013.0$\pm$5.8 & 1206.1$\pm$5.4 & 1071.4$\pm$3.2 & 874.91$\pm$4.07 & 753.27$\pm$3.54 & 3$\times$3 \\
\rownumber & M74 & \nodata & \nodata & 16.19$\pm$0.91 & \nodata & 20.79$\pm$0.25 & \nodata & 8.32$\pm$0.36 & 3$\times$3 \\
\rownumber & NGC891 & \nodata & \nodata & 32.97$\pm$1.16 & \nodata & 45.68$\pm$0.27 & \nodata & 351$\pm$28\tablenotemark{a} & 3$\times$3 \\

\cline{1-10} \\
\multicolumn{10}{c}{\bf No lines detected by PACS}  \\
\cline{1-10}
\setcounter{magicrownumbers}{0}
\rownumber & 3C33 & \nodata & \nodata & \nodata & \nodata & \nodata & 0.31$\pm$0.08 & \nodata & 3$\times$3 \\
\rownumber & PG1114+445 & \nodata & \nodata & $<$0.34 & \nodata & \nodata & \nodata & $<$0.25 & 3$\times$3 \\
\rownumber & ESO506-G27 & \nodata & \nodata & \nodata & \nodata & \nodata & 1.34$\pm$0.05 & \nodata & 3$\times$3 \\

\enddata

\tablenotetext{a}{\textit{Herschel}/PACS continuum at $160\, \rm{\micron}$ taken from \citet{hug14}.}
\tablecomments{Continuum flux values derived for the far-IR fine structure lines observed by \textit{Herschel}/PACS. Table\,\ref{tbl_pacs_cont} is published in its entirety in the machine readable format. A portion is shown here for guidance regarding its form and content.}

\end{deluxetable}

\clearpage

\begin{deluxetable}{rlcccccccc}
\tabletypesize{\footnotesize}
\tablecolumns{10} 
\tablewidth{0pt}
\tablecaption{Gaussian sigma for far-IR fine structure lines from \textit{Herschel}/PACS.\label{tbl_pacs_sigma}}
\tablehead{
 & & \multicolumn{7}{c}{Gaussian Sigma ($\rm{km\,s^{-1}}$)} & \\
\cline{3-9} \\
 \colhead{\#} & \colhead{Name} & \colhead{\oiii} & \colhead{\niii}  & \colhead{\oi} & \colhead{\oiii} & \colhead{\nii} & \colhead{\oi} & \colhead{\cii} & \colhead{Notes} \\ [1pt]
 & & $51.81\, \rm{\micron}$ & $57.32\, \rm{\micron}$ & $63.18\, \rm{\micron}$ & $88.36\, \rm{\micron}$ & $121.90\, \rm{\micron}$ & $145.52\, \rm{\micron}$ & $157.74\, \rm{\micron}$ & }

\startdata
\multicolumn{10}{c}{\bf AGN} \\
\cline{1-10}
\setcounter{magicrownumbers}{0}
\rownumber & Mrk334 & \nodata & \nodata & \nodata & \nodata & \nodata & \nodata & 151.8$\pm$0.8 & 3$\times$3 \\
\rownumber & Mrk938 & \nodata & \nodata & 165.9$\pm$2.3 & 195.1$\pm$8.0 & \nodata & 158.4$\pm$6.3 & 185.8$\pm$0.7 & 3$\times$3 \\
\rownumber & IRAS00182-7112 & \nodata & \nodata & 632.6$\pm$42.7 & \nodata & \nodata & \nodata & \nodata & c \\

\cline{1-10} \\
\multicolumn{10}{c}{\bf Starburst galaxies} \\
\cline{1-10}
\setcounter{magicrownumbers}{0}
\rownumber & NGC253 & 142.0$\pm$13.0 & 202.3$\pm$10.8 & 145.4$\pm$0.8 & 125.9$\pm$1.5 & 169.0$\pm$1.2 & 148.9$\pm$2.2 & 136.8$\pm$0.2 & 3$\times$3 \\
\rownumber & M74 & \nodata & \nodata & 114.0$\pm$14.1 & \nodata & 260.4$\pm$134.8 & \nodata & 97.9$\pm$1.4 & 3$\times$3 \\
\rownumber & NGC891 & \nodata & \nodata & 101.8$\pm$3.8 & \nodata & 159.4$\pm$1.8 & \nodata & \nodata & 3$\times$3 \\

\enddata

\tablecomments{Sigma values derived from the Gaussian fit to the profile of the far-IR fine structure lines observed by \textit{Herschel}/PACS. Table\,\ref{tbl_pacs_sigma} is published in its entirety in the machine readable format. A portion is shown here for guidance regarding its form and content.}

\end{deluxetable}


\floattable
\begin{deluxetable}{rlccccccccccl}
\rotate
\renewcommand{\thefootnote}{\fnsymbol{footnote}}
\tabletypesize{\footnotesize}
\tablecolumns{13} 
\tablewidth{0pt}
\tablecaption{Mid-IR fine structure line fluxes from \textit{Spitzer}/IRS.\label{tbl_irs_flux}}
\tablehead{
 & & \multicolumn{5}{c}{Line fluxes (10$^{-17}\, \rm{W\,m^{-2}}$) in SH ($4\farcs7 \times 11\farcs3$)} & \multicolumn{4}{c}{Line fluxes (10$^{-17}\, \rm{W\,m^{-2}}$) in LH ($11\farcs1 \times 22\farcs3$)} & Mode & Ref.\tablenotemark{a}\\
  \cmidrule(r){3-7} \cmidrule(l){8-11}
\colhead{\#} & \colhead{Name} & \colhead{\siv} & \colhead{\neii}  & \colhead{\nev} & \colhead{\neiii} & \colhead{\siii} & \colhead{\nev} & \colhead{\oiv} & \colhead{\siii}   & \colhead{\,\hbox{[\ion{Si}{2}]}}  & \colhead{} \\
 & & $10.51\, \rm{\micron}$ & $12.81\, \rm{\micron}$ & $14.32\, \rm{\micron}$  & $15.56\, \rm{\micron}$  & $18.71\, \rm{\micron}$ & $24.32\, \rm{\micron}$  & $25.89\, \rm{\micron}$ & $33.48\, \rm{\micron}$ & $34.82\, \rm{\micron}$ & }

\startdata
\multicolumn{13}{c}{\bf AGN}  \\
\cline{1-13}
\setcounter{magicrownumbers}{0}
\rownumber &   Mrk334	    	 &  	  11.0$\pm$2.0	 &  30.0$\pm$8.0     &  13.0$\pm$2.0    &    26.0$\pm$3.0    &	23.0$\pm$5.0    & 11.0$\pm$2.0   &   15.0$\pm$3.0   &   90.0$\pm$8.0 &    \nodata	  & LR    & 11  		 \\
\rownumber &   Mrk938    	      	 &  	    $<$1.46      &  52.10$\pm$1.45    &  $<$2.19	        &    6.37$\pm$0.55   &	7.56$\pm$0.64    & $<$0.37	  &   $<$0.66	     &   $<$10.7      &    40.50$\pm$4.21  & HR+LR & 41,11		 \\
\rownumber &   IRAS00182-7112   	 &  	    $<$0.08      &   6.3$\pm$0.3     &  $<$0.13	        &    2.1 $\pm$0.2    &	0.23$\pm$0.03    & $<$0.20	  &   $<$0.35	     &    \nodata     &    \nodata	  & HR    & 39  		 \\

\cline{1-13}  \\
\multicolumn{13}{c}{\bf Starburst galaxies}  \\
\cline{1-13}
\setcounter{magicrownumbers}{0}
\rownumber & NGC253  & $<$10.5 & 2832.3$\pm$64.2 & $<$20.5 & 204.6$\pm$9.6 & 666.4$\pm$14.9 & $<$73.36 & 154.7$\pm$26.9 & 1538.0$\pm$30.1 & 2412.0$\pm$48.0 & HR & 5 \\
\rownumber & M74     & 0.22 & 2.31$\pm$0.36 & $<$0.15 & $<$0.08 & 0.23 & $<$0.46 & $<$0.13 & 12.35$\pm$0.94 & 4.78$\pm$0.40 & HR & 20,47 \\
\rownumber & NGC891  & 0.88 & 8.57$\pm$0.78 & $<$0.04 & 0.84$\pm$0.07 & 1.99$\pm$0.17 & $<$0.62 & $<$1.03 & 10.74$\pm$2.02 & 28.11$\pm$0.92 & HR & 20,47 \\

\cline{1-13} \\
\multicolumn{13}{c}{\bf No lines detected by PACS}  \\
\cline{1-13}
\setcounter{magicrownumbers}{0}
\rownumber &   3C33	            	 &  	   1.2$\pm$0.1   &  3.9$\pm$0.2     &   2.0$\pm$0.3   &    5.3$\pm$0.2    &	2.5$\pm$0.4     &  1.6$\pm$0.2  &   8.1$\pm$0.2   &   \nodata      &    \nodata	  & LR    & 24  		 \\
\rownumber &   PG1114+445       	 &	0.90$\pm$0.18   &     \nodata       &     \nodata  	&      $<$1.53       &       \nodata     &  $<$1.70	  &   2.15$\pm$0.49  &     \nodata    &        \nodata    & LR    & 47  		 \\
\rownumber &   ESO506-G27       	 &	$<$1.02   	 &  5.18$\pm$1.25    &     \nodata  	&    6.02$\pm$0.89   &    1.99$\pm$1.12  &     \nodata    &   3.80$\pm$0.71  &     \nodata    &     \nodata	  & LR    & 47,35		 \\

\enddata

\tablenotetext{a}{
References are coded as follows:
(1)\,\citet{alo12}; 
(2)\,\citet{arm04}; 
(3)\,\citet{arm06}; 
(4)\,\citet{arm07}; 
(5)\,\citet{bs09}; 
(6)\,\citet{bre06}; 
(7)\,\citet{dal09}; 
(8)\,\citet{das08}; 
(9)\,\citet{das11}; 
(10)\,\citet{deo06}; 
(11)\,\citet{deo07}; 
(12)\,\citet{d-s09}; 
(13)\,\citet{dic12}; 
(14)\,\citet{dj11}; 
(15)\,\citet{don11}; 
(16)\,\citet{dud07}; 
(17)\,\citet{dud09}; 
(18)\,\citet{far07}; 
(19)\,\citet{gor07}; 
(20)\,\citet{ga09}; 
(21)\,\citet{gui12}; 
(22)\,\citet{ina13}; 
(23)\,\citet{ker09}; 
(24)\,\citet{ogl06}; 
(25)\,\citet{ogl10}; 
(26)\,\citet{p-s10}; 
(27)\,\citet{p-s10b}; 
(28)\,\citet{p-b11}; 
(29)\,\citet{per07}; 
(30)\,\citet{pet11}; 
(31)\,\citet{pri12}; 
(32)\,\citet{ram13}; 
(33)\,\citet{rou07}; 
(34)\,\citet{sal10}; 
(35)\,\citet{sar11}; 
(36)\,\citet{sat07}; 
(37)\,\citet{sch06}; 
(38)\,\citet{s15}; 
(39)\,\citet{spo09}; 
(40)\,\citet{tom08}; 
(41)\,\citet{tom10}; 
(42)\,Tommasin upublished; 
(43)\,\citet{vei09}; 
(44)\,\citet{wea10}; 
(45)\,\citet{wil10}; 
(46)\,\citet{wil11}; 
(47)\,This work.\\ 
Polarimetry:
(48)\,\citet{ale00}; 
(49)\,\citet{mor01}; 
(50)\,\citet{per03}. 
}
\tablecomments{Mid-IR fine structure line fluxes for the galaxies in our sample from \textit{Spitzer}/IRS observations. Table\,\ref{tbl_irs_flux} is published in its entirety in the machine readable format. A portion is shown here for guidance regarding its form and content.}
\end{deluxetable}

\clearpage

\begin{deluxetable}{lccccccccccl}
\renewcommand{\thefootnote}{\fnsymbol{footnote}}
\tabletypesize{\footnotesize}
\tablecolumns{9} 
\tablewidth{0pt}
\tablecaption{\nii$_{205\, \rm{\micron}}$ and \ci$_{371, 609\, \rm{\micron}}$ line fluxes.\label{tbl_n2c1}}
\tablehead{
  &  & & & & \multicolumn{3}{c}{Line fluxes ($10^{-17}\, \rm{W\,m^{-2}}$)} &   \\ 
  \cline{6-8}
\colhead{Name} & \colhead{R.A.} & \colhead{Dec.} & \colhead{Redshift} & \colhead{Class} & \colhead{\nii} & \colhead{\ci} & \colhead{\ci} & \colhead{RefNC\tablenotemark{a}} \\ [1pt]
 & & & & & $205.2\, \rm{\micron}$ & $371\, \rm{\micron}$ & $609\, \rm{\micron}$ & }

\startdata
Mrk938 		& 00:11:06.5412 & -12:06:27.6624 & 0.019617 & S2 & 3.47$\pm$0.24 & 1.16$\pm$0.13 & 0.81$\pm$0.26 & 6 \\
Haro11 		& 00:36:52.4544 & -33:33:16.7652 & 0.020598 & Dwarf & 2.31$\pm$0.46 & $<$0.81 & \nodata & 6 \\
NGC253 		& 00:47:33.0727 & -25:17:18.9960 & 0.000811 & \textsc{H\,ii} & 175.93$\pm$11.21 & 107.82$\pm$2.70 & 42.69$\pm$2.13 & 6 \\
IZw1 		& 00:53:34.9236 & +12:41:35.9232 & 0.0612 & S1n & 1.11$\pm$0.19 & $<$0.55 & $<$1.17 & 6 \\
IRAS01003-2238  & 01:02:49.9894 & -22:21:57.2616 & 0.117835 & S2 & \nodata & $<$0.40 & \nodata & 6 \\
IIIZw35 	& 01:44:30.5400 & +17:06:08.8056 & 0.027436 & S2 & $<$1.07 & 0.44$\pm$0.10 & $<$0.63 & 6 \\
Mrk1014 	& 01:59:50.2483 & +00:23:40.7436 & 0.16311 & S1.5 & 0.46$\pm$0.15 & $<$0.63 & \nodata & 6 \\
NGC891 		& 02:22:32.8464 & +42:20:52.6740 & 0.001761 & \textsc{H\,ii} & 78.46$\pm$1.95 & 4.93$\pm$0.40 & 3.60$\pm$0.55 & 6 \\
NGC1068 	& 02:42:40.7071 & -00:00:48.0204 & 0.003793 & S1h & 186.65$\pm$6.83 & 29.65$\pm$0.81 & 14.04$\pm$0.69 & 6 \\
NGC1097 	& 02:46:18.9775 & -30:16:28.9344 & 0.00424 & LINb & 92.11$\pm$2.44 & 6.68$\pm$0.26 & 4.07$\pm$0.47 & 6 \\

\enddata

\tablenotetext{a}{References are coded as follows:
(1)\,\citet{gp00}; 
(2)\,\citet{ib02}; 
(3)\,\citet{isr09}; 
(4)\,\citet{isr15}; 
(5)\,\citet{kam14}; 
(6)\,\citet{kam15}; 
(7)\,\citet{rig13} 
}
\tablecomments{Line fluxes for nitrogen \nii$_{205\, \rm{\micron}}$ and neutral carbon \ci$_{371, 609\, \rm{\micron}}$ lines in our sample, collected from \textit{Herschel}/SPIRE and ground-based observations published in the literature. Table\,\ref{tbl_n2c1} is published in its entirety in the machine readable format. A portion is shown here for guidance regarding its form and content.}

\end{deluxetable}

\clearpage

\begin{deluxetable}{lccccccccc}
\tabletypesize{\footnotesize}
\tablecolumns{10} 
\tablewidth{0pt}
\tablecaption{Density determinations.\label{tbl_denstrat}}
\tablehead{
\colhead{Name} & \colhead{Class} & \colhead{\nii} & \colhead{$\log(n_{\rm e})$} & \colhead{\siii} & \colhead{$\log(n_{\rm e})$} & \colhead{\oiii} & \colhead{$\log(n_{\rm e})$} & \colhead{\nev} & \colhead{$\log(n_{\rm e})$} \\ [1pt]
 &  & 205.18/121.89 & $\rm{cm^{-3}}$  & 33.48/18.71 & $\rm{cm^{-3}}$ & 88.36/51.82 & $\rm{cm^{-3}}$ & 24.32/14.32 & $\rm{cm^{-3}}$}

\startdata
Haro11 & Dwarf & 0.66$\pm$0.18 & 1.49 (1.25-1.75) & \nodata & \nodata & \nodata & \nodata & \nodata & \nodata \\
Mrk334 & S1.8 & \nodata & \nodata & 3.91$\pm$1.20 & $<$1.0 & \nodata & \nodata & 0.85$\pm$0.28 & 3.30 (2.60-3.75) \\
IRAS00198-7926 & S2 & \nodata & \nodata & 3.17$\pm$0.49 & $<$1.0 & \nodata & \nodata & 0.93$\pm$0.05 & 3.15 (3.04-3.25) \\
ESO012-G21 & S1.5 & \nodata & \nodata & 1.99$\pm$0.26 & $<$1.0 & \nodata & \nodata & 1.45$\pm$0.15 & $<$2.0 \\
Mrk348 & S1h & \nodata & \nodata & 1.74$\pm$0.30 & $<$2.23 & \nodata & \nodata & 0.85$\pm$0.11 & 3.29 (3.08-3.47) \\
IRAS00521-7054 & S1h & \nodata & \nodata & \nodata & \nodata & \nodata & \nodata & 0.42$\pm$0.07 & 4.00 (3.88-4.13) \\
ESO541-IG12 & S2 & \nodata & \nodata & \nodata & \nodata & \nodata & \nodata & 0.52$\pm$0.21 & 3.81 (3.48-4.22) \\
3C33 & S1h & \nodata & \nodata & \nodata & \nodata & \nodata & \nodata & 0.80$\pm$0.22 & 3.38 (2.95-3.72) \\
NGC454E & S2 & \nodata & \nodata & \nodata & \nodata & \nodata & \nodata & 1.49$\pm$0.14 & $<$2.0 \\
IRAS01364-1042 & LIN & \nodata & \nodata & 13.19$\pm$3.11 & $<$1.0 & \nodata & \nodata & \nodata & \nodata \\

\enddata

\tablecomments{Density determinations from fine-structure emission lines from \textit{Herschel}/PACS and SPIRE, and \textit{Spitzer}/IRS observations. Table\,\ref{tbl_denstrat} is published in its entirety in the machine readable format. A portion is shown here for guidance regarding its form and content.}

\end{deluxetable}

\clearpage

\begin{deluxetable}{rlcccccccc}
\tabletypesize{\scriptsize}
\setlength{\tabcolsep}{0.3em}
\tablecolumns{10} 
\tablewidth{0pt}
\tablecaption{Central pixel, $3 \times 3$ array, and $5 \times 5$ array line fluxes.\label{tbl_pacs_flux_c35}}
\tablehead{
 & & \multicolumn{7}{c}{Line fluxes ($10^{-17}\, \rm{W\,m^{-2}}$)} & \\
\cline{3-9}
\colhead{\#} & \colhead{Name} & \colhead{\oiii} & \colhead{\niii}  & \colhead{\oi} & \colhead{\oiii} & \colhead{\nii} & \colhead{\oi} & \colhead{\cii} & \colhead{Notes} \\ [1pt]
 & & $51.81\, \rm{\micron}$ & $57.32\, \rm{\micron}$ & $63.18\, \rm{\micron}$ & $88.36\, \rm{\micron}$ & $121.90\, \rm{\micron}$ & $145.52\, \rm{\micron}$ & $157.74\, \rm{\micron}$ &}

\startdata
\multicolumn{10}{c}{\bf AGN} \\
\cline{1-10}
\setcounter{magicrownumbers}{0}
\rownumber & Mrk334 & \nodata & \nodata & \nodata & \nodata & \nodata & \nodata & 66.56$\pm$0.88 & c \\
 &  & \nodata & \nodata & \nodata & \nodata & \nodata & \nodata & 76.66$\pm$1.77 & 3$\times$3 \\
 &  & \nodata & \nodata & \nodata & \nodata & \nodata & \nodata & 68.98$\pm$1.37 & 5$\times$5 \\
\rownumber & Mrk938 & \nodata & \nodata & 65.41$\pm$4.45 & 12.58$\pm$2.38 & \nodata & 5.26$\pm$0.38 & 65.95$\pm$1.86 & c \\
 &  & \nodata & \nodata & 78.12$\pm$6.36 & 19.24$\pm$3.58 & \nodata & 5.26$\pm$0.92 & 82.15$\pm$1.51 & 3$\times$3 \\
 &  & \nodata & \nodata & 71.14$\pm$5.27 & 21.49$\pm$5.71 & \nodata & 3.88$\pm$1.13 & 78.59$\pm$2.67 & 5$\times$5 \\
\rownumber & IRAS00182-7112 & \nodata & \nodata & 7.10$\pm$1.67 & $<$2.04 & \nodata & \nodata & \nodata & c \\
 &  & \nodata & \nodata & $<$9.73 & $<$1.5 & \nodata & \nodata & \nodata & 3$\times$3 \\
 &  & \nodata & \nodata & 10.49$\pm$6.69 & $<$2.12 & \nodata & \nodata & \nodata & 5$\times$5 \\

\cline{1-10} \\
\multicolumn{10}{c}{\bf Starburst galaxies} \\
\cline{1-10}
\setcounter{magicrownumbers}{0}
\rownumber & NGC253 & $<$478.9 & 408.33$\pm$121.33 & 1409.54$\pm$70.02 & 332.54$\pm$48.72 & 413.11$\pm$41.42 & 181.22$\pm$13.58 & 1277.46$\pm$18.81 & c \\
 &  & 389.21$\pm$183.61 & 599.54$\pm$178.33 & 2947.62$\pm$85.69 & 625.21$\pm$53.54 & 802.72$\pm$28.14 & 389.61$\pm$30.99 & 3392.69$\pm$22.89 & 3$\times$3 \\
 &  & 438.08$\pm$169.61 & 626.13$\pm$150.87 & 3214.70$\pm$94.83 & 678.20$\pm$47.67 & 824.30$\pm$24.25 & 393.91$\pm$36.99 & 3985.12$\pm$25.32 & 5$\times$5 \\
\rownumber & M74 & \nodata & \nodata & 10.63$\pm$2.16 & $<$4.55 & $<$3.38 & \nodata & 6.31$\pm$0.93 & c \\
 &  & \nodata & \nodata & 20.58$\pm$10.84 & $<$14.16 & 5.85$\pm$2.70 & \nodata & 43.12$\pm$1.82 & 3$\times$3 \\
 &  & \nodata & \nodata & $<$46.58 & $<$10.16 & 12.60$\pm$2.57 & \nodata & 80.18$\pm$2.35 & 5$\times$5 \\
\rownumber & NGC891 & \nodata & \nodata & 18.82$\pm$6.38 & \nodata & 19.35$\pm$0.80 & \nodata & \nodata & c \\
 &  & \nodata & \nodata & 85.54$\pm$11.08 & \nodata & 73.28$\pm$2.19 & \nodata & \nodata & 3$\times$3 \\
 &  & \nodata & \nodata & 139.03$\pm$10.01 & \nodata & 103.16$\pm$5.08 & \nodata & \nodata & 5$\times$5 \\

\cline{1-10} \\
\multicolumn{10}{c}{\bf No lines detected by PACS} \\
\cline{1-10}
\setcounter{magicrownumbers}{0}
\rownumber & 3C33 & \nodata & \nodata & \nodata & \nodata & \nodata & $<$0.7 & \nodata & c \\
 &  & \nodata & \nodata & \nodata & \nodata & \nodata & $<$1.54 & \nodata & 3$\times$3 \\
 &  & \nodata & \nodata & \nodata & \nodata & \nodata & $<$1.54 & \nodata & 5$\times$5 \\
\rownumber & PG1114+445 & \nodata & \nodata & $<$1.71 & \nodata & \nodata & \nodata & $<$0.9 & c \\
 &  & \nodata & \nodata & $<$3.06 & \nodata & \nodata & \nodata & $<$1.3 & 3$\times$3 \\
 &  & \nodata & \nodata & $<$5.85 & \nodata & \nodata & \nodata & $<$2.18 & 5$\times$5 \\
\rownumber & ESO506-G27 & \nodata & \nodata & \nodata & \nodata & \nodata & $<$0.96 & \nodata & c \\
 &  & \nodata & \nodata & \nodata & \nodata & \nodata & $<$0.96 & \nodata & 3$\times$3 \\
 &  & \nodata & \nodata & \nodata & \nodata & \nodata & $<$1.9 & \nodata & 5$\times$5 \\

\enddata

\tablecomments{Far-infrared fine structure line fluxes of the active, starburst, and dwarf galaxies observed by \textit{Herschel}/PACS, extracted from the central pixel, the $3 \times 3$ array, and the full $5 \times 5$ array. Note that no point-source correction has been applied to the $5 \times 5$ fluxes. Table\,\ref{tbl_pacs_flux_c35} is published in its entirety in the machine readable format. A portion is shown here for guidance regarding its form and content.}
\end{deluxetable}

\clearpage

\appendix

\section{Aperture effects}\label{app}

\begin{figure}[h]
  \centering
  \includegraphics[width=0.7\textwidth]{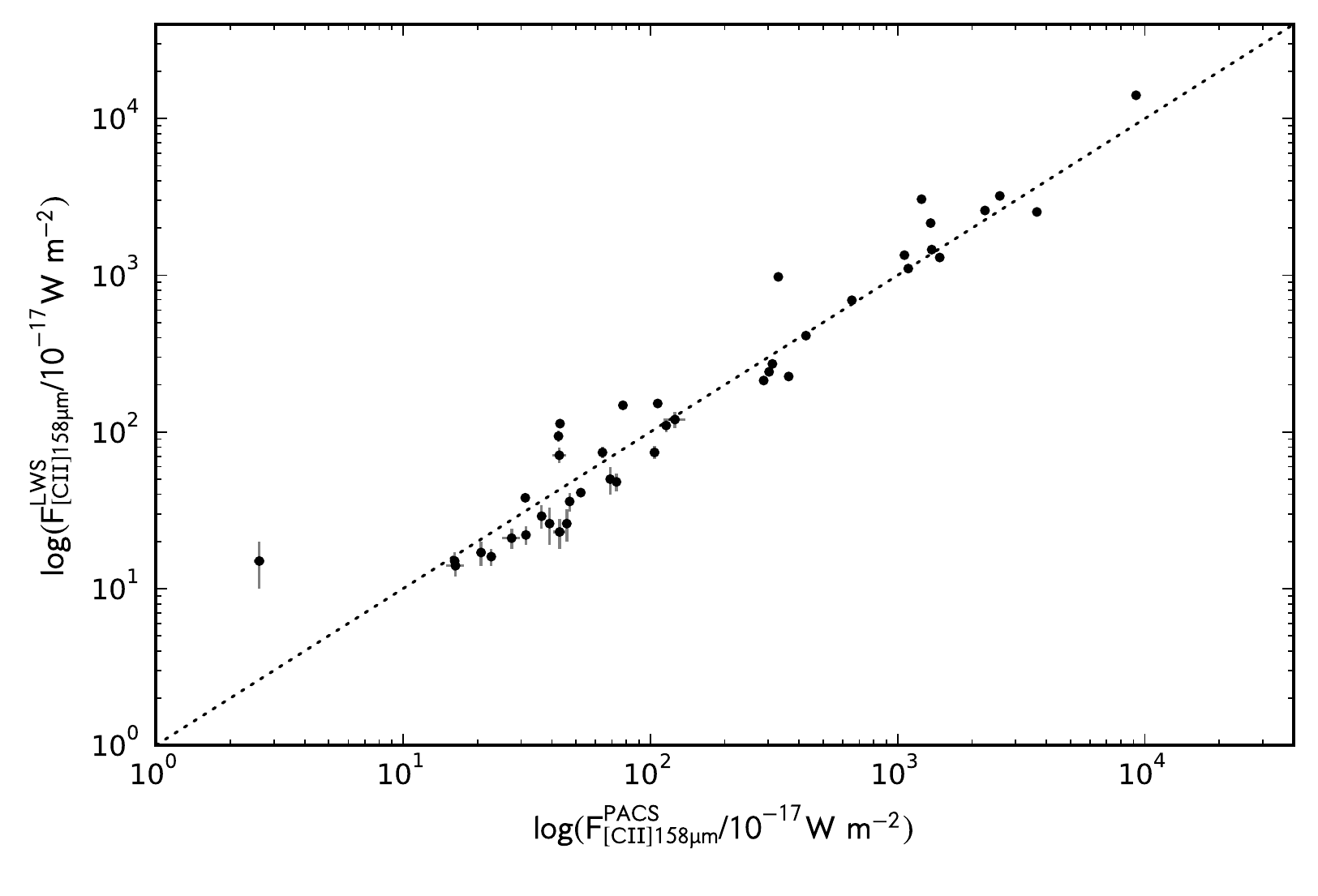}
\caption{[\textsc{C\,ii}]$_{158\, \rm{\micron}}$ line flux for the subsample of AGN and starburst galaxies measured with both \textit{Herschel}/PACS in this work and \textit{ISO}/LWS in \citet{bra08}. The dotted line corresponds to the ratio $F^{\rm LWS}_{\rm [CII]158}/F^{\rm PACS}_{\rm [CII]158} = 1.0$.\label{fig:pacs_iso}}
\end{figure}

In this section we test the possible aperture effects between \textit{Herschel}/PACS and \textit{ISO}/LWS line flux measurements reported in \citet{bra08}. To illustrate this we show in Fig.\,\ref{fig:pacs_iso} the comparison of PACS {\it vs} LWS measurements for the \cii$_{158\, \rm{\micron}}$ line, which is the worst case scenario in terms of the contribution of the extended emission from the galaxy, as can be seen in the line maps of Fig.\,Set\,\ref{fig_mapIC4329A}. For a subsample of 45 AGN and starburst galaxies with both PACS and LWS \cii$_{158\, \rm{\micron}}$ measurements, we find a median value and a median absolute deviation of $F^{\rm LWS}_{\rm [CII]158}/F^{\rm PACS}_{\rm [CII]158} = 0.95 \pm 0.37$. Despite the smaller aperture involved, PACS fluxes are $\sim 5\%$ higher than LWS fluxes, which suggests that most of the emission in our sample is compact within the $3 \times 3$ spaxel aperture ($28\farcs2 \times 28\farcs2$).

\begin{figure*}[b]
  \centering
  \includegraphics[width=0.8\textwidth]{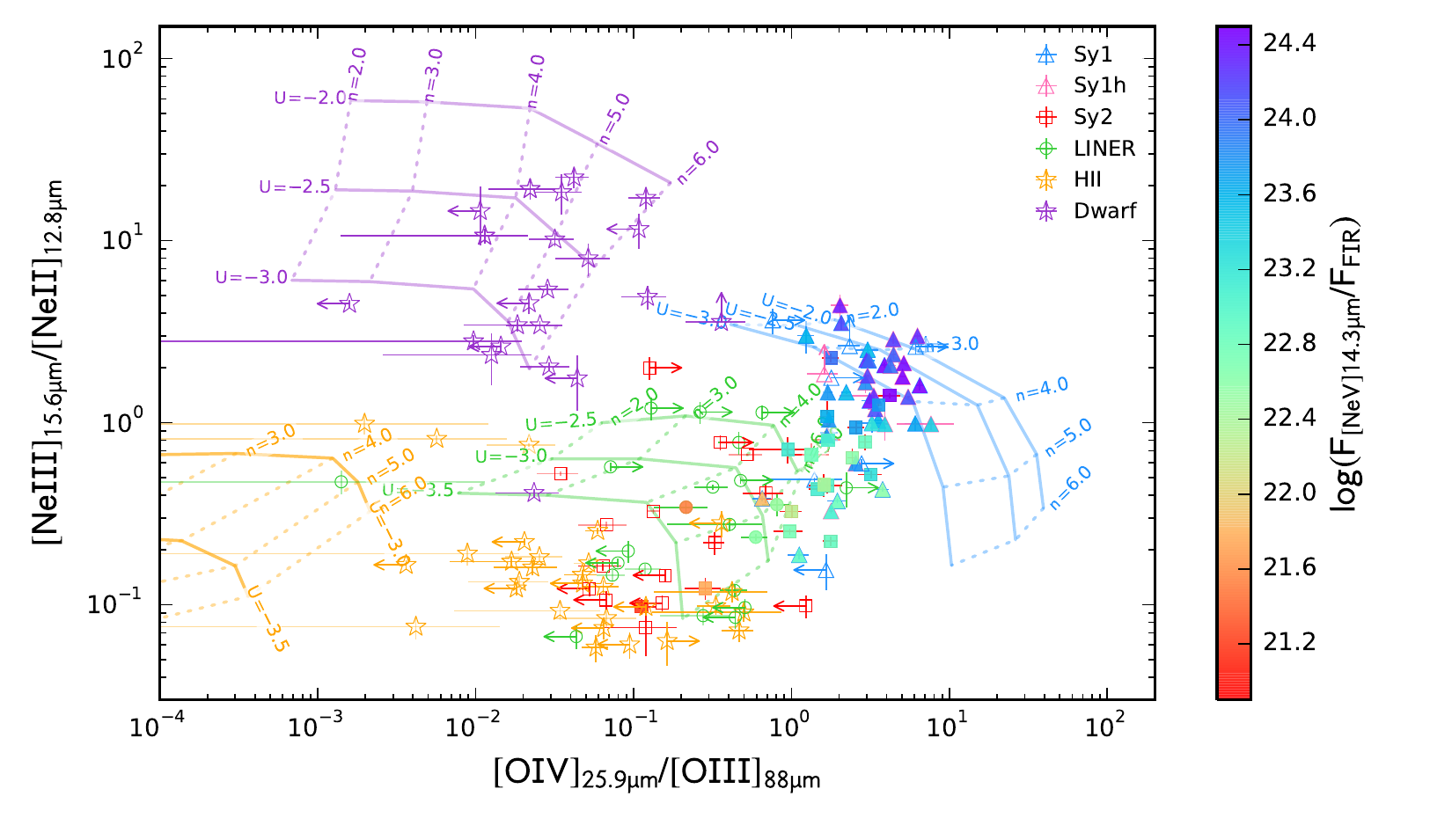}
\caption{[Ne\,\textsc{iii}]$_{15.6}$/[Ne\,\textsc{ii}]$_{12.8}$ line ratio {\it vs} the [\textsc{O\,iv}]$_{25.9}$/[\textsc{O\,iii}]$_{88}$ ratio (same notations as in Fig.\,\ref{fig:o4_o3vsne3_ne2}) using fluxes extracted from the \textit{Herschel}/PACS central spaxel for the [\textsc{O\,iii}]$_{88\, \rm{\micron}}$ line. Symbols are colour-coded according to their $\rm F_{\rm [Ne\,V]14.3}/F_{\rm FIR}$ flux ratio, when available (see colour bar). \label{fig:o4_o3vsne3_ne2_C}}
\end{figure*}

On the other hand, in order to show the impact of using different apertures in line ratios involving \textit{Spitzer}/IRS and \textit{Herschel}/PACS lines, we reproduce in Fig.\,\ref{fig:o4_o3vsne3_ne2_C} the same diagram shown in Fig.\,\ref{fig:o4_o3vsne3_ne2}, but using only central spaxel fluxes for the \oiii$_{88\, \rm{\micron}}$ line. In Fig.\,\ref{fig:o4_o3vsne3_ne2} the \oiv$_{25.9\, \rm{\micron}}$ is extracted from a $11\farcs1 \times 22\farcs3$ aperture, that is a factor of $\sim 3$ smaller when compared to the \oiii$_{88\, \rm{\micron}}$ aperture ($28\farcs2 \times 28\farcs2$), thus extended \oiii$_{88\, \rm{\micron}}$ emission could lower the line \oiv$_{25.9}$/\oiii$_{88}$ line ratio. Fig.\,\ref{fig:o4_o3vsne3_ne2_C} shows that, by using central spaxel fluxes in \oiii$_{88\, \rm{\micron}}$, thus an aperture $\sim 3$ times smaller than in the \oiv$_{25.9\, \rm{\micron}}$ line, the variations in the diagram with respect to Fig.\,\ref{fig:o4_o3vsne3_ne2} are within the errors for the majority of the galaxies. Therefore, we conclude that aperture effects in this study are not relevant and do not affect our results.

\end{document}